\documentclass[prd,twocolumn,amssymb,nofootinbib,12,showpacs]{revtex4-1}
\pdfoutput=1
\usepackage{units}
\usepackage[xcolor=x11names]{xcolor}
\usepackage{amsmath}
\usepackage{amssymb}
\usepackage{stmaryrd}
\usepackage{esint}
\usepackage{filecontents}
\usepackage{siunitx}
\usepackage{comment}
\usepackage[bottom]{footmisc}
\usepackage{diagbox}
\usepackage{float}
\usepackage[justification=raggedright,
  singlelinecheck=on]{caption}
\restylefloat{table}
\usepackage[unicode=true,pdfusetitle,
 bookmarks=true,bookmarksnumbered=false,bookmarksopen=false,
 breaklinks=false,pdfborder={0 0 0},backref=false,colorlinks=true,linkcolor=blue!50!black,urlcolor=red!50!black,citecolor=blue!50!black]
 {hyperref}

\makeatother
\begin{document}
\title{Astrophysical Gravitational Waves in Conformal Gravity}
\author{Chiara Caprini} \email{caprini@apc.in2p3.fr} \affiliation{Laboratoire Astroparticule et Cosmologie, CNRS UMR 7164, Universit\'e
Paris-Diderot, 10 rue Alice Domon et L\'eonie Duquet, 75013 Paris, France} 
\author{Patric H\"olscher} \email{patric.hoelscher@physik.uni-bielefeld.de} 
\author{Dominik J. Schwarz} \email{dschwarz@physik.uni-bielefeld.de} 
\affiliation{Fakult\"at f\"ur Physik, Universit\"at Bielefeld, Postfach 100131, 33501 Bielefeld, Germany}
\begin{abstract}
We investigate the gravitational radiation from binary systems
in conformal gravity (CG) and massive conformal gravity (MCG). 
CG might explain observed galaxy rotation curves without dark matter, and 
both models are of interest in the context of quantum gravity. 
Here we show that gravitational radiation emitted 
by compact binaries allows us to strongly constrain both models. 

We work in Weyl gauge,
which fixes the rescaling invariance of the models, and derive the linearized fourth-order 
equation of motion for the metric, which describes massless and massive modes of propagation. 
In the limit of a large graviton mass, MCG reduces to general relativity (GR), whereas CG does 
not. Coordinates are fixed by Teyssandier gauge to show that for a conserved energy-momentum
tensor the gravitational radiation is due to the time-dependent quadrupole moment of a 
non-relativistic source and we derive the gravitational energy-momentum tensor for both models. 
We apply our findings to the case of close binaries on circular orbits, which have been used to 
indirectly infer the existence of gravitational radiation prior to the direct observation of 
gravitational waves. 
As an example, we analyze the binary system PSR J1012+5307, chosen for its small eccentricity. 
When fixing the graviton mass in CG such that observed galaxy rotation curves could be explained 
without dark matter,  the gravitational radiation of a binary system is much smaller than in GR. 
The same holds for MCG for small masses of the graviton. Thus gravitational radiation cannot 
explain the orbital decay of binary systems and replace dark matter simultaneously. 
We also analyse MCG for large graviton masses and conclude that MCG can describe the 
orbital periods of compact binaries in agreement with data, as it reduces to GR in that limit.
\end{abstract}

\maketitle

\section{Introduction\label{sec:Introduction}}

The standard model of gravity, general relativity (GR), is tested very well.  
The equivalence principle has been probed for a large region of the relevant 
parameter space and GR passes all Solar system tests (see e.g.~\cite{Will2014}).
Also the orbits of relativistic compact binaries show no deviation from the GR prediction 
and provide indirect evidence for the existence of gravitational waves
\cite{2004WeisbergTaylorRelativisticbinarypulsarapa,2006Sci...314...97K}. 

In 2015, the aLIGO interferometers recorded the first direct observation of gravitational waves, as 
they observed the very last moment of a binary black-hole merger 
\cite{abbott2016observation,2016_Abbott_GW151226_Observationofgravitationalwavesfroma22-solar-massbinaryblackholecoalescence_PRL}.  
So far, five binary black hole mergers have been reported by the LIGO/VIRGO collaboration \cite{2017_Scientific_GW170104_Observationofa50-Solar-MassBinaryBlackHoleCoalescenceatRedshift0.2_PRL,2017_Abbott_GW170608_Observationofa19-solar-massBinaryBlackHoleCoalescence_,2017_Abbott_GW170814_Athree-detectorobservationofgravitationalwavesfromabinaryblackholecoalescence_PRL}. 
Very recently, the aLIGO and VIRGO interferometers detected a gravitational wave signal from the merger of two 
neutron stars (GW170817) with follow-up measurements across the 
electromagnetic spectrum coming from GRB 170817A \cite{2017_Abbott_GW170817_PRL,2017_Abbott_Multi-messengerobservationsofabinaryneutronstarmerger_AJL}. Strong constraints on the speed of 
gravitational waves follow from the detected difference in arrival time of the gravitational the 
electromagnetic signal, which in turn allows to constrain modified models of gravity
\cite{2017_Abbott_GravitationalWavesandGammaRaysfromaBinary,
2017_Lombriser_ChallengestoSelfAccelerationinModifiedGravity,
2017_Baker_StrongConstraintsonCosmologicalGravityfromGW170817andGRB170817A_Prl,2017_Creminelli_DarkEnergyAfterGW170817andGRB170817A_PRL,2017_Ezquiaga_DarkEnergyAfterGW170817DeadEndsandtheRoadAhead_PRL,
2017_Sakstein_ImplicationsoftheNeutronStarMergerGW170817forCosmologicalScalarTensorTheories_Prl,
2018_Nersisyan_GravitationalWaveSpeedImplicationsforModelswithoutaMassScale_,
2018_Akrami_NeutronStarMergerGW170817StronglyConstrainsDoublyCoupledBigravity_apa}.

But GR also faces shortcomings. We do not understand how to combine quantum physics 
with the principles of GR. In the ultra-violet regime, GR does not lead to a renormalizable model and 
therefore is non-predictive. In the infrared regime, cosmological and astrophysical observations 
interpreted in the context of GR imply the existence of dark energy and dark matter, and 
the observed smallness of the cosmological constant is not understood.

This motivates our study of modified models of gravity that change the gravitational field equations, 
instead of introducing a dark sector. Most physical models employ second-order 
equations of motion, which insures that the theory is free from Ostrogradski instabilities 
\cite{1850OstrogradskiMASP,2007WoodardAvoidingDarkEnergy}. 
However higher-derivative theories, albeit suffering from ghost instabilities, can improve renormalization issues. 

In this work we consider a unique higher-order derivative theory of gravity: a 
conformal model that reduces either to conformal gravity (CG) or to massive conformal gravity (MCG). 
The difference between those models is encoded in a parameter $\epsilon$, with $\epsilon = -1$ corresponding to 
CG and $\epsilon = +1$ to MCG (see Sec.~\ref{sec:Conformal-Gravity} for details). 
These models are not only invariant under general coordinate transformations, but also under Weyl rescaling 
of the metric and the matter fields. The purpose of this work is to study
gravitational waves in conformal models of gravity. 

Conformal models of
gravity have been considered for the first time shortly after the introduction of GR, especially CG by 
Weyl and Bach \citep{1919WeylEineneueErweiterungAdPL,1921BachZurWeylschenrelativitatstheorieMZ}.
The approach of Weyl has been dropped briefly after its publication, because of non-integrability. On 
the other hand, the theory of Bach has been the precursor of CG, introduced by Mannheim and Kazanas 
\citep{1989MannheimKazanasExactvacuumsolutionTAJ,1990MannheimConformalcosmologynoGRaG,1991KazanasMannheimGeneralstructuregravitationalTAJSS,1991MannheimKazanasSolutionsReissnerNordstromPRD,1993MannheimLinearpotentialsgalacticapa,Mannheim2006,Mannheim2011}.  
More recently 't Hooft, who considered a non-perturbative
approach of the path-integral formalism for quantum gravity, found
connections between GR and CG. Terms of the same form as in the Weyl action appear as the 
only divergent term after a dimensional regularization \citep{2010HooftConformalConstraintCanonical,2010HooftProbingsmalldistance,2011HooftclasselementaryparticleFP,mannheim2012making}.
Maldacena considered CG as a possible UV-completion of GR by using
specific boundary conditions, which separate out the Einstein-Hilbert
solutions from the larger set of solutions in CG \citep{2011MaldacenaEinsteinGravityConformal}.

The early approach to CG has been discarded. Firstly, the fourth-order structure 
of the theory made it mathematically uncomfortable. Secondly, from an experimental point of view 
there was no need to modify GR. Last but not least, the theory did not allow for bare mass 
terms in the matter action and our every day experience is strongly 
against the concept of scaling invariance. 

It is now clear that masses in particle physics arise dynamically. In that light  
CG has been revived by Mannheim and Kazanas in 1989, masses arise only 
after a spontaneous breaking of the conformal symmetry 
\cite{Mannheim2006}. Besides that, Mannheim and Kazanas got some 
remarkable results, which made CG interesting again. CG was demonstrated to be 
renormalizable \cite{mannheim2012making,Mannheim2011} and
they solved the field equations for a static and spherically symmetric
system in the Newtonian limit. They found a modified Newtonian
potential which contains a term that grows linearly with distance
\cite{mannheim1989exact,1993MannheimLinearpotentialsgalacticapa,1994MannheimKazanasNewtonianlimitconformalGrag,2015DelidumanKasikciYapiskanFlatGalacticRotation}.
This modified potential makes it possible to fit rotation curves of a huge class
of galaxies \citep{1997MannheimAregalacticrotationTAJ,mannheim2012fitting,2013MannheimOBrienGalacticrotationcurves}.

It was shown that CG contains viable cosmological solutions, which fit the Hubble diagram and
solve the singularity and cosmological constant problems \cite{1990MannheimConformalcosmologynoGRaG,1992MannheimConformalgravityflatnessTAJ,1999_Mannheim_ConformalGravityandaNaturalSolutiontotheCosmologicalConstantProblem_apa,
Mannheim2006,2011_Diaferio_GammaRayBurstsAsCosmologicalProbes,2012MannheimCosmologicalperturbationsconformalPRD}. However, in \citep{2017_Roberts_TestsofLambdaCDMandConformalGravity} it has been argued that the 
$\Lambda$ cold dark matter model is favoured by data from gamma-ray bursts and quasars. 
Besides checking the Hubble diagram, much work is left to be done. There is no analysis of the cosmic 
microwave background yet. Primordial nucleosynthesis has been analyzed in conformal models of gravity \cite{1993KnoxKosowskyPrimordialnucleosynthesisconformal,1994_Elizondo_CanConformalWeylGravityBeConsideredaViableCosmologicalTheory} and it seems that 
there is a tension with the deuterium and lithium abundances, the latter being also at odds with the 
cosmological standard model. But most importantly, structure formation has not been investigated in any 
detail. 

Several authors claimed that light deflection is problematic in CG, but possibly there is a way out \citep{1998EderyParanjapeClassicaltestsWeylPRD,2001EderyMethotParanjapeGaugechoicegeodeticGRG,2010SultanaKazanasBendinglightconformalPRD,2012SultanaKazanasSaidConformalWeylgravityPRD,2015DelidumanKasikciYapiskanFlatGalacticRotation,2013CattaniScaliaLaserraBochicchioNandiCorrectlightdeflectionPRD,2016LimWangExactgravitationallensing,2017SultanaKazanasGaugeChoiceConformalapa,2017_Campigotto_ConformalGravity_LightDeflectionRevisitedandtheGalacticRotationCurveFailure_apa}. 
The work of Perlick and Xu \citep{1995_Perlick_MatchingExteriortoInteriorSolutionsinWeylGravity_Commenton``ExactVacuumSolutiontoConformalWeylGravityandGalacticRotationCurves_TAJ}
represented the major criticism on CG for a long time. They have shown that pure CG
without a Weyl invariant energy-momentum tensor of matter is ruled out and contributed to 
fundamental advances in the understanding of conformally invariant theories. For a detailed 
discussion of this work, see also
\citep{2001_Wood_Solutionsofconformalgravitywithdynamicalmassgenerationinthesolarsystem_apg}.

Previous investigations of gravitational waves in CG presented first steps like the linearization of the equations of 
motion and the calculation of the gravitational energy-momentum tensor in pure CG \citep{2014_Myung_PrimordialGravitationalWavesfromConformalGravity_apa,2011FabbriParanjapeMonochromaticplanefrontedPRD,2017_Yang}.

The present work goes well beyond these first studies. 
We include matter, discuss in depth the choice of gauge and we derive 
the formalism for analyzing the gravitational radiation by binary systems in CG and MCG.

Until recently, there were only indirect measurements of gravitational
radiation from pulsar binary systems, see e.g. \citep{2004WeisbergTaylorRelativisticbinarypulsarapa}. 
Gravitational radiation was indirectly detected through the measurement of the decreasing 
orbital period of the system. The recent direct detection, as already discussed above, and especially 
the observation of the merger of two neutron stars opens new possibilities to test GR and its alternatives.  
Several models of modified gravity like $f(R)$ gravity, Horndeski's theories, vector theories or bimetric theories 
have been tested and constrained \cite{2017_Abbott_GravitationalWavesandGammaRaysfromaBinary,2017_Baker_StrongConstraintsonCosmologicalGravityfromGW170817andGRB170817A_Prl,2017_Creminelli_DarkEnergyAfterGW170817andGRB170817A_PRL,2017_Ezquiaga_DarkEnergyAfterGW170817DeadEndsandtheRoadAhead_PRL,2018_Nersisyan_GravitationalWaveSpeedImplicationsforModelswithoutaMassScale_}.

Here we study linearized gravity for non-relativistic binaries, thus we can 
compare our findings to systems long before the merger. This allows us to demonstrate that CG, when fixing the free parameters to explain galaxy rotation 
curves, cannot at the same time reproduce the gravitational radiation from 
binaries (observed indirectly via their orbital period decay). For MCG there is a region of parameter space 
that is in concordance with observations. 

Similar analyses study generalizations of CG and MCG \cite{2015_Stabile_Post-MinkowskianLimitandGravitationalWavessolutionsofFourthOrderGravity_acompletestudy_apa,
2010_Nelson_GravitationalWavesintheSpectralActionofNoncommutativeGeometry,
2010_Nelson_ConstrainingtheNoncommutativeSpectralActionViaAstrophysicalObservations}. In these works an incomplete gravitational 
energy-momentum tensor has been used to calculate the radiated energy from a binary system. 
It is assumed that the expression for the radiated energy is approximately the same as in GR and 
hence the result differs significantly from ours. As we show, there are important additional contributions. 

Section \ref{sec:Conformal-Gravity} gives an introduction to CG and MCG with the basic assumptions and
equations. In section \ref{sec:Linearisation} we show how to obtain
the linearized field equations for the gravitational field and obtain their general solutions.   
We calculate the decay of the orbital period of coalescing binaries in the early inspiraling phase 
for CG and MCG in section \ref{sec:signal_from_a_binary} and in section 
\ref{sec:Energy-Momentum-Tensor-of} we derive the gravitational
energy-momentum tensor in CG and MCG. In section \ref{sec:Energy-Loss-Due-grav-rad}
we evaluate the radiated energy from a binary system, and in the last section we conclude 
and summarize our findings.

For the Weyl and Riemann tensor we use definitions and sign conventions of 
Weinberg \citep{weinberg1972gravitation}, see Appendix \ref{appendix A}. 
We use natural units in which $c = \hbar = 1$, unless stated otherwise. Greek letters 
denote spacetime indices ($0 \ldots 3$) and latin letters are spatial indices ($1 \ldots 3$).

\section{Conformal Gravity \label{sec:Conformal-Gravity}}

Conformal and massive conformal gravity are based on a Weyl invariant action. The spacetime metric $g_{\mu\nu}$ is rescaled by a Weyl transformation (conformal transformation) according to
\begin{equation}
g_{\mu\nu}(x) \rightarrow \Omega^{2}(x)g_{\mu\nu}(x),
	\label{eq: Weyl transformation}
\end{equation}
where $\Omega > 0$ is a real and smooth function called
the conformal factor and $x$ denotes the spacetime coordinates. To model gravity, 
the Einstein-Hilbert action is replaced by the Weyl action $I_{W}$ and the action for the Universe 
is given by
\begin{align}
I &= I_{W} + I_{M} \nonumber \\
& = -\alpha_{g}\int d^{4}x\sqrt{-g}\,C_{\lambda\mu\nu\kappa}C^{\lambda\mu\nu\kappa} + I_{M}
	\label{eq: Weyl action}\\
 & = -\alpha_{g}\int d^{4}x\sqrt{-g}\left[2\left(R_{\mu\kappa}R^{\mu\kappa}-\frac{1}{3}R^{2}\right) + L_L\right] + I_{M},
 	\label{eq:Weyl_action_gauss_bonnet}
\end{align}
where $I_M$ is the matter action. $\alpha_{g}$ is a dimensionless coupling constant, 
$g =\mbox{det}(g_{\mu\nu})$, and  $C_{\lambda\mu\nu\kappa}$, $R_{\mu\nu}$ and $R$ are 
the Weyl and Ricci tensors and the Ricci scalar, defined in Appendix \ref{appendix A}. 
To obtain expression \eqref{eq:Weyl_action_gauss_bonnet} the Gauss-Bonnet term 
(Lanczos lagrangian), which is a total derivative in four spacetime dimensions, has been used \citep{lanczos1938} 
\begin{equation}
\sqrt{-g}L_{L} = \sqrt{-g}(R_{\lambda\mu\nu\kappa} R^{\lambda\mu\nu\kappa} - 4R^{\mu\nu} R_{\mu\nu} + R^{2}),
\end{equation}
where $R_{\lambda\mu\nu\kappa}$ denotes the Riemann tensor.
Hence, it does not contribute to the field equations and can be discarded.
Let us note that it is forbidden to introduce a cosmological constant
term in Eq. \eqref{eq: Weyl action}, because of the
Weyl symmetry. 

The Weyl tensor has some outstanding properties. It is the
traceless part of the Riemann tensor 
\begin{equation}
g^{\mu\kappa} C_{\,\mu\nu\kappa}^{\lambda} = 0
\end{equation}
and under the transformation \eqref{eq: Weyl transformation}
it behaves like
\begin{align}
C_{\,\mu\nu\kappa}^{\lambda}\left(x\right) & \rightarrow C_{\,\mu\nu\kappa}^{\lambda}\left(x\right),\\
C^{\lambda\mu\nu\kappa}C_{\lambda\mu\nu\kappa} & \rightarrow \Omega^{-4}C^{\lambda\mu\nu\kappa}C_{\lambda\mu\nu\kappa}.
\end{align}
Variation of the action \eqref{eq:Weyl_action_gauss_bonnet} leads to the equation for the gravitational field \citep{1921BachZurWeylschenrelativitatstheorieMZ}
\begin{equation}
4 \alpha_{g} W^{\mu\nu} = 4\alpha_{g} \left[2C_{\,\,\;;\lambda;\kappa}^{\mu\lambda\nu\kappa}-C^{\mu\lambda\nu\kappa}R_{\lambda\kappa}\right] = T_{M}^{\mu\nu},
	\label{Bach equation}
\end{equation}
where
\begin{widetext} 
\begin{equation}
W^{\mu\nu} = -\frac{1}{6} g^{\mu\nu} R_{\;\;;\beta}^{;\beta}+R_{\;\;\;\;\;\;\;;\beta}^{\mu\nu;\beta} - R_{\;\;\;\;\;\;\;;\beta}^{\mu\beta;\nu} - R_{\;\;\;\;\;\;\;;\beta}^{\nu\beta;\mu} - 2R^{\mu\beta}R_{\:\,\beta}^{\nu} + \frac{1}{2}g^{\mu\nu}R_{\alpha\beta}R^{\alpha\beta} + \frac{2}{3}R^{;\mu;\nu} + \frac{2}{3}RR^{\mu\nu} - \frac{1}{6}g^{\mu\nu}R^{2}
	\label{eq: Bach Tensor}
\end{equation}
is the Bach tensor and 
\begin{equation}
T_{M}^{\mu\nu} \equiv \frac{2}{(-g)^{1/2}} \frac{\delta I_M}{\delta g_{\mu\nu}}
\end{equation}
is the matter energy-momentum
tensor.

The matter energy-momentum tensor should also be Weyl invariant. Then the most general local matter action for a  generic scalar and spinor field coupled conformally to gravity is \cite{1990MannheimConformalcosmologynoGRaG}
\begin{equation}
I_{M}=-\int d^{4}x\sqrt{-g}\left[\epsilon\left(-\frac{S^{,\mu}S_{,\mu}}{2}+\frac{S^{2}R}{12}\right)+\lambda S^{4}+i\bar{\psi}\gamma^{\mu}\left(x\right)\left[\partial_{\mu}+\Gamma_{\mu}\left(x\right)\right]\psi-\xi S\bar{\psi}\psi\right]. 
	\label{eq: matter action}
\end{equation}
\end{widetext}
$S(x)$ represents a self-interacting scalar field and $\psi(x)$ is a generic 
spin-$\nicefrac{1}{2}$ fermion field.
$\xi$ and $\lambda$ are dimensionless coupling constants, $\gamma^{\mu}(x)$ are the 
vierbein-dependent Dirac-gamma matrices, $\bar{\psi} = \psi^{\dagger} \gamma^{0}$ and 
$\Gamma_{\mu}(x)$ is the fermion spin connection \cite{simplifiednotation}. To be invariant under local 
Weyl transformations the matter fields have to transform as 
$S(x) \rightarrow \Omega^{-1}(x) S(x)$,
$\psi(x) \rightarrow \Omega^{-3/2}(x) \psi(x)$
and $g_{\mu\nu}(x) \rightarrow \Omega^{2}(x) g_{\mu\nu}(x)$. 
The exponent of the conformal factor is called conformal weight.

In \eqref{eq: matter action}
we introduce the parameter $\epsilon$, which can assume values of $-1$ or
$+1$. In the first case, the theory corresponds to CG, while in the second it corresponds to MCG \cite{faria2014massive, 2014_Faria_CosmologyinMassiveConformalGravity_apa, footnoteMCG}. 
Note that only the combination
of the two terms in parenthesis is Weyl invariant.

For $\epsilon R < 0$ and $\lambda > 0$ the potential 
$V(S) = \epsilon\, S^{2}R/12 + \lambda S^{4}$ can lead to a spontaneous breaking of 
Weyl symmetry. 

We find the field equations for the scalar and fermion field 
\begin{align}
\epsilon \left(-S_{\;\;;\mu}^{,\mu} - \frac{1}{6}SR\right)-4\lambda S^{3} + \xi\bar{\psi}\psi &= 0,
	\label{eq: scalar field equation-1}\\
i\gamma^{\mu}(x)\left[\partial_{\mu} + \Gamma_{\mu}(x)\right]\psi -  \xi S\psi &= 0.	
	\label{eq: fermion field equation-1}
\end{align}

Variation of the action \eqref{eq: matter action} and using the equation of motion 
\eqref{eq: fermion field equation-1} leads to the matter energy-momentum tensor 
\begin{widetext}
\begin{equation}
T_{\mu\nu}^{M} = T_{\mu\nu}^{f} + \epsilon\left[-\frac{2S_{,\mu}S_{,\nu}}{3} + \frac{g_{\mu\nu}S^{,\alpha}S_{,\alpha}}{6} + \frac{SS_{,\mu;\nu}}{3} - \frac{g_{\mu\nu}SS_{\;\;\;;\alpha}^{,\alpha}}{3} + \frac{1}{6}S^{2}\left(R_{\mu\nu} - \frac{1}{2}g_{\mu\nu}R\right)\right] - g_{\mu\nu}\lambda S^{4},
\end{equation}
\end{widetext}
where 
\begin{equation}
T_{\mu\nu}^{f} \equiv \frac{1}{2} \left[ 
i\bar{\psi}\gamma_{\mu}(x)\left[\partial_{\nu}+\Gamma_{\nu}(x)\right]\psi + 
(\mu \leftrightarrow \nu) \right]
	\label{emt fermion}
\end{equation}
is the energy-momentum tensor of the fermion.

Since the action $I$ given in Eq.~\eqref{eq: Weyl action} is invariant under a Weyl transformation, it is always possible to choose a frame in which
the scalar field is constant 
\begin{equation}
S(x)\rightarrow S'(x) = \Omega^{-1}(x) S(x) = S_{0} = const.,
\end{equation}
with $\Omega(x) = S(x)/S_{0}.$
This is called the Higgs  or unitary gauge
\citep{horne2016conformal,faria2014massive}. In order to make connection to
GR, one chooses 
\begin{align}
8 \pi\tilde{G} & \equiv \frac{6}{S_{0}^{2}}, 
	\label{eq:relation to GR-1}\\
\Lambda & \equiv 6 \lambda S_{0}^{2}, 
\end{align}
where $\tilde{G}$ denotes an effective Newton's constant. As we will see in the following, 
in all cases of interest we will set $\tilde{G} = G$, the Newton's constant.

Since the scalar field $S(x)$ can always be fixed to a constant by choosing a specific Weyl gauge, it is just an auxiliary field and does not represent a dynamical degree of freedom \citep{flanagan2006fourth,oda2015conformal}. Therefore, we do not need to worry about its stability properties. We nevertheless discuss them in Appendix \ref{sub:Ghosts and Tachyons}, where we follow the analysis of \citep{2014SbisaClassicalquantumghostsEJoP}.

In this gauge (fixing the Weyl invariance), there is a constant mass for the fermions given by
$m_{f} = \xi S_{0}$. Since we know from experiments that fermions have masses, one should 
choose $\xi S_{0} > 0$. Consequently, \eqref{eq: scalar field equation-1}
and \eqref{eq: fermion field equation-1} become 
\begin{align}
- \frac{\epsilon R + 4 \Lambda}{8\pi\tilde{G}} + m_{f}\bar{\psi}\psi &= 0,
	\label{eq: scalar field equation - 2}\\
T_{f}-m_{f}\bar{\psi}\psi &= 0,
	\label{eq: fermion field equation - 2}
\end{align}
where $T_{f}$ denotes the trace of the fermion energy-momentum tensor. 
These two equations can be combined to
\begin{equation}
\epsilon R + 4 \Lambda = 8 \pi \tilde{G} T_{f}.
	\label{Bach_trace}
\end{equation}

With the energy-momentum tensor introduced above, the equation for the gravitational field becomes \citep{1985SchmidtstaticsphericalsymmetricAN,1985_Schmidt_SolutionsoftheLinearizedBachEinsteinEquationintheStaticSphericallySymmetricCase_AN}, 
\begin{equation}
4\alpha_g W_{\mu\nu} = T_{\mu\nu}^{f} + 
\frac{1}{8\pi\tilde{G}} \left[ \epsilon G_{\mu\nu} - g_{\mu\nu} \Lambda \right],
	\label{eq: energy-momentum tensor in higgs gauge-1}
\end{equation}
where $G_{\mu\nu}$ denotes the Einstein tensor. Note that the fermion energy-momentum tensor 
is covariantly conserved,
\begin{equation}
T_{f\:\:;\nu}^{\mu\nu} = 0,
\end{equation}
due to the Bianchi identities for the Bach and Einstein tensors.

Before we continue to discuss solutions of the field equation, we observe that it is 
convenient to introduce a ``graviton mass'' $m_g$ via
\begin{equation}
m_g^2 \equiv  \frac{1}{32 \pi \tilde{G} \alpha_g}.
\end{equation}
We can then write 
\begin{equation}
 - \epsilon G_{\mu\nu} + g_{\mu\nu} \Lambda + \frac{1}{m_g^2} W_{\mu\nu} = 8\pi\tilde{G} T_{\mu\nu}^{f},
 	\label{eq: Modified Bach equation}
\end{equation}
and observe that in the limit $m_g \to \infty$, the Einstein equation is recovered for  $\epsilon = 1$. 
This is the case of MCG. The case of CG ($\epsilon = -1$) does not contain general relativity as a limit. 
Note that the trace of \eqref{eq: Modified Bach equation} reproduces Eq.~\eqref{Bach_trace}.

For conformally flat space-times, $W_{\mu\nu} = 0$ and thus, independently of the value of 
$m_g$ the solutions agree with those of GR for MCG (but not for CG, where the relative 
sign between the Einstein and the energy-momentum tensors is reversed). In particular, MGC leaves the 
isotropic and homogeneous Friedmann-Lemaitre models untouched. 
\vspace{0.2cm}

\section{Weak Gravitational Field in Teyssandier Gauge \label{sec:Linearisation}}

\subsection{Equation of motion}

Let us now turn to the study of gravitational waves in CG and MCG. In the following we drop, 
for simplicity, the cosmological constant ($\Lambda = 0$) and linearize around flat 
Minkowski space-time $g_{\mu\nu} = \eta_{\mu\nu} + h_{\mu\nu},$ where $h_{\mu\nu}$ is a 
small metric perturbation. 
For consistency we have to assume that the energy-momentum tensor vanishes 
at zeroth order.

The second term of the Bach tensor in Eq.~\eqref{Bach equation}
is at least of second order in $h_{\mu\nu}$, hence we only need to consider the first term
\begin{equation}
C_{\,\;\;;\lambda;\kappa}^{\mu\lambda\nu\kappa} = \partial_{\kappa}\partial_{\lambda}C_{(1)}^{\mu\lambda\nu\kappa}+\ensuremath{{\mathcal{O}}}\left(h^{2}\right),
\end{equation}
where $\ldots_{(1)}$ denotes terms of first order in $h_{\mu\nu}$. This term can
be rewritten as 
\begin{equation}
\partial_{\kappa}\partial_{\lambda}C_{(1)}^{\mu\lambda\nu\kappa} = \frac{1}{2}\Box R_{\left(1\right)}^{\mu\nu}-\frac{1}{12}\eta^{\mu\nu}\partial_{\rho}\partial^{\rho}R_{\left(1\right)}-\frac{1}{6}\partial^{\mu}\partial^{\nu}R_{\left(1\right)},
\end{equation}
where we have used the Bianchi identities 
\begin{align}
\partial_{\kappa}\partial_{\lambda}R^{\lambda\mu\nu\kappa}_{\left(1\right)} & = \Box R^{\mu\nu}_{\left(1\right)}-\partial_{\lambda}\partial^{\mu}R^{\lambda\nu}_{\left(1\right)},\\
\partial_{\lambda}R^{\lambda\mu}_{\left(1\right)} & = \frac{1}{2}\partial^{\mu}R_{\left(1\right)}.
\end{align}
The d'Alembert operator is defined as $\Box \equiv \partial_\mu \partial^\mu$. 
This leads us to the linearized field equations for the metric 
\begin{widetext}
\begin{equation}
- \epsilon \left(R_{\left(1\right)}^{\mu\nu}-\frac{1}{2}\eta^{\mu\nu}R_{\left(1\right)}\right)
+ \frac{1}{m_g^2} \left(\Box R_{\left(1\right)}^{\mu\nu} - \frac{1}{6}\eta^{\mu\nu}\Box R_{\left(1\right)} -
\frac{1}{3}\partial^{\mu}\partial^{\nu}R_{\left(1\right)}\right) 
= 
8\pi\tilde{G}T_{f \left(1\right)}^{\mu\nu}.
	\label{linearized_eom_unexpanded}
\end{equation}
The linearized energy-momentum tensor satisfies 
\begin{equation}\label{matter energy-momentum conservation to first order}
\partial_\mu T^{\mu\nu}_{f (1)} = 0.
\end{equation}
From now on, all quantities are of first order and we write $T^f_{\mu\nu} = T_{\mu\nu}$.

Using \eqref{Bach_trace} and the expressions from Appendix \ref{appendix A} we can rewrite
\eqref{linearized_eom_unexpanded} as  
\begin{equation}
m_{g}^{-2}\left(\Box-\epsilon m_{g}^{2}\right)
\left(\frac{1}{2}\Box h_{\mu\nu}-\frac{1}{6}\eta_{\mu\nu}R\right) + 
\frac{1}{2}\left(\partial_{\mu} Z_{\nu} + \partial_{\nu}Z_{\mu}\right)
=
8\pi \tilde{G}\left(T_{\mu\nu}-\frac{1}{3}\eta_{\mu\nu}T\right),
	\label{linearized_eom_expanded}
\end{equation}
\end{widetext}
where $Z_{\mu} \equiv - m_{g}^{-2}
\left[\left(\Box-\epsilon m_{g}^{2}\right)\partial_{\rho}\bar{h}_{\mu}^{\rho} + (1/3) \partial_{\mu}R\right]$
and 
\begin{equation}
\bar{h}_{\mu\nu} \equiv h_{\mu\nu} - \frac12 \eta_{\mu\nu}h
	\label{tracereverse}
\end{equation} 
is the trace-reversed metric perturbation.

It turns out to be convenient to choose the gauge condition
\begin{equation}
Z_\mu = 0.
	\label{Tgauge}
\end{equation}
This is called the Teyssandier gauge \citep{1989_Teyssandier_LinearisedR+R2gravity_anewgaugeandnewsolutions_CaQG} (see also Appendix \ref{sub:TT-gauge}).
Then Eq.~\eqref{linearized_eom_expanded} simplifies to
\begin{equation}
\frac{1}{m_g^2} \left(\Box-\epsilon m_{g}^{2}\right)
\left(\Box h_{\mu\nu} - \frac{1}{3}\eta_{\mu\nu}R\right)
=
16 \pi  \tilde{G}\left(T_{\mu\nu}-\frac{1}{3}\eta_{\mu\nu}T\right).
	\label{eom_Teyssandier}
\end{equation}

Before we proceed to solve Eqs.~\eqref{eom_Teyssandier}, let us analyze the forms this equation can take in different limits of CG and MCG. We denote with $L$ the typical variation scale of the metric perturbation and with $d$ the typical size of the source. 

To analyze the various limits of this theory it is useful to rewrite Eq.~\eqref{eom_Teyssandier} to
\begin{equation}
\left[\Box^2- \epsilon m_g^2\Box\right] h_{\mu\nu} = 
16 \pi \tilde{G} m_g^2 \bar{T}_{\mu\nu},
	\label{eom Teyssandier rewritten}
\end{equation}
where 
\begin{equation}
\bar{T}_{\mu\nu} = \left(T_{\mu\nu} - 1/2 \eta_{\mu\nu}T\right) + \epsilon / (6 m_g^2)\eta_{\mu\nu}\Box T.
\end{equation}
By writing \eqref{eom Teyssandier rewritten} approximately as $[L^{-4}~-~\epsilon\,~m_g^2\,~ L^{-2}]\,h \sim \tilde{G}\, m_g^2\,T + \epsilon\, \tilde{G} \, d^{-2}\, T $, it appears that there are four relevant cases. If $ L\, m_g \ll 1 $, one recovers the limits of pure CG and MCG, since the term with higher derivatives dominates the left-hand side. If instead $ L\,m_g \gg 1 $, the wave equation is of second order. Depending on the relation among $m_g$ and $d$, different terms dominate the right-hand side. Note that, if both $L\,m_g \gg 1$ and $d\,m_g \gg 1 $, MCG reduces to GR, while CG provides the same equation as in GR, but with a flip of the sign. The limiting cases are summarized in Table \ref{tab:limits}. 

\begin{center}
\begin{table*}
\global\long\def\arraystretch{1.2}
\begin{tabular}{|c||c||c|c|}
\hline 
 & \textbf{CG $\left(\epsilon = -1\right)$}  & \textbf{MCG $\left(\epsilon = +1\right)$} & remarks\tabularnewline
\hline 
\hline 
$m_g L \ll 1,\ m_g d \ll 1$ & 
$3 \Box^{2} h^{\mu\nu} = - 8\pi \tilde{G} \eta^{\mu\nu}\Box T$    
&
$3 \Box^{2}h^{\mu\nu}= 8\pi \tilde{G} \eta^{\mu\nu}\Box T$ 
& light $m_g$ \tabularnewline 
$m_g L \ll 1,\ m_g d \gg 1$ & $\Box^{2}\bar{h}^{\mu\nu} = 16\pi\tilde{G}m_g^2 T^{\mu\nu}$  & $\Box^{2}\bar{h}^{\mu\nu} = 16\pi\tilde{G}m_g^2 T^{\mu\nu}$ & irrelevant for $L > d$ \tabularnewline 
$m_g L \gg1,\ m_g d \ll 1$ & 
$3 \Box h^{\mu\nu}= - 8\pi \tilde{G} m_g^{-2} \eta^{\mu\nu}\Box T$ 
&
$-3 \Box h^{\mu\nu}= 8\pi \tilde{G} m_g^{-2} \eta^{\mu\nu}\Box T$ 
& intermediate $m_g$
\tabularnewline 
$m_g L \gg 1,\ m_g d \gg 1$ & $\Box\bar{h}^{\mu\nu} = 16 \pi \tilde{G} T^{\mu\nu}$  & $- \Box\bar{h}^{\mu\nu} = 16 \pi \tilde{G} T^{\mu\nu} \Leftrightarrow$ GR 
& heavy $m_g$
\tabularnewline
\hline 
\end{tabular}
\par
\caption{\label{tab:limits} Eq.~\eqref{eom Teyssandier rewritten} for different limits of CG and MCG.}
\end{table*}
\end{center}

\subsection{Gravitational wave propagator\label{sub:Ghosts}}

The solution to the inhomogeneous Eq.~\eqref{eom Teyssandier rewritten} is given by 
\begin{equation}
h_{\mu\nu} = 16 \pi \tilde{G} \int d^{4}x^\prime \mathcal{G} (x-x^\prime) \bar{T}_{\mu\nu}(x^\prime).
	\label{eq: metric perturbation}
\end{equation}
The Green's function $\mathcal{G}(x)$ is defined by 
\begin{equation}
\left(\Box-\epsilon m_{g}^{2}\right)\Box \mathcal{G}\left(x-x^\prime\right) = m_g^2 \delta^{4}\left(x-x^\prime\right).
\end{equation}
For the Fourier transformed Green's function one finds 
\begin{equation}
\tilde{\mathcal{G}}\left(k\right) = 
\frac{m_g^2}{\left(\omega^{2}-k^{2}-\epsilon m_{g}^{2}\right)\left(\omega^{2}-k^{2}\right)}.
\end{equation}
This can be rewritten as
\begin{align}
\tilde{\mathcal{G}}(k) &= 
\epsilon \left[-\frac{1}{\left(\omega^{2}-k^{2}\right)}+\frac{1}{\left(\omega^{2}-k^{2}-\epsilon m_{g}^{2}\right)}\right],
	\label{eq: propagator spin-2 Fourier space}
\end{align}
where $k^2 \equiv \mathbf{k}^2$.
In the propagator for the spin-2 metric perturbation $h_{\mu\nu}$ above, either the massless term $(\epsilon = -1)$ or the massive
term $(\epsilon = +1)$ comes with the wrong sign: the so-called Weyl ghost (see e.g.~\citep{2013BiswasKoivistoMazumdarNonlocaltheoriesgravityapa}). 
Note that the spin-2 ghost excitation around the Minkowski vacuum is present
independently of $\epsilon$ \cite{footnoteghost}. However, CG and MCG have different stability properties and relations to GR.

For CG $(\epsilon = -1)$ we have demonstrated previously (cf. Table \ref{tab:limits}) that there is no limit leading to the action or equations of GR, since the sign of the Einstein-Hilbert term in the matter action is opposite to GR (the Newtonian limit of this theory is studied in section \ref{sec:signal_from_a_binary}). As a consequence, the massless part of the propagator has the wrong sign, representing a ghost instability. Additionally, the massive part of the gravitational wave represents a tachyon, i.e. it travels faster than the speed of light. The ghost instability in CG has been widely discussed by Mannheim \& Bender \cite{2000MannheimDavidsonFourthordertheoriesaph,2007BenderMakingsensenonRoPiP,2008BenderMannheimExactlysolvablePPRD,2008BenderMannheimGivingghostJoPAMaT,2008BenderMannheimNoghosttheoremPRL,2013MannheimPTsymmetryasPTotRSoLAMPaES,2015MannheimAntilinearityRatherthanapa,2016MannheimExtensionCPTtheoremPLB}: they analyzed in a toy model the Pais-Uhlenbeck fourth-order oscillator  \cite{1950PaisUhlenbeckfieldtheoriesnonPR},
which was believed to suffer from ghost instabilities too, and in a series
of papers they claimed that this is not the case and that an
explicit quantization and construction of the Hilbert space is necessary
in order to judge whether a theory suffers from instabilities or not.

The case of MCG is different, since it has the correct sign for the Einstein-Hilbert term and thus it includes GR as a limiting case (cf. Table \ref{tab:limits}). The Newtonian gravitational potential can be recovered (the Newtonian limit of MCG is also studied in section \ref{sec:signal_from_a_binary} and Appendix \ref{appendix C}) and the massless excitation represents a healthy graviton travelling at the speed of light. In this case the wrong sign in the propagator appears for the massive graviton, which is however subluminal and does not propagate at all in the GR limit.

Note that for both theories the shortcoming of the appearance of the Weyl-ghost come along with the major advantage of being better behaved in the UV-limit, i.e. being renormalizable \citep{1977StelleRenormalizationhigherderivativePRD} and hence viable theories of quantum gravity \citep{Mannheim2011,mannheim2012making,2015MannheimLivingSupersymmetryConformalapa}. See also \citep{2014_Salvio_Agravity_JoHEP} for a similar theory presenting the Weyl ghost.

\subsection{Massive and massless mode \label{massive and massless mode}}

Let us proceed to solve Eq.~\eqref{eom_Teyssandier} now. It is possible to reduce the order of the wave equation by splitting the metric perturbation
\begin{equation}
h_{\mu\nu} = \epsilon \left(H_{\mu\nu} + \Psi_{\mu\nu}\right),
	\label{ansatz}
\end{equation}
where $H_{\mu\nu}$ and $\Psi_{\mu\nu}$ are symmetric tensors, and making the ansatz
\begin{equation}
\Psi_{\mu\nu} = \frac{1}{m_g^2} \left(\Box h_{\mu\nu} - \frac{1}{3}\eta_{\mu\nu}R\right).
	\label{psi}
\end{equation}
Then, (\ref{eom_Teyssandier}) turns into the equation of motion for a massive mode
\begin{equation}
\left(\Box - \epsilon m_{g}^{2}\right) \Psi_{\mu\nu} = 
16\pi \tilde{G}\left(T_{\mu\nu} - \frac{1}{3}\eta_{\mu\nu}T\right).
	\label{massive_eom}
\end{equation}
We now use \eqref{massive_eom}, eliminate the term $m_g^2 \Psi_{\mu\nu}$ by means of (\ref{psi}) and 
replace the Ricci scalar by means of \eqref{Bach_trace}. Finally we use \eqref{ansatz} and \eqref{tracereverse} to arrive at a massless equation of motion that looks familiar,
\begin{equation}
\Box\bar{H}_{\mu\nu} = -16\pi \tilde{G}T_{\mu\nu},
	\label{massless_eom}
\end{equation}
where $\bar{H}_{\mu\nu}$ is the trace-reversed massless mode.
In the last step we exploit the gauge condition \eqref{Tgauge}. Using \eqref{massless_eom}, 
\eqref{massive_eom} and \eqref{Bach_trace} we find  
\begin{equation}
Z_\mu = - \partial_{\rho}\bar{H}^\rho_\mu = 0, 
	\label{massless_gauge}
\end{equation}
the condition for the massless mode to be transverse.
But there is one more condition that is fixed in the Teyssandier gauge. 
From the expression for the Ricci scalar, condition (\ref{massless_gauge}), the trace of 
\eqref{massless_eom} and \eqref{Bach_trace} it follows that 
\begin{equation}
\partial_{\rho}\partial_{\sigma}\Psi^{\rho\sigma} = \Box\Psi.
	\label{massive_gauge}
\end{equation}
Defining $\hat{\Psi}_{\mu\nu} \equiv \Psi_{\mu\nu}-\eta_{\mu\nu}\Psi$, equations \eqref{massive_eom} 
and \eqref{massive_gauge} are equivalent to
\begin{equation}
\left(\Box-\epsilon m_{g}^{2}\right)\hat{\Psi}_{\mu\nu} 
= 
16\pi \tilde{G} T_{\mu\nu},\;\;\partial_{\rho}\partial_{\sigma}\hat{\Psi}^{\rho\sigma} = 0 
	\label{massive_eom_hat}.
\end{equation}
Hence, the general solution $h_{\mu\nu}$ is decomposed into a transverse massless mode $H_{\mu\nu}$ and a massive mode $\Psi_{\mu\nu}$.

It is interesting to note that in the limit $m_g \rightarrow 0$, Eq.~\eqref{massive_eom_hat} becomes a massless wave equation, differing only by a sign from Eq.~\eqref{massless_eom}. Hence, in this limit and under the assumption that the traces of both modes vanish the total metric perturbation vanishes, too. 

In the homogeneous case, Eqs.~\eqref{massless_eom} and \eqref{massive_eom_hat} take the form 
\begin{align}
\Box\bar{H}_{\mu\nu} & =0, 
	\label{massless_eom_vacuum}\\
\left(\Box-\epsilon m_{g}^{2}\right)\hat{\Psi}_{\mu\nu} & = 0.
	\label{massive_eom_vacuum}
\end{align}
The solutions to \eqref{massless_eom_vacuum} and \eqref{massive_eom_vacuum} are a massless plane wave  and a massive plane wave 
\begin{align}
\bar{H}_{\mu\nu} & = a_{\mu\nu}e^{ik_{\rho}x^{\rho}},\;k_{\rho}k^{\rho} = 0,\\
\hat{\Psi}_{\mu\nu} & = b_{\mu\nu}e^{il_{\rho}x^{\rho}},\;l_{\rho}l^{\rho} = -\epsilon m_{g}^{2},				\label{eq:massive graviton in vacuum}
\end{align}
where $a_{\mu\nu}$ and $b_{\mu\nu}$ are constant and symmetric.
Depending on the values of $\epsilon$, the wave vector $l^{\rho}$ is time-like or space-like, corresponding to a wave which travels slower than the speed of light for MCG $(\epsilon=+1)$, and a tachyon that is faster than the speed of light for CG $(\epsilon=-1)$. For more details, see Appendix \ref{appendix E}.

In the next subsections, we derive the solutions of Eqs.~\eqref{massless_eom} and \eqref{massive_eom_hat}.

\subsection{Solution with a source: massive part\label{sub:Non-Vacuum-Solution}}

In the following, we only analyze the massive wave equation \eqref{massive_eom_hat}, since the massless part is known from GR. The most convenient way to analyze the inhomogeneous solutions is to keep real
space while switching to $\omega$ dependence. We define
\begin{equation}
	\label{general form of massive solution}
\hat{\Psi}_{\mu\nu} = 
16\pi \tilde{G} \int d^{4} x' \mathcal{G} \left(x-x'\right) T_{\mu\nu}\left(x'\right),
\end{equation}
\begin{widetext}
\noindent with the frequency-domain Green's function
\begin{equation}
\mathcal{G} \left(\omega,\mathbf{x}-\mathbf{x'}\right) = 
\frac{1}{\left(2\pi\right)^{3}} \int d^{3} k \frac{e^{i\mathbf{k}\cdot\left(\mathbf{x}-\mathbf{x'}\right)}}{\omega^{2}-k^{2}-\epsilon m_{g}^{2}}
  =  \frac{-i}{2\left(2\pi\right)^{2}\left|\mathbf{x}-\mathbf{x}'\right|}\int_{-\infty}^{\infty}dk\frac{k}{\omega^{2}-k^{2}-\epsilon m_{g}^{2}}\left(e^{ik\left|\mathbf{x}-\mathbf{x'}\right|}-e^{-ik\left|\mathbf{x}-\mathbf{x}'\right|}\right),
\end{equation}
\end{widetext}
where we have integrated over the angles and extended
the $k$-integral to $-\infty$ to find the last expression. The poles of the integrand are at 
\begin{equation}
k = \pm\sqrt{\omega^{2}-\epsilon m_{g}^{2}}.
\end{equation}
In MCG with $\epsilon = +1$ we have to distinguish two cases, $\omega^{2}>m_{g}^{2}$ and $\omega^{2}<m_{g}^{2}$, while CG with $\epsilon = -1$ always leads to a positive radicand. 

\subsubsection{Propagator for small graviton mass}

For CG and MCG with a small graviton mass ($m_{g}^{2}~<~\omega^{2}$) the radicand is positive, so by finding the residues of these poles we get 
\begin{equation}
\mathcal{G} \left(\omega,\mathbf{x}-\mathbf{x'}\right) =
 -\frac{e^{ik_{\omega,\epsilon}\left|\mathbf{x}-\mathbf{x'}\right|}+e^{-ik_{\omega,\epsilon}\left|\mathbf{x}-\mathbf{x'}\right|}}{4\pi \left|\mathbf{x}-\mathbf{x'}\right|}\theta(m_g - |\omega|),
	\label{propagator MCG small mass}
\end{equation}
where $k_{\omega,\epsilon} \equiv \sqrt{\omega^{2} - \epsilon m_{g}^{2}}$. In the far zone approximation ($r \gg |\mathbf{x'}|$) we get $|\mathbf{x} - \mathbf{x'}| = r - \mathbf{x'} \cdot \mathbf{n} + O(d^2/r)$, where $r$ denotes the distance between the observer and the 
source and $\mathbf{n}$ the spatial unit vector pointing from the source to the observer.
Keeping only the first order yields
\begin{equation}
\mathcal{G} (\omega,\mathbf{x}-\mathbf{x'}) =
-\frac{e^{ik_{\omega,\epsilon}(r - \left|\mathbf{x'}\cdot\mathbf{n}\right|)} + e^{-ik_{\omega,\epsilon}(r - \left|\mathbf{x'}\cdot\mathbf{n}\right|)}}{4\pi r} \theta(m_g - |\omega|).
	\label{propagator MCG small mass far field}
\end{equation}
Note that this result also holds for CG with a large graviton mass ($m_g^2 > \omega^2$). However, we do not consider this case in this work, because the reason for proposing CG was that it can fit galaxy rotation curves without dark matter in the small mass case. Furthermore, it is not obvious that the case of a large graviton mass exhibits a valid Newtonian limit (the gravitational potential oscillates).

\subsubsection{Propagator for large graviton mass}

For MCG with a large graviton mass ($m_{g}^{2} > \omega^{2}$) the radicand is negative, thus $k = \pm i\sqrt{m_{g}^{2} - \omega^{2}}$. The Green's function becomes
\begin{align}
\mathcal{G} \left(\omega,\mathbf{x} - \mathbf{x'}\right) &= 
-\frac{e^{-k_{\omega,>}\left|\mathbf{x}-\mathbf{x'}\right|}}{8\pi \left|\mathbf{x}-\mathbf{x'}\right|} \theta(|\omega| - m_g)\nonumber\\
&= -\frac{e^{-k_{\omega,>}(r + \left|\mathbf{x'}\cdot\mathbf{n}\right|)}}{8\pi r} \theta(|\omega| - m_g),
	\label{propagator MCG large mass}
\end{align}
where $k_{\omega,>} \equiv \sqrt{m_{g}^{2}-\omega^{2}}$. In the second line the far zone approximation has been applied.

Let us remark that \eqref{general form of massive solution} together with \eqref{propagator MCG small mass far field} and \eqref{propagator MCG large mass} is valid for relativistic and non-relativistic sources.

\section{Gravitational Waves from a Binary System\label{sec:signal_from_a_binary}}

We now consider binary systems with masses $m_1$ and $m_2$ on circular orbits moving at 
a speed small compared to the speed of light. This means we can treat the source in the 
non-relativistic and weak field limits. Hence, we can neglect contributions of the 
gravitational potential and the kinetic energy to the energy-momentum tensor $T_{\mu\nu}$ in 
Eq.~\eqref{general form of massive solution}. In general these approximations do not hold 
true for binaries consisting of compact objects like neutron stars. However, for binary systems 
in the inspiralling phase of their evolution, where the objects are still far apart, these 
assumptions are adequate for analyzing the gravitational radiation behavior. Moreover, here 
we do not consider the back reaction on the binaries motion due to its gravitational wave emission.

In particular, we look at the binary system PSR~J1012+5307 \citep{2009LazaridisWexJessnerKramerStappersJanssenDesvignes,1995NicastroLyneLorimerHarrisonBailesSkidmorePSRJ10125307MNotRAS,1998CallananGarnavichKoestermassneutronstarMNotRAS},
which is a neutron star-white dwarf system in quasi-circular motion,
cf. Table \ref{tab: PSR J1012+5307-1}. The orbital frequency of this system is given
by 
\begin{equation}
\omega_{s} \approx \SI{1.3d-20}{eV} \approx \SI{1.9d-5}{Hz}
	\label{frequency_source}.
\end{equation}
The system is picked for its small eccentricity of the orbit, such that we can apply the results of 
our study of circular orbits. Its orbital speed is of order $10^{-5}$c, which justifies the 
low-velocity approximation. The orbital period $P$ of the binary system PSR J1012+5307 and its time derivative $\dot{P}$ have 
been derived from data collected over 15 years and are in excellent agreement 
with the assumption that its decay of the orbital period is due to gravitational radiation as predicted by GR. 

\renewcommand{\arraystretch}{1.2}
\begin{table}
\noindent \begin{centering}
\begin{tabular}{lr}
\hline \hline
period $P$ (days)  & $0.60467271355(3)$\tabularnewline

period derivative (observed) $\dot{P}_{obs}$  & $5.0(1.4)\times10^{-14}$\tabularnewline

period derivative (intrinsic) $\dot{P}_{intr}$  & $-1.5(1.5)\times10^{-14}$\tabularnewline

mass ratio $q$  & $10.5(5)$\tabularnewline

neutron star mass $m_{1}\,(M_{\odot})$  & $1.64(22)$\tabularnewline
 
white dwarf mass $m_{2}\,(M_{\odot})$  & $0.16(2)$\tabularnewline
 
eccentricity $e\,(10^{-6})$  & $1.2(3)$\tabularnewline

projected semimajor axis $a$  & $\SI{0.581872(2)}{ls} $\tabularnewline

distance $r$  & $\SI[separate-uncertainty = true]{840(90)}{pc}$\tabularnewline
\hline \hline
\end{tabular}
\par\end{centering}
\caption{\label{tab: PSR J1012+5307-1}Orbital data for the binary system PSR J1012+5307 \citep{1995NicastroLyneLorimerHarrisonBailesSkidmorePSRJ10125307MNotRAS,1998CallananGarnavichKoestermassneutronstarMNotRAS,2009LazaridisWexJessnerKramerStappersJanssenDesvignes} consisting of a neutron star and a white dwarf in quasi circular motion. The semimajor axis is given in light seconds (ls).}
\end{table}

\subsection{Newtonian limit and Kepler's Third Law}\label{NewtonianLimit}

In general the analysis of the gravitational wave emission  proceeds as follows. The first step is to calculate the decay of the orbital period $\dot{P}/P$ ($P = 2\pi/\omega_s$) via Kepler's third law for two objects of mass $m_1$ and $m_2$ in the Newtonian limit for a circular orbit in the center or mass frame, where $\mu = m_1\, m_2/(m_1 + m_2)$ is the reduced mass. Since the gravitational potential is modified in CG and MCG, we have to rederive Kepler's third law in these theories. In general, we can write
\begin{equation}
\frac{\dot{P}}{P} = \frac{\dot{R}}{2R} - \frac{\dot{V}'}{2V'},
	\label{radiated_energy_general}
\end{equation}
where $V'(R) = \mu^{-1}\partial_R E_{pot}(R)$ is the derivative of the gravitational potential $V$ with respect to distance between the objects $R$ and $E_{pot}$ the gravitational potential energy.

Note that in GR it is assumed that the total decrease of the orbital period occurs due to the emission of energy in gravitational radiation. The result of this chapter is that the theories we investigate in this work should predict the same amount of energy that is radiated by gravitational waves (within the precision of the measurements) as GR in order to explain the decrease of the orbital period of binary systems without using any other mechanism than gravitational wave emission.

\subsubsection{Conformal gravity}

In \citep{2007MannheimSchwarzschildlimitconformalPRD} it has been claimed that in CG $(\epsilon = -1)$ with a small graviton mass the line element in a static, spherically symmetric geometry exterior to a source of one solar mass with a non-vanishing scalar field $S_0$ can be written in the form
\begin{equation}
ds^2 = -B(r)dt^2 + \frac{dr^2}{B(r)} + r^2 d\Omega^2,
	\label{line element CG}
\end{equation}
where $B(r) = 1 - \beta(2-3\beta\gamma)/r - 3\beta\gamma + \gamma r - k r^2.$
Here $\beta$, $\gamma$ and $k$ are constants of integration and are used to fit galaxy rotation curves. $k$ has an influence on the outer parts of galaxies, but is much smaller than $\beta$ and $\gamma$ and we can neglect the $k$-term in the following. Also terms proportional to $\beta \gamma \ll 1$ are negligible on the distance scale which corresponds to our binary system. For a source of one solar mass $M_{\odot}$, the parameters are given by \citep{2007MannheimSchwarzschildlimitconformalPRD,2013MannheimOBrienGalacticrotationcurves}
\begin{align}
S_{0}^{2} & = \SI{9.7d34}{kg.s^{-1}},
	\label{S0 parameter}
\\
\alpha_{g} & = \SI{3.3d75}{kg.m^{2}.s^{-1}},
\\
\gamma & = \SI{5.4d-39}{m^{-1}},
\\
2\beta & = \SI{3d3}{m},
	\label{beta parameter}
\\
k & = \SI{9.5d-50}{m^{-2}}
	\label{k parameter}.
\end{align}
For the graviton mass this yields 
\begin{equation}
m_{g,CG} = \SI{1.9d-58}{kg} = \SI{1.1d-22}{eV}.															\label{gravitonmass_CG}
\end{equation} 
But note that in literature there is some criticism on using the line element \eqref{line element CG}, see \citep{horne2016conformal, flanagan2006fourth, 1999BarabashShtanovNewtonianlimitconformalPRD,2007BarabashPyatkovskaWeakfieldlimitapa,2017_Campigotto_ConformalGravity_LightDeflectionRevisitedandtheGalacticRotationCurveFailure_apa}. 
Nevertheless, in the following we show that it does not matter for our analysis of the gravitational radiation whether these additional terms are there or not. For the parameter values that are needed to fit galaxy rotation curves (corresponding to a small graviton mass) the additional terms do not affect the gravitational radiation of the system under study.

To be consistent with solar system tests we have to choose $\tilde{G} = G = \beta /M_{\odot}$ and $\gamma_{\odot} = \gamma/M_{\odot}$. The gravitational potential energy and its time derivative for CG is given by \citep{1989MannheimKazanasExactvacuumsolutionTAJ,1993MannheimLinearpotentialsgalacticapa,1994MannheimKazanasNewtonianlimitconformalGrag,2007MannheimSchwarzschildlimitconformalPRD}
\begin{align}
E_{pot} & = -\frac{G \mu M}{R} + \frac{\gamma_{\odot} \mu M }{2} R,
	\label{eq:grav_pot_CG}
\\
\dot{E}_{pot} & = 
G\mu M \frac{\dot{R}}{R^{2}} \left(1 + \frac{\gamma_{\odot} R^{2}}{2G}\right),
	\label{radiated_energy_CG}	
\end{align}
where $M = m_{1} + m_{2}$ is the total mass of the system. Inserting this into \eqref{radiated_energy_general}, we find
\begin{equation}
\frac{\dot{P}}{P} \approx - \frac{\vert E_{GR} \vert^{\boldsymbol{\cdot}}}{\vert E_{GR} \vert} \left(\frac{3}{2} -\frac{\gamma_{\odot}}{2G}R^{2}\right)
	\label{eq:decayorbitCG},
\end{equation}
where $\vert E_{GR} \vert = G M \mu / (2R)$ and $\gamma_{\odot}\, R^{2}/G \ll 1$. To verify that this combination is indeed small, we insert the distance between the sources of the binary system under study, cf. Table \ref{tab: PSR J1012+5307-1} and assume that for a binary system in circular motion, we have $R \approx a \approx \SI{8.9d14}{eV^{-1}}$, where $a$ is the semimajor axis of PSR J1012+5307. Also we use the parameters determined by the analysis of galaxy rotation curves in \eqref{S0 parameter}-\eqref{k parameter}, which shows that the second term in Eq.~\eqref{eq:decayorbitCG} is indeed negligible since $ \gamma_\odot \, a^{2} /G  \approx 10^{-26}$. This demonstrates that in CG with a small graviton mass the orbital energy, which is lost by the system, is, up to small modifications, the same as in GR, because on Solar System distance scales the second term in \eqref{eq:grav_pot_CG} can be neglected with respect to the first one. Therefore, we can treat the 
binary system in the Newtonian limit.

\subsubsection{Massive conformal gravity}

Now, let us apply the same analysis for MCG $\left(\epsilon = +1\right)$ in the case of a large graviton mass $\left(m_{g,>}^{2} > \omega^{2}\right)$, where $m_{g,>}$ denotes the graviton mass for this case. 

In \citep{1999BarabashShtanovNewtonianlimitconformalPRD,2007BarabashPyatkovskaWeakfieldlimitapa}  or in Appendix \ref{appendix C} it is pointed out that this model cannot fit galaxy
rotation curves without dark matter, but it is still interesting because of its GR limit. 

In this case the massive part of the graviton becomes damped and we are left with a theory that is just GR modified by exponentially suppressed contributions. Nevertheless, there is a profound difference to GR, since it is claimed that this theory is power-counting renormalizable \citep{1977StelleRenormalizationhigherderivativePRD,2016FariaQuantummassiveconformalTEPJC}.

In Appendix \ref{appendix C} it is shown that the gravitational potential in the Newtonian limit is given by
\begin{equation}
\Phi(r) = -\frac{G M}{r}\left( 1 - \frac{4}{3} e^{-m_{g,>} r}\right),
	\label{grav potential MCG}
\end{equation}
where $\tilde{G} = G$ has been chosen.

We have to constrain the graviton mass with data from short range tests of the inverse square law. From \citep{2009AdelbergerGundlachHeckelHoedlSchlammingerTorsionbalanceexperimentsPiPaNP}, we get
\begin{equation}
m_{g,>} > \SI{d-38}{kg} \approx \SI{d-2}{eV}.
\end{equation}  
This means that the Yukawa term in \eqref{grav potential MCG} becomes important only on submillimeter distance scales. For binary systems in the inspiral phase the distance between the objects is always macroscopic $(m_{g,>}a \approx \SI{3.5d12}{})$ and hence we can completely neglect this term for the analysis of gravitational radiation. 
The result for the decay of the orbital period is
\begin{equation}
\frac{\dot{P}}{P} \approx - \frac{3\vert E_{GR} \vert^{\boldsymbol{\cdot}}}{2\vert E_{GR} \vert}			\label{eq:decay of the orbital period_MCG3}.
\end{equation}

Further, we can look at the case of a small graviton mass $(m_{g,<}^{2} < \omega^{2})$ in MCG. Let us first assume the same potential as for the case of large graviton mass in Eq.~\eqref{grav potential MCG}. 

From the constraint $m_{g,<}^{2} < \omega^{2}$ it is clear that one cannot make the graviton mass large enough to push the Yukawa contribution to the sub-millimeter scale. Rather we get an upper bound on the graviton mass from solar system tests on the inverse square law of the gravitational force \citep{2009AdelbergerGundlachHeckelHoedlSchlammingerTorsionbalanceexperimentsPiPaNP}
\begin{equation}
m_{g,<} < \SI{d-58}{kg} \approx \SI{d-22}{eV}.
\end{equation}
With this bound even on galactic distance scales the Yukawa contribution is too small and has the wrong sign to compensate for dark matter. 

Here we get for the decay of the orbital period 
\begin{equation}
\frac{\dot{P}}{P} \approx - \frac{\vert E_{GR} \vert^{\boldsymbol{\cdot}}}{\vert E_{GR} \vert} \left(\frac{3}{2} - \frac{2}{3} m_{g,<}^2 R^2 e^{-m_{g,<} R}  \right). 
	\label{eq:decay of the orbital period_MCG2}
\end{equation}
The second term in the bracket is negligible, since $\left(m_{g,<}a\right)^{2}\lesssim10^{-36}$ for $R \approx a$, where $a$ is the semimajor axis of the system.
\\
\\
\\

We have shown that in all cases of interest the choice $\tilde{G} = G$ leads to the Newtonian limit and 
hence this relation should hold from now on.
Note however that $\tilde{G}$ is not a free parameter of the theory. The theory is independent of $\tilde{G}$, 
because of the Weyl invariance. The only free parameter is $\alpha_g$  (or equivalently $m_g$). 
Hence, the choice of $\tilde{G} = G$ is just convenient to recover expressions that look familiar and to 
compare to GR.

\subsection{Gravitational waves from binary systems}\label{Radiated Energy}

We discuss the GW solutions for an explicit binary system in circular motion and in the Newtonian limit.

But before doing so, we show that for a small graviton mass monopole and dipole radiation can be neglected. For the massless part, since it is the same as in GR, there is no monopole and dipole radiation and the leading contribution comes from the quadrupole term. The reason for this is that the metric perturbation is a massless spin-2 field and that the matter energy-momentum tensor is conserved far away from the source. But for a small graviton mass there are non-vanishing contributions from the monopole and dipole radiation. Nevertheless, in the following we will show that in the quadrupole approximation these do not contribute to the radiated energy and that only two of the five additional degrees of freedom of the massive mode are excited by a conserved matter energy-momentum tensor.

The quadrupole approximation requires that the typical velocities of the source are much smaller than the velocity 
of the gravitational waves such that $k_{\omega}\,d \ll 1$ is fulfilled. In GR this holds true for non-relativistic 
sources, since the gravitational waves travel with the speed of light.

For a small graviton mass we can apply the quadrupole approximation in 
\eqref{propagator MCG small mass far field} because 
$k_{\omega,\epsilon} \approx \omega (1-\epsilon m_g^2/(2\omega^2))$ for $m_g^2/\omega^2 \ll 1$ and, 
as we will verify later in this section, $\omega = 2\omega_s$. Thus the speed of the massive mode of 
the gravitational waves is nearly the speed of light and hence much higher than the orbital speed of the source.

However, in the case of MCG with a large graviton mass we have 
$k_{\omega,>} \approx m_g (1 - \omega^2/(2 m_g^2))$ for $\omega^2/m_g^2~\ll~1$, which 
leads to $k_{\omega,>} d \gg 1$. This shows that the quadrupole approximation cannot be used in 
\eqref{propagator MCG large mass}. Nevertheless, the term $\exp{(-m_g r)}$ in 
\eqref{propagator MCG large mass} leads to an exponential suppression of the massive mode anyway. 
Hence, we do not need the quadrupole approximation and keep only the leading order term of the far 
field approximation. For more details to the multipole expansion, see e.g. \citep{maggiore2008gravitational}.

\newpage
Before we apply the quadrupole approximation let us define the mass-energy moments
\begin{align}
M(t) &= 
\int d^3x\, T^{00}(t,\mathbf{x}),\\
D^i(t) &= 
\int d^3x\, x^{i} T^{00}(t,\mathbf{x}), \\
M^{ij}(t) &= 
\int d^3x\, x^i x^j T^{00}(t,\mathbf{x}).
\end{align}
These quantities are called monopole, dipole and quadrupole moments and we denote their time Fourier transformations as $\tilde{M}(\omega)$, $\tilde{D}^i(\omega)$ and $\tilde{M}^{ij}(\omega)$.
We further introduce  relations between the energy-momentum tensor and the mass-energy moments using energy-momentum conservation in flat space time
\begin{align}
\int d^3x\tilde{T}^{ij}\left(\omega,\mathbf{x}\right) &= -\frac{\omega^2}{2}\int d^3 x x^i x^j\tilde{T}^{00}(\omega,\mathbf{x}) = -\frac{\omega^2}{2}\tilde{M}^{ij}\left(\omega\right),\\
\int d^3 x \tilde{T}^{0i}(\omega,\mathbf{x}) &= -i\omega\int d^3 x x^i \tilde{T}^{00}(\omega,\mathbf{x}) = -i\omega \tilde{D}^i(\omega),\\
\int d^3 x \tilde{T}^{ij}(\omega,\mathbf{x}) &= -i\omega \int d^3 x x^i \tilde{T}^{j0}(\omega,\mathbf{x}) = -\frac{\omega^2}{2} \tilde{M}^{ij}(\omega).
\end{align}

Now, we transform \eqref{propagator MCG small mass far field} back to real space, insert it into \eqref{general form of massive solution}, expand in $k_{\omega}\left|\mathbf{x}^\prime\cdot \mathbf{n}\right| \ll 1$ and keep terms up to the quadrupole contribution. This yields
\begin{widetext}
\begin{align}
\hat{\Psi}_{\mu\nu}\left(t,\mathbf{x}\right) &= -\frac{4 G}{r}\int\frac{d\omega}{2\pi}\int d^{3}x^\prime e^{-i\omega t} \left[ e^{i k_{\omega}r} \left(1-ik_{\omega,\epsilon}\mathbf{x^\prime} \cdot \mathbf{n}-\frac{k_{\omega,\epsilon}^2}{2}\left(\mathbf{x^\prime} \cdot 
\mathbf{n}\right)^2\right) + c.c. \right]\tilde{T}_{\mu\nu}\left(\omega,\mathbf{x^\prime}\right),
	\label{multipole expansion h_solution_case1}
\end{align}
where $k_{\omega,\epsilon} = \sqrt{\omega^2-\epsilon m_g^2}$ and $c.c.$ denotes the complex conjugate of the first term in the square bracket. The equal sign here means that this expression is exact up to the quadrupole contribution.
For the components we find
\begin{align}
\hat{\Psi}^{00} &= -\frac{4 G}{r}\int \frac{d\omega}{2 \pi} e^{-i\omega t} \left[e^{i k_{\omega,\epsilon}r} \left(\tilde{M}(\omega)-ik_{\omega,\epsilon}n^k \tilde{D}^{k}(\omega)-\frac{k_{\omega,\epsilon}^2}{2}n^k n^l \tilde{M}^{kl}(\omega)\right) + c.c. \right],						
	\label{00_solution_fourier}\\
\hat{\Psi}^{0i} &= -\frac{4 G}{r}\int \frac{d\omega}{2 \pi} e^{-i\omega t} \left[e^{ik_{\omega,\epsilon}r} \left(-i\omega \tilde{D}^{i}(\omega)-\frac{\omega}{2}k_{\omega,\epsilon}n^k \tilde{M}^{ki}(\omega)\right) + c.c. \right],
	\label{0i_solution_fourier}\\
\hat{\Psi}^{ij} &= \frac{2 G}{r}\int \frac{d\omega}{2 \pi} e^{-i\omega t} \left[e^{ik_{\omega,\epsilon}r} \left(\omega^2 \tilde{M}^{ij}(\omega)\right) + c.c. \right].
	\label{ij_solution_fourier}
\end{align}
We will later see that for the radiated energy we only need time derivatives of these components, because all spatial derivatives can be translated into time derivatives. They are given by
\begin{align}
\dot{\hat{\Psi}}^{00} &= -\frac{4 G}{r}\int \frac{d\omega}{2 \pi} e^{-i\omega t} \left[e^{ik_{\omega,\epsilon}r} \left(-i\omega \tilde{M}(\omega) - \omega k_{\omega,\epsilon} n_k \tilde{D}^{k}(\omega) + i\omega\frac{k_{\omega,\epsilon}^2}{2} n_k n_l \tilde{M}^{kl}(\omega)\right) + c.c. \right],
	\label{time_derivative_h_00}\\
\dot{\hat{\Psi}}^{0i} &= -\frac{4 G}{r}\int \frac{d\omega}{2 \pi}  e^{-i\omega t}\left[e^{ik_{\omega,\epsilon}r} \left(-\omega^2 \tilde{D}^{i}(\omega) + i\frac{\omega^2}{2}k_{\omega,\epsilon} n_k \tilde{M}^{ki}(\omega)\right) + c.c. \right],\\
\dot{\hat{\Psi}}^{ij} &= -\frac{2\tilde{G}}{r}\int \frac{d\omega}{2 \pi} e^{-i\omega t} \left[e^{ik_{\omega,\epsilon}r} \left(i\omega^3 \tilde{M}^{ij}(\omega)\right) + c.c. \right].
	\label{time derivative h_ij}
\end{align}
\end{widetext}

We note that we can expand $k_{\omega,\epsilon} = \omega  \sqrt{1-\epsilon \frac{m_g^2}{\omega^2}} \approx \omega \left(1-\epsilon\frac{m_g^2}{2 \omega^2}\right)$ for $\frac{m_g^2}{\omega^2} \ll 1$. The validity of this expansion will be shown in sec. \ref{sec:Energy-Loss-Due-grav-rad}. Using this expansion Eqs.~\eqref{time_derivative_h_00}-\eqref{time derivative h_ij} simplify to
\begin{widetext}
\begin{align}
\dot{\hat{\Psi}}^{00} &\approx -\frac{4 G}{r}\int \frac{d\omega}{2 \pi} e^{-i\omega t} \left[e^{ik_{\omega,\epsilon}r} \left( -i\omega \tilde{M}(\omega) + \omega^2 n_k \tilde{D}^k + i \frac{\omega^3}{2} n_k n_l \tilde{M}^{kl}(\omega) \right) + c.c. \right],
	\label{time derivative massive wave 00}\\
\dot{\hat{\Psi}}^{0i} &\approx -\frac{4 G}{r}\int \frac{d\omega}{2 \pi} e^{-i\omega t} \left[e^{ik_{\omega,\epsilon}r} \left( \omega^2 \tilde{D}^i - i\frac{\omega^3}{2} n_k \tilde{M}^{ki}(\omega) + c.c.\right) \right],
	\label{time derivative massive wave 0i}\\
\dot{\hat{\Psi}}^{ij} &\approx -i\frac{2 G}{r}\int \frac{d\omega}{2 \pi} e^{-i\omega t} \left[e^{ik_{\omega,\epsilon}r} \omega^3 \tilde{M}^{ij}(\omega) + c.c. \right].
	\label{time derivative massive wave ij}
\end{align}
\end{widetext}

In GR, one can go to the Transverse-Traceless (TT) gauge $\left(h_\mu^{TT\,\mu} = 0, h_{TT}^{0\mu} = 0, \partial^{j}h^{TT}_{ij} = 0\right)$ in vacuum and one only needs to calculate the spatial components of the metric perturbation. In CG/MCG, the choice of the gauge is more subtle, since there are more degrees of freedom than in GR. In principle, the massive graviton contributes five additional degrees of freedom.  Hence, by using the additional coordinate freedom left over after choosing the Teyssandier gauge, we can find the analog to the TT-gauge (see Appendix \ref{sub:TT-gauge} for details),
\begin{align}
H^{TT \mu}_{\mu} &= 0, \; \partial_{\mu}H_{TT}^{\mu\nu} = 0,\;  H^{TT}_{\mu0} = 0,\\
\Psi^{TT i}_{i} &= 0,\; \Psi^{TT}_{0i} = 0, \; \partial_{i}\partial_{j}\Psi_{TT}^{ij} = -\partial_{i}\partial^{i}\Psi^{TT}_{00}.
\end{align}
Note that for the massive part only the spatial trace is zero and the $00-$component does not vanish.

Nevertheless, we show that these additional modes are not excited by a conserved energy-momentum tensor.
Contracting \eqref{general form of massive solution} with a partial derivative yields
\begin{align}
\partial^\mu\hat{\Psi}_{\mu\nu}  = &\int_{V} d^4 x^\prime \left(\frac{\partial}{\partial x_\mu} \mathcal{G}(x-x^\prime)\right) T_{\mu\nu}(x^\prime)\nonumber\\
 = &- \int_V d^4 x^\prime \left(\frac{\partial}{\partial x^\prime_\mu} \mathcal{G}(x-x^\prime)\right) T_{\mu\nu}(x^\prime)\nonumber\\
 = &- \mathcal{G}(x-x^\prime) T_{\mu\nu}(x^\prime)\vert_{\partial V}\nonumber\\
&+ \int_{V} d^4 x^\prime  \mathcal{G}(x-x^\prime) \left(\frac{\partial}{\partial x^\prime_\mu}T_{\mu\nu}(x^\prime)\right)\nonumber\\
= \,&0,
	\label{harmonic gauge for massive part}
\end{align}
where we have used $\frac{\partial}{\partial x^\mu} \mathcal{G}(x-x^\prime)= - \frac{\partial}{\partial x^{\prime\mu}} \mathcal{G}(x-x')$ for the second equal sign and integration by parts for the third equal sign. Furthermore, we have chosen an integration volume $V$ that is larger than the source, such that $T_{\mu\nu}(x)$ vanishes on the boundary $\partial V$. The last expression vanishes due to matter energy-momentum conservation, see \eqref{matter energy-momentum conservation to first order}. Hence, although the Teyssandier gauge (see Appendix \ref{sub:TT-gauge}) does not lead to the harmonic gauge for the massive mode of the metric perturbation, a conserved energy-momentum tensor only excites the transverse modes and we get the harmonic gauge condition for the massive part automatically. By applying a further coordinate transformation in an analogous way as in GR without spoiling the harmonic gauge, we can bring both parts of the wave to the standard GR-TT-gauge
\begin{align}
H^{TT \mu}_{\mu} = 0, \; \partial_{\mu}H_{TT}^{\mu\nu} &= 0,\;  H^{TT}_{\mu0} = 0,\\
\Psi^{TT \mu}_{\mu} = 0,\; \partial_{\mu}\Psi_{TT}^{\mu\nu} &= 0,\; \Psi^{TT}_{\mu 0} = 0.
\end{align}
Note that in TT-gauge $\bar{H}_{\mu\nu} = H_{\mu\nu}$ and $\hat{\Psi}_{\mu\nu} = \Psi_{\mu\nu}$, since the traces vanish. 

Inserting \eqref{00_solution_fourier}-\eqref{ij_solution_fourier} explicitly into \eqref{harmonic gauge for massive part} leads to 
\begin{align}
\int \frac{d\omega}{2\pi} e^{i\omega t} \left[ e^{i k_{\omega,\epsilon} r} \left(-i \omega \tilde{M}(\omega) + cc \right)\right] &= 0, \\
\int \frac{d\omega}{2\pi} e^{i\omega t} \left[ e^{i k_{\omega,\epsilon} r} \left(- \omega^2 \tilde{D}^i(\omega) + cc \right)\right] &= 0.
\end{align}
This shows that the monopole and dipole contributions in \eqref{time derivative massive wave 00}-\eqref{time derivative massive wave ij}, which are the quantities that enter into the radiated energy, drop out and we are left only with the quadrupole contribution as for the massless mode. However, there is a phase difference between the massless and the massive mode varying with the distance to the source. This becomes obvious by the factor $\exp(-ik_\omega r)$ in \eqref{time derivative massive wave 00}-\eqref{time derivative massive wave ij}.

Let us calculate the explicit solution for the gravitational wave that is generated by a simple binary system in circular motion in the Newtonian limit, which can be described in the center of mass frame as one particle with the reduced mass $\mu$.
We choose the orbit such that it lies in the xy-plane and get for the relative coordinates
\begin{align}
x_{0}^{1}(t) &= -R\sin(\omega_{s}t),\\
x_{0}^{2}(t) &= R\cos(\omega_{s}t),\\
x_{0}^{3}(t) &= 0,
\end{align}
where $R$ is the radius of the source. We do not need to calculate the $0\mu$-components, because our aim is to calculate the radiated energy far away from the source, where we can use the TT-gauge. Therefore, we restrict here to calculate only the spatial components in the harmonic gauge and project the solutions into the TT-gauge when needed.

For a point particle of reduced mass $\mu$ in the non-relativistic limit we get for the quadrupole moment
\begin{equation}
M^{ij} = \mu x_{0}^{i}x_{0}^{j}.
\end{equation}
In components this reads
\begin{align}
M_{11} &= \mu R^{2}\frac{1 - \cos\left(2\omega_{s}t\right)}{2},\\
M_{22} &= \mu R^{2}\frac{1 + \cos\left(2\omega_{s}t\right)}{2},\\
M_{12} &= -\mu R^{2}\frac{\sin\left(2\omega_{s}t\right)}{2},\\
M_i^i &= \mu R^{2},
\end{align}
where $M_i^i$ is the spatial trace of the mass moment.
The time Fourier transform of these expressions is given by
\begin{align}
\tilde{M}_{11}(\omega) &= 
\frac{\mu R^{2}\pi}{2} \left[\delta\left(\omega\right) - \delta\left(\omega + 2\omega_{s}\right) - \delta\left(\omega - 2\omega_{s}\right)\right],
	\label{mass moment 11 omega}\\
\tilde{M}_{22}(\omega) &= \frac{\mu R^{2}\pi}{2} \left[\delta\left(\omega\right) + \delta\left(\omega + 2\omega_{s}\right) + \delta\left(\omega - 2\omega_{s}\right)\right],
	\label{mass moment 22 omega}\\
\tilde{M}_{12}(\omega) &= \frac{\mu R^{2}\pi}{2i} \left[\delta\left(\omega-2\omega_{s}\right) -\delta\left(\omega + 2\omega_{s}\right)\right],
	\label{mass moment 12 omega}\\
\tilde{M}^i_i(\omega) &= \mu R^{2}\pi \delta\left(\omega\right)
	\label{mass moment trace omega}.
\end{align} 
For CG and MCG with a small graviton mass, inserting \eqref{mass moment 11 omega}-\eqref{mass moment 12 omega} into \eqref{ij_solution_fourier},  we find the non-vanishing components for the massive mode, 
%\begin{widetext}
\begin{align}
\hat{\Psi}_{11}(t,r) = -\hat{\Psi}_{22}\left(t,r\right) &= -\frac{4 G \mu R^{2}\omega_{s}^{2}}{r}\cos\left(2\omega_{s}t_{m}\right) ,
	\label{solution_MG_h11_c13} \\
\hat{\Psi}_{12}(t,r) = \hat{\Psi}_{21}(t,r) &= -\frac{4 G \mu R^{2}\omega_{s}^{2}}{r}\sin\left(2\omega_{s}t_{m}\right),
	\label{solution_MG_h12_c13}\\
\hat{\Psi}_i^i(t,r) &= 0,
	\label{solution_MG_trace_c13}
\end{align}
%\end{widetext}
where $t_{m} = t - v_{g,\epsilon}r$ is the travel time and $v_{g,\epsilon} = \sqrt{1-\epsilon m_g^2/(4\omega_s^2)}$ is the speed of the massive gravitational wave.

To get the full solution to Eq.~\eqref{eom_Teyssandier}, we now add the massless mode of the metric perturbation to the massive mode in \eqref{solution_MG_h11_c13} and \eqref{solution_MG_h12_c13}. The derivation of the solution for the massless mode can be found in nearly every standard textbook about GR and gravitational waves, see e.g \citep{maggiore2008gravitational,weinberg1972gravitation} (it is analogous to the derivation of the massive mode in the small mass case of MCG, but just replacing $k_\omega$ by $\omega$ in sec. \ref{sub:Non-Vacuum-Solution}). We find 
\begin{widetext}
\begin{align}
h_{11}(t,r)  = -h_{22}(t,r) &= \frac{4 G \epsilon\mu R^{2}\omega_{s}^{2}}{r} \left[ \cos\left(2\omega_{s}t_{ret}\right) - \cos\left(2\omega_{s}t_{m}\right) \right],
	\label{eq: solution Nieuwenhuizen 1-1-1}\\
h_{12}(t,r) = h_{21}(t,r) &= \frac{4 G \epsilon\mu R^{2}\omega_{s}^{2}}{r} \left[\sin\left(2\omega_{s}t_{ret}\right) - \sin\left(2\omega_{s}t_{m}\right)\right],
	\label{eq: solution Nieuwenhuizen 2-1-1}
\end{align}
where $t_{ret} = t-r$ is the retarded time.

For MCG with a large graviton mass, we find
\begin{align}
\hat{\Psi}_{11}(t,r) = -\hat{\Psi}_{22}(t,r) &= \frac{-4 G \mu R^{2}\omega_{s}^{2}}{r}e^{-k_{\omega,>}r}\cos\left(2\omega_{s}t\right) ,
	\label{solution Psi 11}\\
\hat{\Psi}_{12}(t,r) = \hat{\Psi}_{21}(t,r) &= -\frac{4 G\mu R^{2}\omega_{s}^{2}}{r}e^{-k_{\omega,>}r}\sin\left(2\omega_{s}t\right), 
	\label{solution Psi 12}
\end{align}
and $\hat{\Psi}_i^i(t,r) = 0$. We combine this with the massless mode and get the final result
\begin{align}
h_{11}\left(t,r\right) = -h_{22}(t,r) &= \frac{4 G \mu R^{2}\omega_{s}^{2}}{r}\left[\cos(2\omega_{s}t_{ret})-e^{-k_{\omega,>}r}\cos(2\omega_{s}t)\right],			
	\label{eq: solution of case 2}\\
h_{12}(t,r) = h_{21}(t,r)  & = \frac{4 G \mu R^{2}\omega_{s}^{2}}{r}\left[\sin(2\omega_{s}t_{ret})-e^{-k_{\omega,>}r}\sin(2\omega_{s}t)\right],
	\label{eq: solution of case 2_2}
\end{align}
which is just the GR solution modified by an exponentially damped term.
\end{widetext}

\section{Energy-Momentum Tensor of Gravitational Waves\label{sec:Energy-Momentum-Tensor-of}}

To analyze the radiation emitted by sources like binary systems, we
need to calculate the explicit form of the gravitational energy-momentum
tensor in CG and MCG. 

We calculate the gravitational energy-momentum tensor via the corresponding Noether current. In order to do so, we have to expand the gravitational part of the total action 
\begin{equation}
I_{GRAV} = \frac{1}{16\pi G} \int d^{4}x\sqrt{-g}\left[ - m_g^{-2}\left(R_{\mu\nu}R^{\mu\nu} - \frac{1}{3}R^{2}\right) - \epsilon R \right]
\end{equation}
to second order in $h_{\mu\nu}$ and apply the TT-gauge. We find 

\begin{widetext}
\begin{equation}
I_{GRAV}^{TT\left(2\right)} = \frac{1}{64 \pi G}\int d^{4}x\left( -m_g^{-2} \Box h_{\rho\sigma}^{TT}\Box h_{TT}^{\rho\sigma} + \epsilon \partial_{\alpha}h_{\rho\sigma}^{TT}\partial^{\alpha}h_{TT}^{\rho\sigma}\right).
\end{equation}
The formula for an energy-momentum tensor of a fourth-order derivative theory is
given by 
\begin{equation}
\left(T^{GRAV}\right)_{\alpha}^{\lambda} = 
\frac{1}{\sqrt{-g}} \left\langle\left(\partial_{\xi}\frac{\partial\mathcal{L}}{\partial g_{\rho\sigma,\lambda\xi}} - \frac{\partial\mathcal{L}}{\partial g_{\rho\sigma,\lambda}}\right)g_{\rho\sigma,\alpha} - \frac{\partial\mathcal{L}}{\partial g_{\rho\sigma,\lambda\xi}}g_{\rho\sigma,\xi\alpha} + \delta_{\alpha}^{\lambda}\mathcal{L}\right\rangle,
\end{equation}
where the angle brackets denote the average over several wavelength or periods of the wave.
This leads to 
\begin{equation}
\left(T_{GRAV}^{\left(2\right)}\right)_{\alpha}^{\lambda} = \frac{1}{32 \pi G} \left\langle 2 m_g^{-2} \Box h^{TT}_{\rho\sigma}\partial_{\alpha}\partial^{\lambda}h_{TT}^{\rho\sigma} + \epsilon \, \partial^{\lambda}h^{TT}_{\rho\sigma} \partial_{\alpha}h_{TT}^{\rho\sigma}\right\rangle.
\end{equation}
Here, we have already discarded terms proportional to $\eta_\alpha^\lambda$, since they do not contribute to the radiated energy, cf. \eqref{emt in TT splitted}.
In vacuum with the help of \eqref{ansatz}, \eqref{massless_eom_vacuum} and \eqref{massive_eom_vacuum} it is possible to write this as
\begin{equation}
\left(T_{GRAV}^{\left(2\right)}\right)_{\alpha}^{\lambda} = \frac{1}{32\pi G} \left\langle 2 \Psi^{TT}_{\rho\sigma} \partial_\alpha \partial^\lambda h_{TT}^{\rho\sigma} + \epsilon\, \partial_\alpha h^{TT}_{ \rho\sigma} \partial^\lambda h_{TT}^{\rho\sigma} \right\rangle.
	\label{gravitational emt in vacuum}
\end{equation}
\end{widetext}

\section{Energy Loss Due to Gravitational Wave Emission\label{sec:Energy-Loss-Due-grav-rad}}

\subsection{Radiated Energy\label{sub:Radiated-Energy}}

In this section we want to calculate the amount of energy
that is radiated by binary systems. We use the conservation of the 
energy-momentum tensor in the far zone $\left(r \gg R\right)$ and set 
$T_{\mu\nu} = 0$. Hence we can go to TT-gauge. We find 
\begin{equation}
\partial_{0}T_{GRAV}^{0\nu}+\partial_{s}T_{GRAV}^{s\nu} = 0.
	\label{eq:  conservation of grav emt}
\end{equation}
%Integration of \eqref{eq:  conservation of grav emt}
%over a volume $V$, which is bigger than the source we look at, yields 
%\begin{equation}
%\int_V d^{3}x\left(\partial_{0}T_{GRAV}^{0\nu} + \partial_{s}T_{GRAV}^{s\nu}\right) = 0.
%	\label{energy momentum conservation in V}
%\end{equation}
The energy carried in $V$ by gravitational waves is given by $E_V = \int d^{3}xT_{GRAV}^{00}$.
By combining with Eq.~\eqref{eq:  conservation of grav emt} we find 
\begin{eqnarray}
\dot{E}_V &=& \int_{V} d^{3}x\,\partial_{0}T_{GRAV}^{00} \nonumber \\
                 &=& -\int_{V} d^{3}x\,\partial_{s}T_{GRAV}^{s0} \nonumber \\
                 & = & -r^2\int_{\partial V} d\Omega\,n_{s}T_{GRAV}^{s0},
	\label{radiated energy}
\end{eqnarray}
%\end{widetext}
where $d\Omega = \sin\theta\, d\theta\, d\phi$ is the differential solid angle and $\partial V$ is the surface of the volume $V$. The minus sign means that gravitational waves carry away energy flux from the volume. Hence, the radiated energy of gravitational waves has the opposite sign and we find 
\begin{equation}
\dot{E} = r^2\int_{\partial V} d\Omega\,n_{s}T_{GRAV}^{s0}.
\end{equation}
Therefore, the quantity of interest is
%\begin{widetext}
\begin{eqnarray}
\lefteqn{T_{GRAV}^{s0}n_s} \nonumber \\
%T_{GRAV}^{s0}n_s
& = & \frac{1}{32\pi G} n_s \left\langle 2 \Psi^{TT}_{\rho\sigma}\partial^s \partial^0 h_{TT}^{\rho\sigma} + \epsilon\, \partial^s h^{TT}_{ \rho\sigma}\partial^0 h_{TT}^{\rho\sigma} \right\rangle \\ 
&=&  \frac{\epsilon}{32\pi G} n_s\! \left\langle -\partial^s \Psi^{TT}_{ij} \partial^0 \Psi_{TT}^{ij} + \partial^s\! H^{TT}_{ij}\! \partial^0 H_{TT}^{ij} \right\rangle\!\! ,
	\label{emt in TT splitted}
\end{eqnarray}
%\end{widetext}
where we used \eqref{ansatz} and integration by parts in the second line. Only the spatial components contribute in TT-gauge.

\subsubsection{Small Graviton Mass \label{Small Graviton Mass}}

For CG and MCG with a small graviton mass ($m_g^2~<~4\omega_s^2$) we find
\begin{widetext}
\begin{align}
T_{GRAV}^{s0}n_s &= \frac{\epsilon}{32\pi G} \left\langle -\partial^0 \Psi^{TT}_{ij} \partial^0 \Psi_{TT}^{ij} + \partial^0 H^{TT}_{ij} \partial^0 H_{TT}^{ij} + \epsilon \frac{m_g^2}{8\omega_s^2} \partial^0 \Psi^{TT}_{ij} \partial^0 \Psi_{TT}^{ij} + \mathcal{O}\left( \frac{m_g^4}{\omega_s^4} \partial^0 \Psi_{ij}^{TT} \partial^0 \Psi_{TT}^{ij} \right)\right\rangle\nonumber\\
&\approx \frac{\epsilon}{32\pi G} \Lambda_{ijkl} \left\langle -\partial^0 \hat{\Psi}_{ij} \partial^0 \hat{\Psi}^{kl} + \partial^0 \bar{H}_{ij} \partial^0 \bar{H}^{kl} + \epsilon \frac{m_g^2}{8\omega_s^2} \partial^0 \hat{\Psi}_{ij} \partial^0 \hat{\Psi}^{kl}\right\rangle,
	\label{small graviton mass energy-momentum tensor}
\end{align}
%\end{widetext}
where we have used $\partial^s \bar{H}^{\rho\sigma} = \partial^0 \bar{H}^{\rho\sigma} n^s + \mathcal{O}(1/r^2) $, $\partial^s \hat{\Psi}^{\rho \sigma} = \partial^0 \hat{\Psi}^{\rho \sigma} [1 - \epsilon \, m_g^2/(8\omega_s^2) + \mathcal{O}(m_g^4/\omega_s^4)] n^s + \mathcal{O}(1/r^2)$ and $n_s n^s~=~1$ to find the second line.
In the third line we introduced the so-called Lambda-tensor $\Lambda_{ijkl}$, which projects $h^{ij}$ into the TT-gauge (see Appendix \ref{appendix A} for details).
Note that the second term is the same as in GR for $\epsilon = + 1$.
This shows that the contribution from the massless and the massive part of the metric perturbation have the same structure, but come with a relative sign.

We insert \eqref{small graviton mass energy-momentum tensor} in \eqref{radiated energy} and use 
\begin{equation}
\int d\Omega \, \Lambda_{ijkl} = \frac{2\pi}{15}\left( 11\delta_{ik}\delta_{jl} - 4\delta_{ij}\delta_{kl} + \delta_{il}\delta_{jk}\right),
	\label{lambda integration}
\end{equation}
to find 
\begin{equation}
\dot{E} \approx \frac{\epsilon\, r^2}{20 G} \left\langle -\partial^0 \hat{\Psi}_{ij} \partial^0 \hat{\Psi}^{ij} + \partial^0 \bar{H}_{ij} \partial^0 \bar{H}^{ij} + \epsilon \frac{m_g^2}{8\omega_s^2} \partial^0 \hat{\Psi}_{ij} \partial^0 \hat{\Psi}^{ij} \right\rangle.
	\label{radiated energy integrated}
\end{equation}
Inserting \eqref{eq: solution Nieuwenhuizen 1-1-1} and \eqref{eq: solution Nieuwenhuizen 2-1-1}  yields
%\begin{widetext}
\begin{equation}
\dot{E} \approx \epsilon\dot{E}_{GR}\left\langle - \sin^2(2\omega_s t_{m}) - \cos^2(2\omega_s t_{m}) + \sin^2(2\omega_s t_{ret}) + \cos^2(2\omega_s t_{ret}) \right\rangle + \frac{m_g^2}{8\omega_s^2} \dot{E}_{GR}
= \frac{m_g^2}{8\omega_s^2}\dot{E}_{GR},
	\label{radiated energy case 1/3}
\end{equation}
\end{widetext}
where 
\begin{equation}
\dot{E}_{GR} = \frac{32 G \mu^2 R^4 \omega_s^6}{5}.
\end{equation}
We note that \eqref{radiated energy case 1/3} is independent of $\epsilon$. 

For CG we have $m_{g,CG}^2/(8\omega_s^2) \approx 9.1 \times 10^{-6} $ and for MCG we have $m_{g,<}^2/(8\omega_s^2) < 10^{-5}$. Hence, the radiated energy is several orders of magnitude smaller than in GR.

Since we have shown in sec. \ref{NewtonianLimit} that a too small radiated energy directly translates into a too small decay of the orbital period, it seems that gravitational radiation cannot explain the measured decrease of the orbital period of binary systems in these theories.

\begin{center}
\begin{table}[b]

\caption{Summary of the radiated energy for CG and MCG with small and large graviton mass.}
\begin{tabular}{|c||c|c|}
  \hline
  $\epsilon$ & $-1$ & $+1$ \\
  \hline
  \hline
  & & \\[-9pt]
  $m_g^2 < 4\omega_s^2$ &  $\frac{m_g^2}{8\omega_s^2}\dot{E}_{GR}$   &  $\frac{m_g^2}{8\omega_s^2}\dot{E}_{GR}$ \\[7pt]
  $m_g^2 > 4\omega_s^2$ &   ---   &  $\dot{E}_{GR}$ \\[5pt]
  \hline
\end{tabular}
\end{table}
\end{center}	

\subsubsection{Large Graviton Mass}

For MCG with a large graviton mass ($\epsilon = +1$, $m_g^2 > 4\omega_s^2$) we use $\partial_r \hat{\Psi}_{\mu\nu} = -k_\omega \hat{\Psi}_{\mu\nu} + \mathcal{O}(1/r^2)$ in \eqref{emt in TT splitted} to find
\begin{equation}
T_{GRAV}^{s0}n_s \approx \frac{1}{32\pi G} \Lambda_{ij,kl}\left\langle k_\omega \hat{\Psi}_{ij} \partial^0 \hat{\Psi}^{kl} + \partial^0 \bar{H}_{ij} \partial^0 \bar{H}^{kl} \right\rangle, 
	\label{emt in TT splitted large mass}
\end{equation}
with $k_{\omega} = \sqrt{m_g^2 - 4\omega_s^2}$.
The second term in \eqref{emt in TT splitted large mass} gives the same contribution as in GR. To calculate the first term, we use Eqs.~\eqref{solution Psi 11}-\eqref{solution Psi 12}. We find $k_\omega \hat{\Psi}_{ij} \partial^0 \hat{\Psi}^{kl} \propto k_\omega e^{-2k_\omega r}\sin{(2\omega_s t)}\cos(2\omega_s t)$, which vanishes in combination with the average over several periods of the wave.  Hence, MCG with a large graviton mass reproduces the GR result exactly. 
We get
\begin{equation}
\dot{E} = \frac{r^2}{20G} \left\langle \partial^0 \bar{H}_{ij} \partial^0 \bar{H}^{ij} \right\rangle = \dot{E}_{GR}.
\end{equation}

Therefore, MCG with a large graviton mass represents a theory that still needs dark matter to explain galaxy rotation curves, but accounts for the decay of the orbital period due to gravitational waves. 
On macroscopic distance scales like $r \gg m_g^{-1}$ it can be split into GR plus small contributions from the higher derivative terms. Therefore, it is reasonable to expect that also the other tests of gravity can be passed. Only on very small scales, where the higher derivative terms become important, a significant deviation from GR is expected. This is the reason, why this theory is renormalizable \citep{1977StelleRenormalizationhigherderivativePRD}.

\section{Summary, Conclusion and Outlook}

In this work we have investigated gravitational radiation from the binary system 
PSR J1012+5307 in CG and MCG. Both theories belong to the class of models 
containing higher derivatives and are invariant under Weyl rescaling. 
The action is given by a $C^2$-term which contributes the higher-derivative 
part and a term that resembles the Einstein-Hilbert term in the Weyl gauge $S(x) = S_{0}$ and $\tilde{G} = G$.
By introducing the parameter $\epsilon = \pm 1$, we distinguished between CG and MCG. The difference 
between these two theories is the sign in front of the Einstein-Hilbert term and both signs are allowed by 
the Weyl symmetry. This choice of sign does not only change the results for gravitational radiation, but also 
changes the properties of the gravitational wave. We have argued that in CG ($\epsilon = -1$) the choice of 
sign leads to metric perturbations which can be written as a massless ghost field and a massive tachyon. 
Whereas in MCG ($\epsilon = +1$), the massless mode is healthy and the massive mode is a ghost, but 
not a tachyon. A ghost field represents a severe problem for a theory, but as we discussed in 
Sec.\ \ref{sub:Ghosts}, 
there seem to be solutions to the ghost problems in CG and MCG \cite{Mannheim2011,2016FariaQuantummassiveconformalTEPJC}. 

In section \ref{sec:Linearisation} we derived the inhomogeneous
linearized field equations in the Teyssandier gauge for the metric perturbation given in Eq.~\eqref{eom_Teyssandier}.
These equations are higher-derivative partial differential equations
for a partially massive field. It was shown that one can divide this equation into a massless and massive mode, see 
\eqref{massless_eom} and \eqref{massive_eom_hat}. Since the solution to the massless part is known from GR, we 
only investigated the massive part. In principle, the massive part contributes five additional degrees of freedom, 
including monopole and dipole radiation. However, in Sec.\ \ref{Radiated Energy} we have shown that these additional 
degrees of freedom are not excited by a conserved energy-momentum tensor (for non-relativistic binaries it is 
mass conservation) and hence monopole and dipole radiation vanish. This means that only the transverse 
modes contribute. 

We found solutions for three different cases. For CG the massive mode has the same form as in GR, but travels 
faster than the speed of light. In the case of MCG with a small graviton mass the solution is the same, but the 
sign is different and the velocity is smaller than the speed of light. For MCG in the case of a large graviton mass, 
the massive terms are damped exponentially, such that in the limit of a large graviton mass GR is recovered.

To calculate the energy radiated by an idealized binary system, we derived the gravitational 
energy-momentum tensor in section \ref{sec:Energy-Momentum-Tensor-of}. It has a 
contribution from the massless mode that is the same as in GR (the sign depends on $\epsilon$)
and additional contributions that depend on the massive mode of the metric perturbation.
Most importantly, there is a relative sign between the various contributions 
which can lead to cancellations and reduces the efficiency of gravitational wave emission 
in certain regions of the parameter space.

Finally, in section \ref{sec:Energy-Loss-Due-grav-rad} we were able to calculate the radiated 
energy in the Newtonian limit for a binary system in circular motion. In CG and MCG with a 
small graviton mass (small compared to the orbital frequency of the binary system), we find 
the radiated energy to be much smaller than in GR. For CG we fixed 
the graviton mass by the analysis of galaxy rotation curves without the introduction of 
dark matter, $m_{g,CG} = \SI{1.1d-22}{eV}$, which turns out to fall into the small mass regime. 

Hence, CG and MCG with a small graviton mass  cannot explain the decay of the orbital period
via gravitational radiation. Nevertheless, one could think of another mechanism to account for 
the shrinkage of the orbits of binary systems. A suggestion in this direction is given 
in \citep{Mannheim2011}. Thus our result does not rule out CG, as we cannot exclude the 
existence of such another mechanism, but it makes CG a less attractive solution to the dark 
matter problem.

MCG cannot fit galaxy rotation curves without dark matter, but experiments on the inverse 
square law of the Newtonian potential constrain the graviton mass to the ranges 
$m_{g,<} < \SI{d-22}{eV}$ or $m_{g,>} > \SI{d-2}{eV}$.
 
The application and extension of our findings to coalescing binaries, as observed by 
gravitational wave interferometers and for compact stars followed up by telescopes at 
various wavebands is most interesting and will be presented in another work.  

Most interestingly, MCG with a large graviton mass (i.e. $m_{g,>} > \SI{d-2}{eV}$) 
shows properties close to GR. As it contains 
GR as a limit, MCG is expected to pass all tests of GR on length scales $r \gg m_g^{-1}$. 
And besides, due to its higher-derivative nature, it seems to be a renormalizable model for gravity
\cite{1977StelleRenormalizationhigherderivativePRD}.
Thus this model seems to offer interesting opportunites for future work.

\begin{acknowledgments}

We wish to thank Felipe F. Faria for valuable comments and discussions. PH and DJS acknowledge financial support from Deutsche
Forschungsgemeinschaft (DFG) under grant RTG 1620 `Models of Gravity'.
We also thank the COST Action CA15117 `Cosmology
and Astrophysics Network for Theoretical Advances and Training Actions (CANTATA)', 
supported by COST (European Cooperation in Science and Technology).
\end{acknowledgments}

\appendix

\section{\label{appendix A}Conventions}

The signature of the metric is 
\begin{equation}
g = \text{diag}\left(-,+,+,+\right).
\end{equation}
The Christoffel Symbols are defined by 
\begin{equation}
\Gamma_{\kappa\mu}^{\lambda} = 
\frac{1}{2}g^{\lambda\rho} \left(\partial_{\kappa}g_{\rho\mu} + \partial_{\mu}g_{\rho\kappa} - 
\partial_{\rho}g_{\kappa\mu}\right)
\end{equation}
and the Riemann tensor is given by 
\begin{equation}
R_{\mu\nu\kappa}^{\lambda} = 
- \left(\partial_{\nu}\Gamma_{\mu\kappa}^{\lambda} - 
	\partial_{\kappa}\Gamma_{\mu\nu}^{\lambda} + 
	\Gamma_{\nu\alpha}^{\lambda}\Gamma_{\mu\kappa}^{\alpha} - 
	\Gamma_{\kappa\alpha}^{\lambda}\Gamma_{\mu\nu}^{\alpha}\right).
\end{equation}
From this we find the Ricci tensor $R_{\mu\kappa} = g^{\lambda\nu} R_{\lambda\mu\nu\kappa}$ and the Ricci scalar $g^{\mu\kappa} R_{\mu\kappa}$. The Einstein equations in the convention used by Mannheim and Weinberg \cite{weinberg1972gravitation} read 
\begin{equation}
G_{\mu\nu} \equiv R_{\mu\nu} - \frac{1}{2}g_{\mu\nu}R =
- 8\pi G T_{\mu\nu} + \Lambda g_{\mu\nu}.
\end{equation}
The Weyl tensor is given by the expression 
\begin{widetext}
\begin{equation}
C_{\lambda\mu\nu\kappa} = R_{\lambda\mu\nu\kappa}+\frac{1}{6} R \left[g_{\lambda\nu}g_{\mu\kappa}-g_{\lambda\kappa}g_{\mu\nu}\right] -\frac{1}{2}\left[g_{\lambda\nu}R_{\mu\kappa}-g_{\lambda\kappa}R_{\mu\nu}-g_{\mu\nu}R_{\lambda\kappa}+g_{\mu\kappa}R_{\lambda\nu}\right].
	\label{eq: Weyl Tensor} 
\end{equation}
In the following we give a list of the curvature tensors, expanded around flat spacetime, at first
order in $h_{\mu\nu}$
\begin{align}
R_{\nu\rho\sigma}^{\mu\left(1\right)} & = \frac{1}{2}\left(-\partial_{\nu}\partial_{\rho}h_{\sigma}^{\mu}-\partial^{\mu}\partial_{\sigma}h_{\nu\rho}+\partial^{\mu}\partial_{\rho}h_{\nu\sigma}+\partial_{\nu}\partial_{\sigma}h_{\rho}^{\mu}\right),\\
R_{\mu\nu}^{\left(1\right)} & = \frac{1}{2}\left(\boxempty h_{\mu\nu}-\partial_{\rho}\partial_{\mu}h_{\nu}^{\rho}-\partial_{\nu}\partial_{\rho}h_{\mu}^{\rho}+\partial_{\mu}\partial_{\nu}h\right),\\
R^{\left(1\right)} & = \boxempty h-\partial_{\mu}\partial_{\nu}h^{\mu\nu}.
\end{align}
At second order in $h_{\mu\nu}$ the Ricci tensor is given by 
\begin{align}
R_{\mu\nu}^{\left(2\right)} = &-\frac{1}{2}h^{\rho\sigma}\left[\partial_{\mu}\partial_{\nu}h_{\rho\sigma}-\partial_{\nu}\partial_{\rho}h_{\mu\sigma}-\partial_{\sigma}\partial_{\mu}h_{\rho\nu} + \partial_{\rho}\partial_{\sigma}h_{\mu\nu}\right]\nonumber\\
&+ \frac{1}{4} \left[2\partial_{\sigma}h_{\rho}^{\sigma}-\partial_{\rho}h\right]\left[\partial_{\nu}h_{\mu}^{\rho}+\partial_{\mu}h_{\nu}^{\rho}-\partial^{\rho}h_{\mu\nu}\right]\nonumber\\
&- \frac{1}{4}\left[\partial_{\rho}h_{\sigma\nu} + \partial_{\nu}h_{\sigma\rho}-\partial_{\sigma}h_{\rho\nu}\right] \left[\partial^{\rho}h_{\mu}^{\sigma}+\partial_{\mu}h^{\sigma\rho} - \partial^{\sigma}h_{\mu}^{\rho} \right].
\end{align}
Since we use the Ricci tensor in the action integral, we can use integration by parts. Using  
TT-gauge we find
\begin{align}
R_{\mu\nu}^{(2) TT} =
\frac{1}{4}\partial_{\nu}h^{TT}_{\sigma\rho}\partial_{\mu}h_{TT}^{\sigma\rho} - \frac{1}{2}\partial_{\rho}h^{TT}_{\sigma\nu}\partial^{\rho}h_{TT \mu}^{\sigma}.
\end{align}
The Ricci scalar is given by
\begin{equation}
R^{(2) TT} = \frac{1}{4}\Box h_{TT}^{\rho\sigma}h^{TT}_{\rho\sigma}.
\end{equation}
\end{widetext}

Let us also define the Lambda tensor, which is the projector into TT-gauge. It is given by
\begin{align}
\Lambda_{ijkl} = 
&\; \delta_{ik} \delta_{jl} - \frac{1}{2} \delta_{ij} \delta_{kl} - n_j n_l\delta_{ik} - n_i n_k \delta_{jl} \nonumber\\
&+\frac{1}{2} n_k n_l \delta_{ij} + \frac{1}{2} n_i n_j \delta_{kl} + \frac{1}{2} n_i n_j n_k n_l,
\end{align}
where $n_i$ denotes the spatial unit vector pointing into the direction of wave propagation.
The Lambda tensor has some useful properties:
\begin{align}
\Lambda_{ijmn} &= \Lambda_{ijkl} \Lambda^{kl}_{\ \ mn},\\
\Lambda^i_{\ ikl} &= \Lambda^{\ \ k}_{ij\ k} = 0,\\
n^i \Lambda_{ijkl} &= 0,\\
n^j \Lambda_{ijkl} &= 0.
\end{align}

\section{\label{appendix B}Generalized Gauge Condition\label{sub:TT-gauge}}

In order to find the physical degrees of freedom of CG and MCG we have to 
choose gauge fixing conditions. In higher-order derivative theories, it is convenient to 
choose a generalization of the harmonic gauge, the so-called Teyssandier gauge 
\citep{1989_Teyssandier_LinearisedR+R2gravity_anewgaugeandnewsolutions_CaQG}. 
To show the usefulness of this gauge, let us start with gauging the theories in a naive way, 
similar to how it is usually done for GR.

To find the number of physical degrees of freedom, it is enough to study gravitational waves 
that propagate in vaccum. 
In GR, the metric perturbation is a symmetric $4\times4$-matrix and has 10 independent components. We are free to perform a coordinate transformation 
\begin{equation}
x_\mu \longrightarrow x^{\prime \mu} = x^\mu + \xi^\mu(x),
\end{equation}
where $\vert\partial_\mu\xi_\nu\vert$ is of order $\vert h_{\mu\nu}\vert$. The trace-reversed metric perturbation transforms like
\begin{equation}
\bar{h}_{\mu\nu}(x)\longrightarrow \bar{h}^\prime_{\mu\nu}(x^\prime)=\bar{h}_{\mu\nu}(x)-(\partial_\mu\xi_\nu + \partial_\nu\xi_\mu - \eta_{\mu\nu}\partial_\rho \xi^\rho)
\end{equation}
and hence
\begin{equation}
\partial^\nu\bar{h}_{\mu\nu}\longrightarrow \left(\partial^\nu \bar{h}_{\mu\nu}\right)^\prime = \partial^\nu\bar{h}_{\mu\nu}-\Box\xi_\mu. 
\end{equation}
So to find the harmonic gauge condition one has to choose 
\begin{equation}
\Box \xi_\mu= \partial^\nu \bar{h}_{\mu\nu}. 
\end{equation}
These four conditions reduce the degrees of freedom to six. Nevertheless, this does not fix the gauge freedom completely. One can do a residual coordinate transformation
\begin{equation}
x^{\prime \mu} \longrightarrow x^{\prime\prime\mu} = x^{\prime\mu} + \zeta^\mu,
\end{equation}
where $\vert\partial_\mu\zeta_\nu\vert$ is again of the order of $\vert h_{\mu\nu}\vert$. This leads to 
\begin{equation}
\bar{h}^\prime_{\mu\nu}(x)\longrightarrow \bar{h}^{\prime\prime}_{\mu\nu}(x^{\prime}) = 
\bar{h}^\prime_{\mu\nu}(x) - (\partial_\mu\zeta_\nu + \partial_\nu\zeta_\mu - \eta_{\mu\nu}\partial_\rho \zeta^\rho).
	\label{2nd coordinate transformation}
\end{equation}
Since we do not want to spoil the harmonic gauge condition, we have to demand
\begin{equation}
\Box \zeta_\mu = 0.
\end{equation}
For simplicity we only look at a single mode and find the plane wave solution to this equation
\begin{equation}
\zeta_\mu = c_\mu e^{ik_\rho x^\rho}+ c.c..
\end{equation}
In the following, we will also suppress the complex conjugate (c.c.). Here $c_\mu$ represents four arbitrary constants for fixed wavenumber $k_\mu$, which is light-like $\left(k_\rho k^\rho = 0\right)$.
Inserting this into \eqref{2nd coordinate transformation} one can explicitly use the four components of $\zeta_\mu$ to set components of $\bar{h}'_{\mu\nu}$ to zero. In the TT-gauge these functions are chosen in order to get 
\begin{align}
\bar{h}_{00} &= 0,\\
\partial^j \bar{h}_{ij} &= 0,\\
\bar{h}_{0i} &= 0, \\
\bar{h}_i^{\;i} &= 0.
\end{align}

For CG/MCG naively one could apply the same procedure. The difference is that there is now also a massive part of the metric perturbation
\begin{equation}
\bar{h}_{\mu\nu} = \epsilon(\bar{H}_{\mu\nu} + \bar{\Psi}_{\mu\nu}),
\end{equation}
where $\bar{H}_{\mu\nu}$ corresponds to the massless and $\bar{\Psi}_{\mu\nu}$ to the massive part of the spin-2 field.
Hence, there are 20 independent components now. To analyze this, let us expand the metric perturbation in Fourier modes
\begin{equation}
\bar{h}_{\mu\nu} = \epsilon (\bar{a}_{\mu\nu} e^{ik_\rho x^\rho} +  \bar{b}_{\mu\nu} e^{il_\rho x^\rho}),
	\label{fourier expansion of h}
\end{equation}
where $l^\rho l_\rho = -\epsilon m_g^2$. $\bar{a}_{\mu\nu}$ and $\bar{b}_{\mu\nu}$ are called the polarization tensors.
Inserting Eq.~\eqref{fourier expansion of h} into the harmonic gauge condition we get
\begin{align}
k^\nu \bar{a}_{\mu\nu} &= 0,\\
l^\nu \bar{b}_{\mu\nu} &= 0.
\end{align}
These 8 conditions reduce the degrees of freedom to 12. Now, there appears a problem. Although it is possible to bring the massless part to the standard TT-gauge, it is not possible to set terms of the massive part to zero, since $\zeta_\mu$ is a light-like vector field, which cannot cancel a massive wave. But since we still have 8 independent components, there has to be one more condition, since a massless and a massive spin-2 field should have only 7 degrees of freedom. This is the reason why it is more convenient to use the Teyssandier gauge. Let us briefly derive this gauge here.

The field equations and gauge conditions for the massless and the massive part of the wave are shown in sec. \ref{sec:Linearisation}.

In Eq. \eqref{Tgauge} we have chosen the Teyssandier gauge condition. But the gauge freedom is not fixed completely and hence we can do another coordinate transformation.
Under a coordinate transformation, $x^\mu \longrightarrow x^{\prime\mu} = x^\mu~+~\zeta^\mu$, this quantity transforms like
\begin{equation}
Z^{\prime\mu} = Z^\mu - \epsilon m_g^{-2}\left(\Box-\epsilon m_g^2\right)\Box\zeta^\mu.
\end{equation}
Again, to not spoil the Teyssandier gauge condition $Z_\mu~=~0$ we have to demand 
\begin{equation}
\left(\Box-\epsilon m_g^2\right)\Box\zeta^\mu = 0.
\end{equation}
The solution to this equation is
\begin{equation}
\zeta^\mu = c^\mu e^{ik_\rho x^\rho} + d^\mu e^{il_\rho x^\rho}.
\end{equation}
We look only at one mode and discard the c.c. for simplicity. $c^\mu$ and $d^\mu$ are arbitrary constants for fixed wavenumbers $k_\mu$ and $l_\mu$ $(k_\rho k^\rho = 0 \text{ and } l_\rho l^\rho = -\epsilon m_g^2)$. The second term describes a massive vector field and hence it is possible to set components of the massive mode of the metric perturbation to zero. 
The massless and massive part of the metric perturbation expanded in Fourier modes transform like
\begin{align}
a^\prime_{\mu\nu} &= a_{\mu\nu} - i(k_\mu c_\nu + k_\nu c_\mu),
	\label{transformation law for a}\\
b^\prime_{\mu\nu} &= b_{\mu\nu} - i(l_\mu d_\nu + l_\nu d_\mu)
	\label{transformation law for a}.
\end{align}
We bring the massless part to the TT-gauge as in GR. With no loss of generality we choose the wave propagating in the z-direction, $k^\mu = \left(k,0,0,k\right)$. From the gauge \eqref{massless_gauge} for the massless part we get
\begin{align}
\bar{a}_{00} &= -\bar{a}_{30},\\
\bar{a}_{01} &= -\bar{a}_{31},\\
\bar{a}_{02} &= -\bar{a}_{32},\\
\bar{a}_{03} &= -\bar{a}_{33},
\end{align}
and hence $\bar{a}_{00} = \bar{a}_{33}$. Using this, one can show that $a = -a_{00} + a_{33}$ and $a_{11} + a_{22} = 0$. For the trace we find 
\begin{equation}
a^\prime = a - 2ik^\rho c_\rho.
\end{equation}
We can set $a^\prime = 0$ if we choose 
\begin{equation}
c_{0} = \frac{-a_{00} + a_{33}}{2ik} - c_3.
\end{equation}
Using \eqref{transformation law for a} we also see that $a^\prime_{11} + a^\prime_{22} = 0$, because $a_{11}$ and $a_{22}$ do not transform under this coordinate transformation. To set $a^\prime_{0i} = 0$, we have to choose
\begin{align}
c_1 &= -\frac{a_{01}}{ik},\\
c_2 &= -\frac{a_{02}}{ik},\\
c_3 &= \frac{a_{03}-a}{2ik}.
\end{align}
Inserting this in the harmonic gauge condition yields
\begin{align}
\bar{a}^\prime_{00} &= 0,\\
\bar{a}^\prime_{33} &= 0,\\
\bar{a}^\prime_{31} &= 0,\\
\bar{a}^\prime_{32} &= 0.
\end{align}
This brings the massless part to the convenient TT-gauge
\begin{align}
H^{TT}_{00} &= 0,\\
\partial^j H^{TT}_{ij} &= 0,\\
H^{TT}_{0i} &= 0, \\
H_i^{TT \,i} &= 0.
\end{align}

For the massive mode we can proceed analogously. 
The gauge condition $\partial_\rho \partial_\sigma \Psi^{\rho\sigma} = \Box \Psi$ yields
\begin{equation}
(l^0)^2 b_{00} + 2 l^0 l_3 b_{03} +(l_3)^2 b_{33} = -\epsilon m_g^2 b,									\label{massive_harmonic_gauge}
\end{equation}
where we again have chosen the massive part to travel in the z-direction with $l^\mu = \left(l,0,0,\sqrt{l^2 - \epsilon m_g^2}\right)$.
We choose	 
\begin{align}
d_1 &= \frac{b_{01}}{il_0},\\
d_2 &= \frac{b_{02}}{il_0},\\
d_3 &= \frac{b_{03}}{il_0} - \frac{l_3}{2i\left(l_0\right)^2} b_{00},
\end{align}
to set $b^{\prime}_{0i} = 0$. Inserting this back into the transformed \eqref{massive_harmonic_gauge} we find the condition
\begin{equation}
\left(l^0\right)^2 b^{\prime}_{00} + (l_3)^2 b^{\prime}_{33} = -\epsilon m_{g}^{2} b_{\mu}^{\prime \mu}.
\end{equation}
We have the freedom to choose $d_0$ to set to zero either $b^{\prime}_{00}$, $b^{\prime}_{33}$, $b^{\prime i}_i$ or $b^{\prime \mu}_\mu$.
Thus, one choice for the completely gauge-fixed massive mode is
\begin{align}
\Psi^{TT}_{0i} &= 0,\\
\Psi_i^{TT \,i} &= 0,\\
\Psi^{TT}_{00} &= -\Psi^{TT}_{33},
\end{align}
which reduce the degrees of freedom of the massive mode to five.

\section{\label{appendix E}Ghosts and Tachyons\label{sub:Ghosts and Tachyons}}

In this appendix we want to study the properties of a free scalar field
$S(x)$ in flat spacetime.
We investigate
\begin{equation}
I_{M} = \int d^{4}x \sqrt{-g} \frac{\epsilon}{2} \left(\nabla_\mu S \nabla^\mu S - m_{s}^{2}S^{2}\right),
	\label{eq:free_scalar_field_action}
\end{equation}
where $m_{s}^{2} = R/6$ represents the mass of the scalar field $S(x)$. The equation of motion is given by 
\begin{equation}
\left(\nabla_\mu \nabla^\mu - m_{s}^{2}\right)S = 0.
	\label{eq:free_scalar_wave_equation}
\end{equation}
By defining the conjugate momentum $\pi_{S} = \epsilon\sqrt{-g}\,\nabla_{0}{S}$
we find the Hamiltonian density 
\begin{equation}
{\cal H} = \pi_{S}\nabla_0 S - {\cal L} = \frac{\epsilon}{2} \sqrt{-g} \left(-\nabla_0 S \nabla_0 S - \left(\nabla S\right)^{2} + m_{s}^{2}S^{2}\right).
	\label{eq:free_scalar_hamilton_density}
\end{equation}
From this we can derive the stability properties of the scalar field, which are summarized in Table \ref{stability of scalar field}.
We see that the sign of $\epsilon$ and the Ricci scalar are crucial for the properties of the scalar field. Nevertheless, since in CG and MCG the scalar field represents no real degree of freedom, it is not necessary that it is a healthy field. We can always choose a Weyl gauge, which sets the scalar field to a constant.

\begin{center}
\begin{table}[h]
\caption{This table shows the stability properties of the scalar field minimizing the action \eqref{eq:free_scalar_field_action}.}
\begin{tabular}{|c|ccc|}
  \hline
  \diagbox{$\epsilon$}{$R$}
                 & $<0$ & $>0$ & \\
  \hline
  -1     &   healthy     &    tachyon      &   \\
  +1     &    tachyonic ghost    &   ghost       &     \\
  \hline
\end{tabular}
	\label{stability of scalar field}
\end{table}
\end{center}

\section{\label{appendix C}Analysis of the Newtonian Limit\label{sub:Analysis-of-the-Newtonian-Limit}}

Let us investigate the Newtonian limit of MCG.
We use the wave equations \eqref{massive_eom} and \eqref{massless_eom}
and make the following assumptions corresponding to the Newtonian limit
\begin{align}
\partial_{t}h_{\mu\nu} &= 0,\\
T_{00}^{Newt} &\approx M\eta_{\mu 0}\eta_{\nu 0}\delta^{(3)}(\mathbf{r}),\\
T^{Newt} &\approx -M \delta^{(3)}(\mathbf{r}),\\
\partial_{t}\rho &= 0,
\end{align}
where $T_{00}^{Newt}$ is the time-time component of the matter energy-momentum
tensor $T_{\mu\nu}^{Newt}$ in the Newtonian limit and $T^{Newt}$ is the trace. $M$ is the mass of the point source at $r = 0$ and we have neglected the pressure $p$.
Inserting this into \eqref{massive_eom} and \eqref{massless_eom}
we find 
\begin{align}
\triangle H_{\mu\nu} & = 
-16\pi \tilde{G} M\left(\eta_{\mu 0}\eta_{\nu 0} + \frac{1}{2}\eta_{\mu\nu}\right)\delta^{(3)}(\mathbf{r}),
	\label{massless eom newt}
\\
\left(\triangle-m_{g}^{2}\right)\Psi_{\mu\nu}\left(r\right)& = 
16\pi \tilde{G} M\left(\eta_{\mu 0}\eta_{\nu 0}+\frac{1}{3}\eta_{\mu\nu}\right)\delta^{(3)}(\mathbf{r}).
	\label{massive eom newt}
\end{align}

First, let us find the vacuum solutions to these equations.
In the Newtonian limit we can write the line element as
\begin{equation}
ds^2 = \left( -1 -2\Phi(r) \right)dt^2 + \left(1 - 2\Theta(r)\right)dr^2 + r^2 d\Omega^2,
\end{equation}
where $h_{00} = -2\Phi$, $h_{rr} = -2\Theta$ and $d\Omega^2 = d\theta^2 + \sin^2\theta d\phi^2$ is the line element of a unit 2-sphere.
The $00$-component of \eqref{massless eom newt} and \eqref{massive eom newt} in the vacuum yields
\begin{equation}
\Phi\left(r\right)=c_{0}+\frac{c_{1}}{r}+\frac{c_{2}e^{-m_{g}r}}{r}+\frac{c_{3}e^{m_{g}r}}{r},
\end{equation}
where $c_0, \,c_1, \,c_2$ and $c_3$ are arbitrary constants.
The constant term has no physical relevance, so we set $c_0 = 0$. The condition of asymptotic flatness demands $c_3 = 0$. For the other constants it is convenient to choose $\,c_{1} = -GM$ and $c_{2} = -GM \alpha$. This leads to
\begin{equation}
\Phi\left(r\right) = -\frac{GM}{r} \left(1 + \alpha e^{-m_{g}r}\right).					\label{grav_pot_MCG}
\end{equation}

The point source solution to \eqref{massless eom newt} and \eqref{massive eom newt} in spatial Fourier space reads
\begin{equation}
\tilde{h}_{\mu\nu}(k) = 
16\pi \tilde{G} M \left[\frac{\eta_{\mu0}\eta_{\nu0} + \frac{1}{2}\eta_{\mu\nu}}{k^2} + \frac{\eta_{\mu0}\eta_{\nu0} + \frac{1}{3} \eta_{\mu\nu}}{k^2 + m_g^2}\right].
\end{equation}
In real space this yields
\begin{equation}
h_{\mu\nu} = \frac{4\tilde{G}}{r} \left(\eta_{\mu0}\eta_{\mu0}+\frac{1}{2}\eta_{\mu\nu}\right)-\frac{4 \tilde{G}}{r} e^{-m_gr} \left(\eta_{\mu0}\eta_{\mu0}+\frac{1}{3}\eta_{\mu\nu}\right),
	\label{metric perturbation in Newtonian limit}
\end{equation}
where we have chosen boundary conditions of asymptotic flatness.
From the $00$-component we find for the Newtonian potential
\begin{equation}
\Phi(r) = -\frac{\tilde{G} M}{r}\left( 1 - \frac{4}{3} e^{-m_g r}\right).
	\label{newtonian point source solution}
\end{equation}
Choosing $\tilde{G} = G$ and comparing with \eqref{grav_pot_MCG} we get $\alpha = -4/3$. The limit $m_g r \gg 1$ yields just the standard Newtonian potential. For $m_g r \ll 1$ one gets $\Phi(r) = G M/ (3 r)$, which leads to a repulsive gravitational force. This points out that the additional Yukawa potential cannot serve to fit galaxy rotation curves without dark matter in any parameter range, since it always comes with the wrong sign, see \citep{2011StabileScelzaRotationcurvesgalaxiesPRD}.

In literature there also exists another choice for the gravitational potential. This is the phenomenological approach by Sanders \citep{1990SandersMassdiscrepanciesgalaxiesTAaAR}, which is also adopted by MCG \citep{faria2014massive,2013MishraSinghFourthordergravityPRD}. 
In this case the gravitational potential exterior to a source, given by
\begin{equation}
\Phi = -\frac{GM}{r \left(1 + \delta\right)} \left[1 + \delta e^{-m_{g,S}\,r}\right],						\label{eq: Sanders solution}
\end{equation}
with parameters $\delta = -0.92$ and $m_{g,S} \approx \SI{1.6d-28}{eV}$ ($m_{g,S}$ is the graviton mass) \citep{1990SandersMassdiscrepanciesgalaxiesTAaAR}, has been used to fit galaxy rotation curves without dark matter.
The standard Newtonian potential is recovered in the limit $m_{g}r \ll 1$. 
Trying to match \eqref{eq: Sanders solution} with \eqref{newtonian point source solution} seems to be impossible, unless the massive part of the metric perturbation couples differently to matter than the massless part.

Assuming that it is possible to derive such a potential in some way, let us calculate the decay of the orbital period of the binary system. 
We find
\begin{equation}
\frac{\dot{P}}{P} \approx - \frac{\vert E_{GR} \vert^{\boldsymbol{\cdot}}}{\vert E_{GR} \vert} \left(\frac{3}{2} + \frac{\delta}{2} m_{g,S}^2 R^2 e^{-m_{g,s} R}  \right),
	\label{eq:decay of the orbital period}
\end{equation}
where we have assumed that $m_{g,S} R \ll 1$, which can be verified using Table $\ref{tab: PSR J1012+5307-1}$. We find $m_{g,S} R \approx 10^{-13}$.
Hence, the contribution from the second term in the bracket in Eq. \eqref{eq:decay of the orbital period} is negligible.

Fourth-order theories, such as $\mathcal{L} = f\left(R,R_{\mu\nu}\right)$, have been criticized for explaining galaxy rotation curves without dark matter, see \citep{2011StabileScelzaRotationcurvesgalaxiesPRD}. In this reference the authors state that for a fourth-order
theory, which includes squares of the Ricci tensor, the additional
Yukawa potential term always comes with the wrong sign, such that
it does not give additional but less attraction.

In the phenomenological approach of Sanders the Yukawa term also appears with the wrong sign, since $\delta$ is negative. Nevertheless, the reason why this approach is able to fit galaxy rotation curves without invoking dark matter is that also the gravitational constant $G$ is changed to an effective gravitational
constant $\tilde{G} =G/\left(1 + \delta\right)$. This procedure
has also been adopted in MCG \citep{faria2014massive} and in SVTG
\citep{2013MishraSinghFourthordergravityPRD}. However, in accordance with our findings, it is also criticized in \citep{2013StabileCapozzielloGalaxyrotationcurvesPRD}
that it is not clear how such a modified gravitational
potential as in \eqref{eq: Sanders solution} can emerge
from a standard matter source.

On top of that it also fails to explain the decrease of the orbital period of binary systems by gravitational radiation.

The ratio between the graviton mass and the orbital frequency of the binary system is 
$m_{g,S}/\omega_{s} \lesssim 10^{-8}$
and the radiated energy is to first-order in $m_g^2/\omega_s^2$
%\begin{widetext}
\begin{align}
\dot{E} \approx 
\frac{m_g^2}{8\omega_s^2 (1 + \delta)}\dot{E}_{GR},
\end{align}
which is much smaller than $\dot{E}_{GR}$.
%\end{widetext}

\vspace{-0.5cm}

\begin{thebibliography}{95}%
\makeatletter
\providecommand \@ifxundefined [1]{%
 \@ifx{#1\undefined}
}%
\providecommand \@ifnum [1]{%
 \ifnum #1\expandafter \@firstoftwo
 \else \expandafter \@secondoftwo
 \fi
}%
\providecommand \@ifx [1]{%
 \ifx #1\expandafter \@firstoftwo
 \else \expandafter \@secondoftwo
 \fi
}%
\providecommand \natexlab [1]{#1}%
\providecommand \enquote  [1]{``#1''}%
\providecommand \bibnamefont  [1]{#1}%
\providecommand \bibfnamefont [1]{#1}%
\providecommand \citenamefont [1]{#1}%
\providecommand \href@noop [0]{\@secondoftwo}%
\providecommand \href [0]{\begingroup \@sanitize@url \@href}%
\providecommand \@href[1]{\@@startlink{#1}\@@href}%
\providecommand \@@href[1]{\endgroup#1\@@endlink}%
\providecommand \@sanitize@url [0]{\catcode `\\12\catcode `\$12\catcode
  `\&12\catcode `\#12\catcode `\^12\catcode `\_12\catcode `\%12\relax}%
\providecommand \@@startlink[1]{}%
\providecommand \@@endlink[0]{}%
\providecommand \url  [0]{\begingroup\@sanitize@url \@url }%
\providecommand \@url [1]{\endgroup\@href {#1}{\urlprefix }}%
\providecommand \urlprefix  [0]{URL }%
\providecommand \Eprint [0]{\href }%
\providecommand \doibase [0]{http://dx.doi.org/}%
\providecommand \selectlanguage [0]{\@gobble}%
\providecommand \bibinfo  [0]{\@secondoftwo}%
\providecommand \bibfield  [0]{\@secondoftwo}%
\providecommand \translation [1]{[#1]}%
\providecommand \BibitemOpen [0]{}%
\providecommand \bibitemStop [0]{}%
\providecommand \bibitemNoStop [0]{.\EOS\space}%
\providecommand \EOS [0]{\spacefactor3000\relax}%
\providecommand \BibitemShut  [1]{\csname bibitem#1\endcsname}%
\let\auto@bib@innerbib\@empty
%</preamble>

\bibitem [{\citenamefont {Will}(2014)}]{Will2014}%
  \BibitemOpen
  \bibfield  {author} {\bibinfo {author} {\bibfnamefont {C.~M.}\ \bibnamefont
  {Will}},\ }\href {http://doi.org/10.12942/lrr-2014-4} {\bibfield  {journal} {\bibinfo  {journal} {Living
  Rev. Rel.}\ }\textbf {\bibinfo {volume} {17}},\ \bibinfo {pages} {4}
  (\bibinfo {year} {2014})}\BibitemShut {NoStop}%
\bibitem [{\citenamefont {Kramer}\ \emph {et~al.}(2006)\citenamefont {Kramer},
  \citenamefont {Stairs}, \citenamefont {Manchester}, \citenamefont
  {McLaughlin}, \citenamefont {Lyne}, \citenamefont {Ferdman}, \citenamefont
  {Burgay}, \citenamefont {Lorimer}, \citenamefont {Possenti}, \citenamefont
  {D'Amico} \emph
  {et~al.}}]{2006Sci...314...97K}%
  \BibitemOpen
  \bibfield  {author} {\bibinfo {author} {\bibfnamefont {M.}~\bibnamefont
  {Kramer}}, \bibinfo {author} {\bibfnamefont {I.~H.}\ \bibnamefont {Stairs}},
  \bibinfo {author} {\bibfnamefont {R.}~\bibnamefont {Manchester}},  \emph {et~al.},\ }\href {http://doi.org/10.1126/science.1132305 } {\bibfield
  {journal} {\bibinfo  {journal} {Science}\ }\textbf {\bibinfo {volume}
  {314}},\ \bibinfo {pages} {97} (\bibinfo {year} {2006})}\BibitemShut
  {NoStop}%
\bibitem [{\citenamefont {Weisberg}\ and\ \citenamefont
  {Taylor}(2004)}]{2004WeisbergTaylorRelativisticbinarypulsarapa}%
  \BibitemOpen
  \bibfield  {author} {\bibinfo {author} {\bibfnamefont {J.~M.}\ \bibnamefont
  {Weisberg}}\ and\ \bibinfo {author} {\bibfnamefont {J.~H.}\ \bibnamefont
  {Taylor}},\ }\href@noop {} {\bibfield  {journal} {\bibinfo  {journal} \Eprint{https://arxiv.org/pdf/astro-ph/0407149.pdf} {astro-ph/0407149}\ }}\BibitemShut {NoStop}%
\bibitem [{\citenamefont {Abbott}\ \emph
  {et~al.}(2016{\natexlab{a}})\citenamefont {Abbott}, \citenamefont {Abbott},
  \citenamefont {Abbott}, \citenamefont {Abernathy}, \citenamefont {Acernese},
  \citenamefont {Ackley}, \citenamefont {Adams}, \citenamefont {Adams},
  \citenamefont {Addesso}, \citenamefont {Adhikari} \emph
  {et~al.}}]{abbott2016observation}%
  \BibitemOpen
  \bibfield  {author} {\bibinfo {author} {\bibfnamefont {B.}~\bibnamefont
  {Abbott}}, \bibinfo {author} {\bibfnamefont {R.}~\bibnamefont {Abbott}},
  \bibinfo {author} {\bibfnamefont {T.}~\bibnamefont {Abbott}}, \bibinfo
  {author} {\bibfnamefont {M.}~\bibnamefont {Abernathy}}, \bibinfo {author}
  {\bibfnamefont {F.}~\bibnamefont {Acernese}}, \bibinfo {author}
  {\bibfnamefont {K.}~\bibnamefont {Ackley}}, \bibinfo {author} {\bibfnamefont
  {C.}~\bibnamefont {Adams}}, \bibinfo {author} {\bibfnamefont
  {T.}~\bibnamefont {Adams}}, \bibinfo {author} {\bibfnamefont
  {P.}~\bibnamefont {Addesso}}, \bibinfo {author} {\bibfnamefont
  {R.}~\bibnamefont {Adhikari}},  \emph {et~al.},\ }\href {http://dx.doi.org/10.1103/PhysRevLett.116.241102} {\bibfield
  {journal} {\bibinfo  {journal} {Phys. Rev. Lett.}\ }\textbf {\bibinfo
  {volume} {116}},\ \bibinfo {pages} {061102} (\bibinfo {year}
  {2016}{\natexlab{a}})}\BibitemShut {NoStop}%
\bibitem [{\citenamefont {Abbott}\ \emph
  {et~al.}(2016{\natexlab{b}})\citenamefont {Abbott}, \citenamefont {Abbott},
  \citenamefont {Abbott}, \citenamefont {Abernathy}, \citenamefont {Acernese},
  \citenamefont {Ackley}, \citenamefont {Adams}, \citenamefont {Adams},
  \citenamefont {Addesso}, \citenamefont {Adhikari} \emph
  {et~al.}}]{2016_Abbott_GW151226_Observationofgravitationalwavesfroma22-solar-massbinaryblackholecoalescence_PRL}%
  \BibitemOpen
  \bibfield  {author} {\bibinfo {author} {\bibfnamefont {B.}~\bibnamefont
  {Abbott}}, \bibinfo {author} {\bibfnamefont {R.}~\bibnamefont {Abbott}},
  \bibinfo {author} {\bibfnamefont {T.}~\bibnamefont {Abbott}}, \bibinfo
  {author} {\bibfnamefont {M.}~\bibnamefont {Abernathy}}, \bibinfo {author}
  {\bibfnamefont {F.}~\bibnamefont {Acernese}}, \bibinfo {author}
  {\bibfnamefont {K.}~\bibnamefont {Ackley}}, \bibinfo {author} {\bibfnamefont
  {C.}~\bibnamefont {Adams}}, \bibinfo {author} {\bibfnamefont
  {T.}~\bibnamefont {Adams}}, \bibinfo {author} {\bibfnamefont
  {P.}~\bibnamefont {Addesso}}, \bibinfo {author} {\bibfnamefont
  {R.}~\bibnamefont {Adhikari}},  \emph {et~al.},\ }\href {https://doi.org/10.1103/PhysRevLett.116.241103} {\bibfield
  {journal} {\bibinfo  {journal} {Phys. Rev. Lett.}\ }\textbf {\bibinfo
  {volume} {116}},\ \bibinfo {pages} {241103} (\bibinfo {year}
  {2016}{\natexlab{b}})}\BibitemShut {NoStop}%
\bibitem [{\citenamefont {Scientific}\ \emph {et~al.}(2017)\citenamefont
  {Scientific}, \citenamefont {Abbott}, \citenamefont {Abbott}, \citenamefont
  {Abbott}, \citenamefont {Acernese}, \citenamefont {Ackley}, \citenamefont
  {Adams}, \citenamefont {Adams}, \citenamefont {Addesso}, \citenamefont
  {Adhikari} \emph
  {et~al.}}]{2017_Scientific_GW170104_Observationofa50-Solar-MassBinaryBlackHoleCoalescenceatRedshift0.2_PRL}%
  \BibitemOpen
  \bibfield  {author} {\bibinfo {author} {\bibfnamefont {L.}~\bibnamefont
  {Scientific}}, \bibinfo {author} {\bibfnamefont {B.}~\bibnamefont {Abbott}},
  \bibinfo {author} {\bibfnamefont {R.}~\bibnamefont {Abbott}}, \bibinfo
  {author} {\bibfnamefont {T.}~\bibnamefont {Abbott}}, \bibinfo {author}
  {\bibfnamefont {F.}~\bibnamefont {Acernese}}, \bibinfo {author}
  {\bibfnamefont {K.}~\bibnamefont {Ackley}}, \bibinfo {author} {\bibfnamefont
  {C.}~\bibnamefont {Adams}}, \bibinfo {author} {\bibfnamefont
  {T.}~\bibnamefont {Adams}}, \bibinfo {author} {\bibfnamefont
  {P.}~\bibnamefont {Addesso}}, \bibinfo {author} {\bibfnamefont
  {R.}~\bibnamefont {Adhikari}},  \emph {et~al.},\ }\href {https://doi.org/10.1103/PhysRevLett.118.221101} {\bibfield
  {journal} {\bibinfo  {journal} {Phys. Rev. Lett.}\ }\textbf {\bibinfo
  {volume} {118}},\ \bibinfo {pages} {221101} (\bibinfo {year}
  {2017})}\BibitemShut {NoStop}%
\bibitem [{\citenamefont {Abbott}\ \emph
  {et~al.}(2017{\natexlab{a}})\citenamefont {Abbott}, \citenamefont {Abbott},
  \citenamefont {Abbott}, \citenamefont {Acernese}, \citenamefont {Ackley},
  \citenamefont {Adams}, \citenamefont {Adams}, \citenamefont {Addesso},
  \citenamefont {Adhikari}, \citenamefont {Adya} \emph
  {et~al.}}]{2017_Abbott_GW170608_Observationofa19-solar-massBinaryBlackHoleCoalescence_}%
  \BibitemOpen
  \bibfield  {author} {\bibinfo {author} {\bibfnamefont {B.}~\bibnamefont
  {Abbott}}, \bibinfo {author} {\bibfnamefont {R.}~\bibnamefont {Abbott}},
  \bibinfo {author} {\bibfnamefont {T.}~\bibnamefont {Abbott}}, \bibinfo
  {author} {\bibfnamefont {F.}~\bibnamefont {Acernese}}, \bibinfo {author}
  {\bibfnamefont {K.}~\bibnamefont {Ackley}}, \bibinfo {author} {\bibfnamefont
  {C.}~\bibnamefont {Adams}}, \bibinfo {author} {\bibfnamefont
  {T.}~\bibnamefont {Adams}}, \bibinfo {author} {\bibfnamefont
  {P.}~\bibnamefont {Addesso}}, \bibinfo {author} {\bibfnamefont
  {R.}~\bibnamefont {Adhikari}}, \bibinfo {author} {\bibfnamefont
  {V.}~\bibnamefont {Adya}},  \emph {et~al.},\ }\href {https://doi.org/10.3847/2041-8213/aa9f0c} {\bibfield
  {journal} {\bibinfo  {journal} {Astrophys. J. Lett.}\ }\textbf {\bibinfo
  {volume} {851}},\ \bibinfo {pages} {L35} (\bibinfo {year}
  {2017}{\natexlab{a}})}\BibitemShut {NoStop}%
\bibitem [{\citenamefont {Abbott}\ \emph
  {et~al.}(2017{\natexlab{b}})\citenamefont {Abbott}, \citenamefont {Abbott},
  \citenamefont {Abbott}, \citenamefont {Acernese}, \citenamefont {Ackley},
  \citenamefont {Adams}, \citenamefont {Adams}, \citenamefont {Addesso},
  \citenamefont {Adhikari}, \citenamefont {Adya} \emph
  {et~al.}}]{2017_Abbott_GW170814_Athree-detectorobservationofgravitationalwavesfromabinaryblackholecoalescence_PRL}%
  \BibitemOpen
  \bibfield  {author} {\bibinfo {author} {\bibfnamefont {B.~P.}\ \bibnamefont
  {Abbott}}, \bibinfo {author} {\bibfnamefont {R.}~\bibnamefont {Abbott}},
  \bibinfo {author} {\bibfnamefont {T.}~\bibnamefont {Abbott}}, \bibinfo
  {author} {\bibfnamefont {F.}~\bibnamefont {Acernese}}, \bibinfo {author}
  {\bibfnamefont {K.}~\bibnamefont {Ackley}}, \bibinfo {author} {\bibfnamefont
  {C.}~\bibnamefont {Adams}}, \bibinfo {author} {\bibfnamefont
  {T.}~\bibnamefont {Adams}}, \bibinfo {author} {\bibfnamefont
  {P.}~\bibnamefont {Addesso}}, \bibinfo {author} {\bibfnamefont
  {R.}~\bibnamefont {Adhikari}}, \bibinfo {author} {\bibfnamefont
  {V.}~\bibnamefont {Adya}},  \emph {et~al.},\ }\href {https://doi.org/10.1103/PhysRevLett.119.141101} {\bibfield
  {journal} {\bibinfo  {journal} {Phys. Rev. Lett.}\ }\textbf {\bibinfo
  {volume} {119}},\ \bibinfo {pages} {141101} (\bibinfo {year}
  {2017}{\natexlab{b}})}\BibitemShut {NoStop}%
\bibitem [{\citenamefont {Abbott}\ \emph
  {et~al.}(2017{\natexlab{c}})\citenamefont {Abbott}, \citenamefont {Abbott},
  \citenamefont {Abbott}, \citenamefont {Acernese}, \citenamefont {Ackley},
  \citenamefont {Adams}, \citenamefont {Adams}, \citenamefont {Addesso},
  \citenamefont {Adhikari}, \citenamefont {Adya} \emph
  {et~al.}}]{2017_Abbott_GW170817_PRL}%
  \BibitemOpen
  \bibfield  {author} {\bibinfo {author} {\bibfnamefont {B.~P.}\ \bibnamefont
  {Abbott}}, \bibinfo {author} {\bibfnamefont {R.}~\bibnamefont {Abbott}},
  \bibinfo {author} {\bibfnamefont {T.}~\bibnamefont {Abbott}}, \bibinfo
  {author} {\bibfnamefont {F.}~\bibnamefont {Acernese}}, \bibinfo {author}
  {\bibfnamefont {K.}~\bibnamefont {Ackley}}, \bibinfo {author} {\bibfnamefont
  {C.}~\bibnamefont {Adams}}, \bibinfo {author} {\bibfnamefont
  {T.}~\bibnamefont {Adams}}, \bibinfo {author} {\bibfnamefont
  {P.}~\bibnamefont {Addesso}}, \bibinfo {author} {\bibfnamefont
  {R.}~\bibnamefont {Adhikari}}, \bibinfo {author} {\bibfnamefont
  {V.}~\bibnamefont {Adya}},  \emph {et~al.},\ }\href {https://doi.org/10.1103/PhysRevLett.119.161101} {\bibfield
  {journal} {\bibinfo  {journal} {Phys. Rev. Lett.}\ }\textbf {\bibinfo
  {volume} {119}},\ \bibinfo {pages} {161101} (\bibinfo {year}
  {2017}{\natexlab{c}})}\BibitemShut {NoStop}%
\bibitem [{\citenamefont {Abbott}\ \emph
  {et~al.}(2017{\natexlab{d}})\citenamefont {Abbott}, \citenamefont {Abbott},
  \citenamefont {Adhikari}, \citenamefont {Ananyeva}, \citenamefont {Anderson},
  \citenamefont {Appert}, \citenamefont {Arai}, \citenamefont {Araya},
  \citenamefont {Barayoga}, \citenamefont {Barish} \emph
  {et~al.}}]{2017_Abbott_Multi-messengerobservationsofabinaryneutronstarmerger_AJL}%
  \BibitemOpen
  \bibfield  {author} {\bibinfo {author} {\bibfnamefont {B.}~\bibnamefont
  {Abbott}}, \bibinfo {author} {\bibfnamefont {R.}~\bibnamefont {Abbott}},
  \bibinfo {author} {\bibfnamefont {R.}~\bibnamefont {Adhikari}}, \bibinfo
  {author} {\bibfnamefont {A.}~\bibnamefont {Ananyeva}}, \bibinfo {author}
  {\bibfnamefont {S.}~\bibnamefont {Anderson}}, \bibinfo {author}
  {\bibfnamefont {S.}~\bibnamefont {Appert}}, \bibinfo {author} {\bibfnamefont
  {K.}~\bibnamefont {Arai}}, \bibinfo {author} {\bibfnamefont {M.}~\bibnamefont
  {Araya}}, \bibinfo {author} {\bibfnamefont {J.}~\bibnamefont {Barayoga}},
  \bibinfo {author} {\bibfnamefont {B.}~\bibnamefont {Barish}},  \emph
  {et~al.},\ }\href {https://doi.org/10.3847/2041-8213/aa91c9} {\bibfield  {journal} {\bibinfo  {journal}
  {Astrophys. J. Lett.}\ }\textbf {\bibinfo {volume} {848}},\
  \bibinfo {pages} {L12} (\bibinfo {year} {2017}{\natexlab{d}})}\BibitemShut
  {NoStop}%
\bibitem [{\citenamefont {Abbott}\ \emph
  {et~al.}(2017{\natexlab{e}})\citenamefont {Abbott}, \citenamefont {Abbott},
  \citenamefont {Abbott}, \citenamefont {Acernese}, \citenamefont {Ackley},
  \citenamefont {Adams}, \citenamefont {Adams}, \citenamefont {Addesso},
  \citenamefont {Adhikari}, \citenamefont {V.~B.~Adya} \emph
  {et~al.}}]{2017_Abbott_GravitationalWavesandGammaRaysfromaBinary}%
  \BibitemOpen
  \bibfield  {author} {\bibinfo {author} {\bibfnamefont {B.~P.}\ \bibnamefont
  {Abbott}}, \bibinfo {author} {\bibfnamefont {R.}~\bibnamefont {Abbott}},
  \bibinfo {author} {\bibfnamefont {T.~D.}\ \bibnamefont {Abbott}}, \bibinfo
  {author} {\bibfnamefont {F.}~\bibnamefont {Acernese}}, \bibinfo {author}
  {\bibfnamefont {K.}~\bibnamefont {Ackley}}, \bibinfo {author} {\bibfnamefont
  {C.}~\bibnamefont {Adams}}, \bibinfo {author} {\bibfnamefont
  {T.}~\bibnamefont {Adams}}, \bibinfo {author} {\bibfnamefont
  {P.}~\bibnamefont {Addesso}}, \bibinfo {author} {\bibfnamefont {R.~X.}\
  \bibnamefont {Adhikari}}, \bibinfo {author} {\bibfnamefont {V.}~\bibnamefont
  {V.~B.~Adya}},  \emph {et~al.},\ }\href
  {http://stacks.iop.org/2041-8205/848/i=2/a=L13} {\bibfield  {journal}
  {\bibinfo  {journal} {Astrophys. J. Lett.}\ }\textbf {\bibinfo
  {volume} {848}},\ \bibinfo {pages} {L13} (\bibinfo {year}
  {2017}{\natexlab{e}})}\BibitemShut {NoStop}%
\bibitem [{\citenamefont {Lombriser}\ and\ \citenamefont
  {Lima}(2017)}]{2017_Lombriser_ChallengestoSelfAccelerationinModifiedGravity}%
  \BibitemOpen
  \bibfield  {author} {\bibinfo {author} {\bibfnamefont {L.}~\bibnamefont
  {Lombriser}}\ and\ \bibinfo {author} {\bibfnamefont {N.~A.}\ \bibnamefont
  {Lima}},\ }\href {https://doi.org/10.1016/j.physletb.2016.12.048} {\bibfield  {journal} {\bibinfo  {journal} {Phys.
  Lett. B}\ }\textbf {\bibinfo {volume} {765}},\ \bibinfo {pages} {382}
  (\bibinfo {year} {2017})},\ \Eprint {https://arxiv.org/abs/1602.07670} {astro-ph.CO/1602.07670}\BibitemShut {NoStop}%  
\bibitem [{\citenamefont {Baker}\ \emph {et~al.}(2017)\citenamefont {Baker},
  \citenamefont {Bellini}, \citenamefont {Ferreira}, \citenamefont {Lagos},
  \citenamefont {Noller},\ and\ \citenamefont
  {Sawicki}}]{2017_Baker_StrongConstraintsonCosmologicalGravityfromGW170817andGRB170817A_Prl}%
  \BibitemOpen
  \bibfield  {author} {\bibinfo {author} {\bibfnamefont {T.}~\bibnamefont
  {Baker}}, \bibinfo {author} {\bibfnamefont {E.}~\bibnamefont {Bellini}},
  \bibinfo {author} {\bibfnamefont {P.~G.}\ \bibnamefont {Ferreira}}, \bibinfo
  {author} {\bibfnamefont {M.}~\bibnamefont {Lagos}}, \bibinfo {author}
  {\bibfnamefont {J.}~\bibnamefont {Noller}}, \ and\ \bibinfo {author}
  {\bibfnamefont {I.}~\bibnamefont {Sawicki}},\ }\href {https://doi.org/10.1103/PhysRevLett.119.251301} {\bibfield
  {journal} {\bibinfo  {journal} {Phys. Rev. Lett.}\ }\textbf {\bibinfo
  {volume} {119}},\ \bibinfo {pages} {251301} (\bibinfo {year}
  {2017})}\BibitemShut {NoStop}%
\bibitem [{\citenamefont {Creminelli}\ and\ \citenamefont
  {Vernizzi}(2017)}]{2017_Creminelli_DarkEnergyAfterGW170817andGRB170817A_PRL}%
  \BibitemOpen
  \bibfield  {author} {\bibinfo {author} {\bibfnamefont {P.}~\bibnamefont
  {Creminelli}}\ and\ \bibinfo {author} {\bibfnamefont {F.}~\bibnamefont
  {Vernizzi}},\ }\href {\doibase 10.1103/PhysRevLett.119.251302} {\bibfield
  {journal} {\bibinfo  {journal} {Phys. Rev. Lett.}\ }\textbf {\bibinfo
  {volume} {119}},\ \bibinfo {pages} {251302} (\bibinfo {year} {2017})},\
  \Eprint {http://arxiv.org/abs/1710.05877} {astro-ph.CO/1710.05877}
  \BibitemShut {NoStop}%
%%CITATION = ARXIV:1710.05877;%%
\bibitem [{\citenamefont {Ezquiaga}\ and\ \citenamefont
  {Zumalacárregui}(2017)}]{2017_Ezquiaga_DarkEnergyAfterGW170817DeadEndsandtheRoadAhead_PRL}%
  \BibitemOpen
  \bibfield  {author} {\bibinfo {author} {\bibfnamefont {J.~M.}\ \bibnamefont
  {Ezquiaga}}\ and\ \bibinfo {author} {\bibfnamefont {M.}~\bibnamefont
  {Zumalacárregui}},\ }\href {\doibase 10.1103/PhysRevLett.119.251304}
  {\bibfield  {journal} {\bibinfo  {journal} {Phys. Rev. Lett.}\ }\textbf
  {\bibinfo {volume} {119}},\ \bibinfo {pages} {251304} (\bibinfo {year}
  {2017})},\ \Eprint {http://arxiv.org/abs/1710.05901} {astro-ph.CO/1710.05901} \BibitemShut {NoStop}%
%%CITATION = ARXIV:1710.05901;%%
\bibitem [{\citenamefont {Sakstein}\ and\ \citenamefont
  {Jain}(2017)}]{2017_Sakstein_ImplicationsoftheNeutronStarMergerGW170817forCosmologicalScalarTensorTheories_Prl}%
  \BibitemOpen
  \bibfield  {author} {\bibinfo {author} {\bibfnamefont {J.}~\bibnamefont
  {Sakstein}}\ and\ \bibinfo {author} {\bibfnamefont {B.}~\bibnamefont
  {Jain}},\ }\href {\doibase https://doi.org/10.1103/PhysRevLett.119.251303}
  {\bibfield  {journal} {\bibinfo  {journal} {Phys. Rev. Lett.}\ }\textbf
  {\bibinfo {volume} {119}},\ \bibinfo {pages} {251303} (\bibinfo {year}
  {2017})},\ \Eprint {https://arxiv.org/abs/1710.05893} {astro-ph.CO/1710.05893} \BibitemShut {NoStop}%
\bibitem [{\citenamefont {Nersisyan}\ \emph {et~al.}(2018)\citenamefont
  {Nersisyan}, \citenamefont {Lima},\ and \citenamefont
  {Amendola}}]{2018_Nersisyan_GravitationalWaveSpeedImplicationsforModelswithoutaMassScale_}%
  \BibitemOpen
  \bibfield  {author} {\bibinfo {author} {\bibfnamefont {H.}~\bibnamefont
  {Nersisyan}}, \bibinfo {author} {\bibfnamefont {N.~A.}\ \bibnamefont {Lima}},\ and \bibinfo {author} {\bibfnamefont {L.}~\bibnamefont {Amendola}},\
  }\Eprint
  {http://arxiv.org/abs/1801.06683} {astro-ph.CO/1801.06683}
  \BibitemShut {NoStop}%
%%CITATION = ARXIV:1801.06683;%%
\bibitem [{\citenamefont {Akrami}\ \emph {et~al.}(2018)\citenamefont {Akrami},
  \citenamefont {Brax}, \citenamefont {Davis},\ and\ \citenamefont
  {Vardanyan}}]{2018_Akrami_NeutronStarMergerGW170817StronglyConstrainsDoublyCoupledBigravity_apa}%
  \BibitemOpen
  \bibfield  {author} {\bibinfo {author} {\bibfnamefont {Y.}~\bibnamefont
  {Akrami}}, \bibinfo {author} {\bibfnamefont {P.}~\bibnamefont {Brax}},
  \bibinfo {author} {\bibfnamefont {A.-C.}\ \bibnamefont {Davis}},\ and\
  \bibinfo {author} {\bibfnamefont {V.}~\bibnamefont {Vardanyan}},\ } \Eprint {https://arxiv.org/abs/1803.09726} {astro-ph.CO/1803.09726} \BibitemShut {NoStop}%
\bibitem [{\citenamefont {Ostrogradski}(1850)}]{1850OstrogradskiMASP}%
  \BibitemOpen
  \bibfield  {author} {\bibinfo {author} {\bibfnamefont {M.}~\bibnamefont
  {Ostrogradski}},\ }\href@noop {} {\bibfield  {journal} {\bibinfo  {journal}
  {Mem. Ac. St. Petersbourg}\ }\textbf {\bibinfo {volume} {VI}},\ \bibinfo
  {pages} {385} (\bibinfo {year} {1850})}\BibitemShut {NoStop}%
\bibitem [{\citenamefont {Woodard}(2007)}]{2007WoodardAvoidingDarkEnergy}%
  \BibitemOpen
  \bibfield  {author} {\bibinfo {author} {\bibfnamefont {R.}~\bibnamefont
  {Woodard}},\ }\enquote {\bibinfo {title} {Avoiding dark energy with 1/r
  modifications of gravity},}\ in\ \href {\doibase
  10.1007/978-3-540-71013-4_14} {\emph {\bibinfo {booktitle} {The Invisible
  Universe: Dark Matter and Dark Energy}}},\ \bibinfo {editor} {edited by\
  \bibinfo {editor} {\bibfnamefont {L.}~\bibnamefont {Papantonopoulos}}}\
  (\bibinfo  {publisher} {Springer Berlin Heidelberg},\ \bibinfo {year} {2007})\ pp.\ \bibinfo {pages}
  {403--433}\BibitemShut {NoStop}%
\bibitem [{\citenamefont {Weyl}(1919)}]{1919WeylEineneueErweiterungAdPL}%
  \BibitemOpen
  \bibfield  {author} {\bibinfo {author} {\bibfnamefont {H.}~\bibnamefont
  {Weyl}},\ }\href {https://doi.org/10.1002/andp.19193641002} {\bibfield  {journal} {\bibinfo  {journal} {Ann.
  Phys. (Leipzig)}\ }\textbf {\bibinfo {volume} {59}},\ \bibinfo {pages} {101}
  (\bibinfo {year} {1919})}\BibitemShut {NoStop}%
\bibitem [{\citenamefont
  {Bach}(1921)}]{1921BachZurWeylschenrelativitatstheorieMZ}%
  \BibitemOpen
  \bibfield  {author} {\bibinfo {author} {\bibfnamefont {R.}~\bibnamefont
  {Bach}},\ }\href {https://link.springer.com/content/pdf/10.1007/BF01378338.pdf} {\bibfield  {journal} {\bibinfo  {journal}
  {Math. Zeitschr.}\ }\textbf {\bibinfo {volume} {9}},\ \bibinfo
  {pages} {110} (\bibinfo {year} {1921})}\BibitemShut {NoStop}%
\bibitem [{\citenamefont {Mannheim}\ and\ \citenamefont
  {Kazanas}(1989{\natexlab{a}})}]{1989MannheimKazanasExactvacuumsolutionTAJ}%
  \BibitemOpen
  \bibfield  {author} {\bibinfo {author} {\bibfnamefont {P.~D.}\ \bibnamefont
  {Mannheim}}\ and\ \bibinfo {author} {\bibfnamefont {D.}~\bibnamefont
  {Kazanas}},\ }\href {http://adsabs.harvard.edu/full/1989ApJ...342..635M} {\bibfield  {journal} {\bibinfo  {journal} {Astrophys. J.}\ }\textbf {\bibinfo {volume} {342}},\ \bibinfo {pages}
  {635} (\bibinfo {year} {1989}{\natexlab{a}})}\BibitemShut {NoStop}%
\bibitem [{\citenamefont
  {Mannheim}(1990)}]{1990MannheimConformalcosmologynoGRaG}%
  \BibitemOpen
  \bibfield  {author} {\bibinfo {author} {\bibfnamefont {P.~D.}\ \bibnamefont
  {Mannheim}},\ }\href {https://doi.org/10.1007/BF00756278} {\bibfield  {journal} {\bibinfo  {journal}
  {Gen. Relativ. Gravit.}\ }\textbf {\bibinfo {volume} {22}},\
  \bibinfo {pages} {289} (\bibinfo {year} {1990})}\BibitemShut {NoStop}%
\bibitem [{\citenamefont {Kazanas}\ and\ \citenamefont
  {Mannheim}(1991)}]{1991KazanasMannheimGeneralstructuregravitationalTAJSS}%
  \BibitemOpen
  \bibfield  {author} {\bibinfo {author} {\bibfnamefont {D.}~\bibnamefont
  {Kazanas}}\ and\ \bibinfo {author} {\bibfnamefont {P.~D.}\ \bibnamefont
  {Mannheim}},\ }\href {http://adsabs.harvard.edu/full/1991ApJS...76..431K_2} {\bibfield  {journal} {\bibinfo  {journal} {Astrophys. J., Suppl. Ser.}\ }\textbf {\bibinfo {volume} {76}},\
  \bibinfo {pages} {431} (\bibinfo {year} {1991})}\BibitemShut {NoStop}%
\bibitem [{\citenamefont {Mannheim}\ and\ \citenamefont
  {Kazanas}(1991)}]{1991MannheimKazanasSolutionsReissnerNordstromPRD}%
  \BibitemOpen
  \bibfield  {author} {\bibinfo {author} {\bibfnamefont {P.~D.}\ \bibnamefont
  {Mannheim}}\ and\ \bibinfo {author} {\bibfnamefont {D.}~\bibnamefont
  {Kazanas}},\ }\href {https://doi.org/10.1103/PhysRevD.44.417} {\bibfield  {journal} {\bibinfo  {journal}
  {Phys. Rev. D}\ }\textbf {\bibinfo {volume} {44}},\ \bibinfo {pages}
  {417} (\bibinfo {year} {1991})}\BibitemShut {NoStop}%
\bibitem [{\citenamefont
  {Mannheim}(1993)}]{1993MannheimLinearpotentialsgalacticapa}%
  \BibitemOpen
  \bibfield  {author} {\bibinfo {author} {\bibfnamefont {P.~D.}\ \bibnamefont
  {Mannheim}},\ }\href {http://adsabs.harvard.edu/full/1993ApJ...419..150M} {\bibfield  {journal} {\bibinfo  {journal}
  {Astrophys. J.}\ }\textbf {\bibinfo {volume}
  {419}},\ \bibinfo {pages} {150} (\bibinfo {year} {1993})},\ \Eprint {https://arxiv.org/pdf/astro-ph/9307004.pdf} {astro-ph/9307004}\BibitemShut {NoStop}%
\bibitem [{\citenamefont {Mannheim}(2006)}]{Mannheim2006}%
  \BibitemOpen
  \bibfield  {author} {\bibinfo {author} {\bibfnamefont {P.~D.}\ \bibnamefont
  {Mannheim}},\ }\href {https://doi.org/10.1016/j.ppnp.2005.08.001} {\bibfield  {journal} {\bibinfo  {journal}
  {Prog. Part. Nucl. Phys.}\ }\textbf {\bibinfo {volume}
  {56}},\ \bibinfo {pages} {340} (\bibinfo {year} {2006})}\BibitemShut
  {NoStop}%
\bibitem [{\citenamefont {Mannheim}(2011)}]{Mannheim2011}%
  \BibitemOpen
  \bibfield  {author} {\bibinfo {author} {\bibfnamefont {P.~D.}\ \bibnamefont
  {Mannheim}},\ }\href {https://doi.org/10.1007/s10714-010-1088-z} {\bibfield  {journal} {\bibinfo  {journal}
  {Gen. Relativ. Gravit.}\ }\textbf {\bibinfo {volume} {43}},\
  \bibinfo {pages} {703} (\bibinfo {year} {2011})}\BibitemShut {NoStop}%
\bibitem [{\citenamefont
  {'t~Hooft}(2010{\natexlab{a}})}]{2010HooftConformalConstraintCanonical}%
  \BibitemOpen
  \bibfield  {author} {\bibinfo {author} {\bibfnamefont {G.}~\bibnamefont
  {'t~Hooft}},\ }\href@noop {}
  \Eprint {http://arxiv.org/abs/1011.0061} {gr-qc/1011.0061}
  \BibitemShut {NoStop}%
%%CITATION = ARXIV:1011.0061;%%
\bibitem [{\citenamefont
  {'t~Hooft}(2010{\natexlab{b}})}]{2010HooftProbingsmalldistance}%
  \BibitemOpen
  \bibfield  {author} {\bibinfo {author} {\bibfnamefont {G.}~\bibnamefont
  {'t~Hooft}},\ }\href@noop {}
  \Eprint {http://arxiv.org/abs/1009.0669} {gr-qc/1009.0669}
  \BibitemShut {NoStop}%
%%CITATION = ARXIV:1009.0669;%%
\bibitem [{\citenamefont
  {'t~Hooft}(2011)}]{2011HooftclasselementaryparticleFP}%
  \BibitemOpen
  \bibfield  {author} {\bibinfo {author} {\bibfnamefont {G.}~\bibnamefont
  {'t~Hooft}},\ }\href {\doibase 10.1007/s10701-011-9586-8} {\bibfield
  {journal} {\bibinfo  {journal} {Found. Phys.}\ }\textbf {\bibinfo {volume}
  {41}},\ \bibinfo {pages} {1829} (\bibinfo {year} {2011})},\ \Eprint
  {http://arxiv.org/abs/1104.4543} {gr-qc/1104.4543} \BibitemShut
  {NoStop}%
%%CITATION = ARXIV:1104.4543;%%
\bibitem [{\citenamefont {Mannheim}(2012{\natexlab{a}})}]{mannheim2012making}%
  \BibitemOpen
  \bibfield  {author} {\bibinfo {author} {\bibfnamefont {P.~D.}\ \bibnamefont
  {Mannheim}},\ }\href@noop {} {\bibfield  {journal} {\bibinfo  {journal}
  {Found. Phys.}\ }\textbf {\bibinfo {volume} {42}},\ \bibinfo
  {pages} {388} (\bibinfo {year} {2012}{\natexlab{a}})}\BibitemShut {NoStop}%
\bibitem [{\citenamefont
  {Maldacena}(2011)}]{2011MaldacenaEinsteinGravityConformal}%
  \BibitemOpen
  \bibfield  {author} {\bibinfo {author} {\bibfnamefont {J.}~\bibnamefont
  {Maldacena}},\ }\href@noop {} \Eprint
  {http://arxiv.org/abs/1105.5632} {hep-th/1105.5632} \BibitemShut
  {NoStop}%
%%CITATION = ARXIV:1105.5632;%%
\bibitem [{\citenamefont {Anastasiou}\ and\ \citenamefont
  {Olea}(2016)}]{2016_Anastasiou_FromConformaltoEinsteinGravity_PRD}%
  \BibitemOpen
  \bibfield  {author} {\bibinfo {author} {\bibfnamefont {G.}~\bibnamefont
  {Anastasiou}}\ and\ \bibinfo {author} {\bibfnamefont {R.}~\bibnamefont
  {Olea}},\ }\href {\doibase https://doi.org/10.1103/PhysRevD.94.086008} {\bibfield  {journal} {\bibinfo  {journal} {Phys.
  Rev. D}\ }\textbf {\bibinfo {volume} {94}},\ \bibinfo {pages} {086008}
  (\bibinfo {year} {2016})},\ \Eprint {https://arxiv.org/abs/1608.07826} {hep-th/1608.07826v2}\BibitemShut {NoStop}%
\bibitem [{\citenamefont {Mannheim}\ and\ \citenamefont
  {Kazanas}(1989{\natexlab{b}})}]{mannheim1989exact}%
  \BibitemOpen
  \bibfield  {author} {\bibinfo {author} {\bibfnamefont {P.~D.}\ \bibnamefont
  {Mannheim}}\ and\ \bibinfo {author} {\bibfnamefont {D.}~\bibnamefont
  {Kazanas}},\ }\href {http://adsabs.harvard.edu/full/1989ApJ...342..635M} {\bibfield  {journal} {\bibinfo  {journal} {Astrophys. J.}\ }\textbf {\bibinfo {volume} {342}},\ \bibinfo {pages}
  {635} (\bibinfo {year} {1989}{\natexlab{b}})}\BibitemShut {NoStop}%
\bibitem [{\citenamefont {Mannheim}\ and\ \citenamefont
  {Kazanas}(1994)}]{1994MannheimKazanasNewtonianlimitconformalGrag}%
  \BibitemOpen
  \bibfield  {author} {\bibinfo {author} {\bibfnamefont {P.~D.}\ \bibnamefont
  {Mannheim}}\ and\ \bibinfo {author} {\bibfnamefont {D.}~\bibnamefont
  {Kazanas}},\ }\href {https://doi.org/10.1007/BF02105226} {\bibfield  {journal} {\bibinfo  {journal}
  {Gen. Relativ. Gravit.}\ }\textbf {\bibinfo {volume} {26}},\
  \bibinfo {pages} {337} (\bibinfo {year} {1994})}\BibitemShut {NoStop}%
\bibitem [{\citenamefont {Deliduman}\ \emph {et~al.}(2015)\citenamefont
  {Deliduman}, \citenamefont {Kasikci},\ and\ \citenamefont
  {Yapiskan}}]{2015DelidumanKasikciYapiskanFlatGalacticRotation}%
  \BibitemOpen
  \bibfield  {author} {\bibinfo {author} {\bibfnamefont {C.} \bibnamefont
  {Deliduman}}, \bibinfo {author} {\bibfnamefont {O.} \bibnamefont {Kasikci}},
  \ and\ \bibinfo {author} {\bibfnamefont {B.} \bibnamefont {Yapiskan}},\
  }\\ \href@noop {} \Eprint
  {http://arxiv.org/abs/1511.07731} {gr-qc/1511.07731} \BibitemShut
  {NoStop}%
%%CITATION = ARXIV:1511.07731;%%
\bibitem [{\citenamefont
  {Mannheim}(1997)}]{1997MannheimAregalacticrotationTAJ}%
  \BibitemOpen
  \bibfield  {author} {\bibinfo {author} {\bibfnamefont {P.~D.}\ \bibnamefont
  {Mannheim}},\ }\href {http://iopscience.iop.org/article/10.1086/303933/pdf} {\bibfield  {journal} {\bibinfo  {journal} {Astrophys. J.}\ }\textbf {\bibinfo {volume} {479}},\ \bibinfo {pages}
  {659} (\bibinfo {year} {1997})}\BibitemShut {NoStop}%
\bibitem [{\citenamefont {Mannheim}\ and\ \citenamefont
  {O'Brien}(2012)}]{mannheim2012fitting}%
  \BibitemOpen
  \bibfield  {author} {\bibinfo {author} {\bibfnamefont {P.~D.}\ \bibnamefont
  {Mannheim}}\ and\ \bibinfo {author} {\bibfnamefont {J.~G.}\ \bibnamefont
  {O'Brien}},\ }\href {https://doi.org/10.1103/PhysRevD.85.124020} {\bibfield  {journal} {\bibinfo  {journal}
  {Phys. Rev. D}\ }\textbf {\bibinfo {volume} {85}},\ \bibinfo {pages}
  {124020} (\bibinfo {year} {2012})}\BibitemShut {NoStop}%
\bibitem [{\citenamefont {Mannheim}\ and\ \citenamefont
  {O'Brien}(2013)}]{2013MannheimOBrienGalacticrotationcurves}%
  \BibitemOpen
  \bibfield  {author} {\bibinfo {author} {\bibfnamefont {P.~D.}\ \bibnamefont
  {Mannheim}}\ and\ \bibinfo {author} {\bibfnamefont {J.~G.}\ \bibnamefont
  {O'Brien}},\ }\bibfield  {booktitle} {\bibinfo {booktitle} {J. Phys. Conf. Ser.}}\\ \href {http://doi.org/10.1088/1742-6596/437/1/012002} {\ \textbf {\bibinfo {volume}
  {437}},\ \bibinfo {pages} {012002} (\bibinfo {year} {2013})}\BibitemShut
  {NoStop}%
\bibitem [{\citenamefont
  {Mannheim}(1992)}]{1992MannheimConformalgravityflatnessTAJ}%
  \BibitemOpen
  \bibfield  {author} {\bibinfo {author} {\bibfnamefont {P.~D.}\ \bibnamefont
  {Mannheim}},\ }\href {https://doi.org/10.1086/171358} {\bibfield  {journal} {\bibinfo  {journal} {Astrophys. J.}\ }\textbf {\bibinfo {volume} {391}},\ \bibinfo {pages}
  {429} (\bibinfo {year} {1992})}\BibitemShut {NoStop}%
\bibitem [{\citenamefont
  {Mannheim}(1999)}]{1999_Mannheim_ConformalGravityandaNaturalSolutiontotheCosmologicalConstantProblem_apa}%
  \BibitemOpen
  \bibfield  {author} {\bibinfo {author} {\bibfnamefont {P.~D.}\ \bibnamefont
  {Mannheim}},\ }\href@noop {} {\bibfield  {journal} {\bibinfo  {journal}
  \Eprint{https://arxiv.org/pdf/astro-ph/9901219.pdf} {astro-ph/9901219}\ }}\BibitemShut
  {NoStop}%
\bibitem [{\citenamefont {Diaferio}\ \emph {et~al.}(2011)\citenamefont
  {Diaferio}, \citenamefont {Ostorero},\ and\ \citenamefont
  {Cardone}}]{2011_Diaferio_GammaRayBurstsAsCosmologicalProbes}%
  \BibitemOpen
  \bibfield  {author} {\bibinfo {author} {\bibfnamefont {A.}~\bibnamefont
  {Diaferio}}, \bibinfo {author} {\bibfnamefont {L.}~\bibnamefont {Ostorero}},
  \ and\ \bibinfo {author} {\bibfnamefont {V.}~\bibnamefont {Cardone}},\
  }\href {http://dx.doi.org/10.1088/1475-7516/2011/10/008} {\bibfield  {journal} {\bibinfo  {journal} {J. Cos. Astropart. Phys.}\ }\textbf {\bibinfo {volume} {2011}},\
  \bibinfo {pages} {008} (\bibinfo {year} {2011})}\BibitemShut {NoStop}%
\bibitem [{\citenamefont
  {Mannheim}(2012{\natexlab{b}})}]{2012MannheimCosmologicalperturbationsconformalPRD}%
  \BibitemOpen
  \bibfield  {author} {\bibinfo {author} {\bibfnamefont {P.~D.}\ \bibnamefont
  {Mannheim}},\ }\href {https://doi.org/10.1103/PhysRevD.85.124008} {\bibfield  {journal} {\bibinfo  {journal}
  {Phys. Rev. D}\ }\textbf {\bibinfo {volume} {85}},\ \bibinfo {pages}
  {124008} (\bibinfo {year} {2012}{\natexlab{b}})}\BibitemShut {NoStop}%
\bibitem [{\citenamefont {Roberts}\ \emph {et~al.}(2017)\citenamefont
  {Roberts}, \citenamefont {Horne}, \citenamefont {Hodson},\ and\ \citenamefont
  {Leggat}}]{2017_Roberts_TestsofLambdaCDMandConformalGravity}%
  \BibitemOpen
  \bibfield  {author} {\bibinfo {author} {\bibfnamefont {C.}~\bibnamefont
  {Roberts}}, \bibinfo {author} {\bibfnamefont {K.}~\bibnamefont {Horne}},
  \bibinfo {author} {\bibfnamefont {A.~O.}\ \bibnamefont {Hodson}}, \ and\
  \bibinfo {author} {\bibfnamefont {A.~D.}\ \bibnamefont {Leggat}},\
  }\href@noop {} {\bibfield  {journal} {\bibinfo  {journal} \Eprint{https://arxiv.org/pdf/1711.10369.pdf} {astro-ph.CO/1711.10369}\ }}\BibitemShut {NoStop}%
\bibitem [{\citenamefont {Knox}\ and\ \citenamefont
  {Kosowsky}(1993)}]{1993KnoxKosowskyPrimordialnucleosynthesisconformal}%
  \BibitemOpen
  \bibfield  {author} {\bibinfo {author} {\bibfnamefont {L.}~\bibnamefont
  {Knox}}\ and\ \bibinfo {author} {\bibfnamefont {A.}~\bibnamefont
  {Kosowsky}},\ }\href@noop {} \Eprint
  {http://arxiv.org/abs/astro-ph/9311006} {astro-ph/9311006}
  \BibitemShut {NoStop}%
%%CITATION = ASTRO-PH/9311006;%%
\bibitem [{\citenamefont {Elizondo}\ and\ \citenamefont
  {Yepes}(1994)}]{1994_Elizondo_CanConformalWeylGravityBeConsideredaViableCosmologicalTheory}%
  \BibitemOpen
  \bibfield  {author} {\bibinfo {author} {\bibfnamefont {D.}~\bibnamefont
  {Elizondo}}\ and\ \bibinfo {author} {\bibfnamefont {G.}~\bibnamefont
  {Yepes}},\ }\href {\doibase 10.1086/174214} {\bibfield  {journal} {\bibinfo
  {journal} {Astrophys. J.}\ }\textbf {\bibinfo {volume} {428}},\ \bibinfo
  {pages} {17} (\bibinfo {year} {1994})},\ \Eprint
  {http://arxiv.org/abs/astro-ph/9312064} {astro-ph/9312064}
  \BibitemShut {NoStop}%
%%CITATION = ASTRO-PH/9312064;%%
\bibitem [{\citenamefont {Edery}\ and\ \citenamefont
  {Paranjape}(1998)}]{1998EderyParanjapeClassicaltestsWeylPRD}%
  \BibitemOpen
  \bibfield  {author} {\bibinfo {author} {\bibfnamefont {A.}~\bibnamefont
  {Edery}}\ and\ \bibinfo {author} {\bibfnamefont {M.~B.}\ \bibnamefont
  {Paranjape}},\ }\href {https://doi.org/10.1103/PhysRevD.58.024011} {\bibfield  {journal} {\bibinfo  {journal}
  {Phys. Rev. D}\ }\textbf {\bibinfo {volume} {58}},\ \bibinfo {pages}
  {024011} (\bibinfo {year} {1998})}\BibitemShut {NoStop}%
\bibitem [{\citenamefont {Edery}\ \emph {et~al.}(2001)\citenamefont {Edery},
  \citenamefont {Methot},\ and\ \citenamefont
  {Paranjape}}]{2001EderyMethotParanjapeGaugechoicegeodeticGRG}%
  \BibitemOpen
  \bibfield  {author} {\bibinfo {author} {\bibfnamefont {A.}~\bibnamefont
  {Edery}}, \bibinfo {author} {\bibfnamefont {A.~A.}\ \bibnamefont {Methot}}, \
  and\ \bibinfo {author} {\bibfnamefont {M.~B.}\ \bibnamefont {Paranjape}},\
  }\href {\doibase 10.1023/A:1013011312648} {\bibfield  {journal} {\bibinfo
  {journal} {Gen. Relativ. Gravit.}\ }\textbf {\bibinfo {volume} {33}},\ \bibinfo
  {pages} {2075} (\bibinfo {year} {2001})},\ \Eprint
  {http://arxiv.org/abs/astro-ph/0006173} {astro-ph/0006173}
  \BibitemShut {NoStop}%
%%CITATION = ASTRO-PH/0006173;%%
\bibitem [{\citenamefont {Sultana}\ and\ \citenamefont
  {Kazanas}(2010)}]{2010SultanaKazanasBendinglightconformalPRD}%
  \BibitemOpen
  \bibfield  {author} {\bibinfo {author} {\bibfnamefont {J.}~\bibnamefont
  {Sultana}}\ and\ \bibinfo {author} {\bibfnamefont {D.}~\bibnamefont
  {Kazanas}},\ }\href {https://doi.org/10.1103/PhysRevD.81.127502} {\bibfield  {journal} {\bibinfo  {journal}
  {Phys. Rev. D}\ }\textbf {\bibinfo {volume} {81}},\ \bibinfo {pages}
  {127502} (\bibinfo {year} {2010})}\BibitemShut {NoStop}%
\bibitem [{\citenamefont {Sultana}\ \emph {et~al.}(2012)\citenamefont
  {Sultana}, \citenamefont {Kazanas},\ and\ \citenamefont
  {Said}}]{2012SultanaKazanasSaidConformalWeylgravityPRD}%
  \BibitemOpen
  \bibfield  {author} {\bibinfo {author} {\bibfnamefont {J.}~\bibnamefont
  {Sultana}}, \bibinfo {author} {\bibfnamefont {D.}~\bibnamefont {Kazanas}}, \
  and\ \bibinfo {author} {\bibfnamefont {J.~L.}\ \bibnamefont {Said}},\
  }\href {https://doi.org/10.1103/PhysRevD.86.084008} {\bibfield  {journal} {\bibinfo  {journal} {Phys. Rev.
  D}\ }\textbf {\bibinfo {volume} {86}},\ \bibinfo {pages} {084008} (\bibinfo
  {year} {2012})}\BibitemShut {NoStop}%
\bibitem [{\citenamefont {Cattani}\ \emph {et~al.}(2013)\citenamefont
  {Cattani}, \citenamefont {Scalia}, \citenamefont {Laserra}, \citenamefont
  {Bochicchio},\ and\ \citenamefont
  {Nandi}}]{2013CattaniScaliaLaserraBochicchioNandiCorrectlightdeflectionPRD}%
  \BibitemOpen
  \bibfield  {author} {\bibinfo {author} {\bibfnamefont {C.}~\bibnamefont
  {Cattani}}, \bibinfo {author} {\bibfnamefont {M.}~\bibnamefont {Scalia}},
  \bibinfo {author} {\bibfnamefont {E.}~\bibnamefont {Laserra}}, \bibinfo
  {author} {\bibfnamefont {I.}~\bibnamefont {Bochicchio}}, \ and\ \bibinfo
  {author} {\bibfnamefont {K.~K.}\ \bibnamefont {Nandi}},\ }\href {https://doi.org/10.1103/PhysRevD.87.047503}
  {\bibfield  {journal} {\bibinfo  {journal} {Phys. Rev. D}\ }\textbf
  {\bibinfo {volume} {87}},\ \bibinfo {pages} {047503} (\bibinfo {year}
  {2013})}\BibitemShut {NoStop}%
\bibitem [{\citenamefont {Lim}\ and\ \citenamefont
  {Wang}(2016)}]{2016LimWangExactgravitationallensing}%
  \BibitemOpen
  \bibfield  {author} {\bibinfo {author} {\bibfnamefont {Y.-K.}\ \bibnamefont
  {Lim}}\ and\ \bibinfo {author} {\bibfnamefont {Q.-h.}\ \bibnamefont {Wang}},\
  }\href@noop {} \Eprint
  {http://arxiv.org/abs/1609.07633} {gr-qc/1609.07633} \BibitemShut
  {NoStop}%
%%CITATION = ARXIV:1609.07633;%%
\bibitem [{\citenamefont {Sultana}\ and\ \citenamefont
  {Kazanas}(2017)}]{2017SultanaKazanasGaugeChoiceConformalapa}%
  \BibitemOpen
  \bibfield  {author} {\bibinfo {author} {\bibfnamefont {J.}~\bibnamefont
  {Sultana}}\ and\ \bibinfo {author} {\bibfnamefont {D.}~\bibnamefont
  {Kazanas}},\ }\href@noop {} {\bibfield  {journal} {\bibinfo
  {journal} {Mon. Not. R. Astron. Soc.}\ }\textbf {\bibinfo {volume} {466}},\ \bibinfo
  {pages} {4847} (\bibinfo {year} {2017})};\ \Eprint {https://arxiv.org/pdf/1701.03192.pdf} {gr-qc/1701.03192} \BibitemShut{NoStop}%
\bibitem [{\citenamefont {Campigotto}\ \emph {et~al.}(2017)\citenamefont
  {Campigotto}, \citenamefont {Diaferio},\ and\ \citenamefont
  {Fatibene}}]{2017_Campigotto_ConformalGravity_LightDeflectionRevisitedandtheGalacticRotationCurveFailure_apa}%
  \BibitemOpen
  \bibfield  {author} {\bibinfo {author} {\bibfnamefont {M.}~\bibnamefont
  {Campigotto}}, \bibinfo {author} {\bibfnamefont {A.}~\bibnamefont
  {Diaferio}}, \ and\ \bibinfo {author} {\bibfnamefont {L.}~\bibnamefont
  {Fatibene}},\ }\href@noop {} {\bibfield  {journal} {\bibinfo  {journal} \Eprint{https://arxiv.org/pdf/1712.03969.pdf}
  {astro-ph/1712.03969}\ }}\BibitemShut
  {NoStop}%
\bibitem [{\citenamefont {Perlick}\ and\ \citenamefont
  {Xu}(1995)}]{1995_Perlick_MatchingExteriortoInteriorSolutionsinWeylGravity_Commenton``ExactVacuumSolutiontoConformalWeylGravityandGalacticRotationCurves_TAJ}%
  \BibitemOpen
  \bibfield  {author} {\bibinfo {author} {\bibfnamefont {V.}~\bibnamefont
  {Perlick}}\ and\ \bibinfo {author} {\bibfnamefont {C.}~\bibnamefont {Xu}},\
  }\href {https://doi.org/10.1086/176030} {\bibfield  {journal} {\bibinfo  {journal} {Astrophys. J.}\ }\textbf {\bibinfo {volume} {449}},\ \bibinfo {pages} {47}
  (\bibinfo {year} {1995})}\BibitemShut {NoStop}%
\bibitem [{\citenamefont {Wood}\ and\ \citenamefont
  {Moreau}(2001)}]{2001_Wood_Solutionsofconformalgravitywithdynamicalmassgenerationinthesolarsystem_apg}%
  \BibitemOpen
  \bibfield  {author} {\bibinfo {author} {\bibfnamefont {J.}~\bibnamefont
  {Wood}}\ and\ \bibinfo {author} {\bibfnamefont {W.}~\bibnamefont {Moreau}},\
  }\href@noop {} {\bibfield  {journal} {\bibinfo  {journal} \Eprint {https://arxiv.org/pdf/gr-qc/0102056.pdf} {gr-qc/0102056}\ }}\BibitemShut {NoStop}%
\bibitem [{\citenamefont {Myung}\ and\ \citenamefont
  {Moon}(2014)}]{2014_Myung_PrimordialGravitationalWavesfromConformalGravity_apa}%
  \BibitemOpen
  \bibfield  {author} {\bibinfo {author} {\bibfnamefont {Y.~S.}\ \bibnamefont
  {Myung}}\ and\ \bibinfo {author} {\bibfnamefont {T.}~\bibnamefont {Moon}},\
  }\href@noop {} {\bibfield  {journal} {\bibinfo  {journal} \Eprint {https://arxiv.org/pdf/1407.0441.pdf} {gr-qc/1407.0441}\ }}\BibitemShut {NoStop}%
\bibitem [{\citenamefont {Fabbri}\ and\ \citenamefont
  {Paranjape}(2011)}]{2011FabbriParanjapeMonochromaticplanefrontedPRD}%
  \BibitemOpen
  \bibfield  {author} {\bibinfo {author} {\bibfnamefont {L.}~\bibnamefont
  {Fabbri}}\ and\ \bibinfo {author} {\bibfnamefont {M.}~\bibnamefont
  {Paranjape}},\ }\href {https://doi.org/10.1103/PhysRevD.83.104046} {\bibfield  {journal} {\bibinfo  {journal}
  {Phys. Rev. D}\ }\textbf {\bibinfo {volume} {83}},\ \bibinfo {pages}
  {104046} (\bibinfo {year} {2011})}\BibitemShut {NoStop}%
\bibitem [{\citenamefont {Yang}(2017)}]{2017_Yang}%
  \BibitemOpen
  \bibfield  {author} {\bibinfo {author} {\bibfnamefont {R.-J.}\ \bibnamefont
  {Yang}},\ }\href@noop {} {\bibfield  {journal} {\bibinfo  {journal} \Eprint {https://arxiv.org/pdf/1710.10961.pdf} {gr-qc/1710.10961}\ }}\BibitemShut {NoStop}%
\bibitem [{\citenamefont {Stabile}\ and\ \citenamefont
  {Capozziello}(2015)}]{2015_Stabile_Post-MinkowskianLimitandGravitationalWavessolutionsofFourthOrderGravity_acompletestudy_apa}%
  \BibitemOpen
  \bibfield  {author} {\bibinfo {author} {\bibfnamefont {A.}~\bibnamefont
  {Stabile}}\ and\ \bibinfo {author} {\bibfnamefont {S.}~\bibnamefont
  {Capozziello}},\ }\href@noop {} {\bibfield  {journal} {\bibinfo  {journal} \Eprint {https://arxiv.org/pdf/1501.02187.pdf}
  {gr-qc/1501.02187}\ } }\BibitemShut
  {NoStop}%
\bibitem [{\citenamefont {Nelson}\ \emph
  {et~al.}(2010{\natexlab{a}})\citenamefont {Nelson}, \citenamefont {Ochoa},\
  and\ \citenamefont
  {Sakellariadou}}]{2010_Nelson_GravitationalWavesintheSpectralActionofNoncommutativeGeometry}%
  \BibitemOpen
  \bibfield  {author} {\bibinfo {author} {\bibfnamefont {W.}~\bibnamefont
  {Nelson}}, \bibinfo {author} {\bibfnamefont {J.}~\bibnamefont {Ochoa}}, \
  and\ \bibinfo {author} {\bibfnamefont {M.}~\bibnamefont {Sakellariadou}},\
  }\href {\doibase https://doi.org/10.1103/PhysRevD.82.085021} {\bibfield  {journal} {\bibinfo  {journal} {Phys. Rev.
  D}\ }\textbf {\bibinfo {volume} {82}},\ \bibinfo {pages} {085021} (\bibinfo
  {year} {2010}{\natexlab{a}})},\ \Eprint {https://arxiv.org/abs/1005.4276} {hep-th/1005.4276} \BibitemShut {NoStop}%
\bibitem [{\citenamefont {Nelson}\ \emph
  {et~al.}(2010{\natexlab{b}})\citenamefont {Nelson}, \citenamefont {Ochoa},\
  and\ \citenamefont
  {Sakellariadou}}]{2010_Nelson_ConstrainingtheNoncommutativeSpectralActionViaAstrophysicalObservations}%
  \BibitemOpen
  \bibfield  {author} {\bibinfo {author} {\bibfnamefont {W.}~\bibnamefont
  {Nelson}}, \bibinfo {author} {\bibfnamefont {J.}~\bibnamefont {Ochoa}}, \
  and\ \bibinfo {author} {\bibfnamefont {M.}~\bibnamefont {Sakellariadou}},\
  }\href {\doibase https://doi.org/10.1103/PhysRevLett.105.101602} {\bibfield  {journal} {\bibinfo  {journal} {Phys. Rev.
  Lett.}\ }\textbf {\bibinfo {volume} {105}},\ \bibinfo {pages} {101602}
  (\bibinfo {year} {2010}{\natexlab{b}})},\ \Eprint {https://arxiv.org/abs/1005.4279} {hep-th/1005.4279v2} \BibitemShut {NoStop}%
\bibitem [{\citenamefont {Weinberg}(1972)}]{weinberg1972gravitation}%
  \BibitemOpen
  \bibfield  {author} {\bibinfo {author} {\bibfnamefont {S.}~\bibnamefont
  {Weinberg}},\ }\href@noop {} {\emph {\bibinfo {title} {Gravitation and
  cosmology: principles and applications of GR}}}\ (\bibinfo  {publisher} {John
  Wiley \& Sons, Inc.},\ \bibinfo {year} {1972})\BibitemShut {NoStop}%
\bibitem [{\citenamefont {Lanczos}(1938)}]{lanczos1938}%
  \BibitemOpen
  \bibfield  {author} {\bibinfo {author} {\bibfnamefont {C.}~\bibnamefont
  {Lanczos}},\ }\href {http://doi.org/10.2307/1968467 } {\bibfield  {journal} {\bibinfo  {journal}
  {Ann. Math.}\,\ \bibinfo {pages} {842}} (\bibinfo {year}
  {1938})}\BibitemShut {NoStop}%
\bibitem [{sim()}]{simplifiednotation}%
  \BibitemOpen
  \href@noop {} {\bibinfo  {journal} {We have used simplified notation. The
  kinetic part of the fermionic energy-momentum tensor is given by
  $i\bar{\psi}\gamma^{\mu}(x)\left[\partial_{\mu} + \Gamma_{\mu}(x)\right]\psi
  - i\bar{\psi}\left[\overleftarrow{\partial_{\mu}} + \Gamma_{\mu}(x)\right]
  \gamma^{\mu}(x)\psi$ in order to be hermitian}\ }\BibitemShut {NoStop}%
\bibitem [{\citenamefont {Horne}(2016)}]{horne2016conformal}%
  \BibitemOpen
\bibfield  {journal} {  }\bibfield  {author} {\bibinfo {author} {\bibfnamefont
  {K.}~\bibnamefont {Horne}},\ }\href {https://doi.org/10.1093/mnras/stw506} {\bibfield  {journal} {\bibinfo
  {journal} {Mon. Not. Roy. Astron. Soc.}\ }\textbf
  {\bibinfo {volume} {458}},\ \bibinfo {pages} {4122} (\bibinfo {year}
  {2016})}\BibitemShut {NoStop}%
\bibitem [{\citenamefont {Faria}(2014)}]{faria2014massive}%
  \BibitemOpen
  \bibfield  {author} {\bibinfo {author} {\bibfnamefont {F.}~\bibnamefont
  {Faria}},\ }\href {http://dx.doi.org/10.1155/2014/520259} {\bibfield  {journal} {\bibinfo  {journal}
  {Adv. High Energy Phys.}\ }\textbf {\bibinfo {volume} {2014}}
  (\bibinfo {year} {2014})}\BibitemShut {NoStop}%
\bibitem [{\citenamefont
  {Faria}(2014{\natexlab{b}})}]{2014_Faria_CosmologyinMassiveConformalGravity_apa}%
  \BibitemOpen
  \bibfield  {author} {\bibinfo {author} {\bibfnamefont {F.}~\bibnamefont
  {Faria}},\ }\href {https://arxiv.org/abs/1410.5104} {\bibfield  {journal} {\bibinfo  {journal} {gr-qc/1410.5104}\ }}\BibitemShut {NoStop}%
\bibitem [{foo({\natexlab{a}})}]{footnoteMCG}%
  \BibitemOpen
  \href@noop {} {\bibfield  {journal} {\bibinfo  {journal} {In \citep{faria2014massive} there is only one scalar field coupled to the gravitational part of the theory, whereas in \citep{2014_Faria_CosmologyinMassiveConformalGravity_apa} a second scalar field coupled to the fermionic matter sector is introduced. Nevertheless, after choosing a Weyl gauge for the first scalar field and after a spontaneous breaking of Weyl symmetry for the second scalar field, they take constant values throughout spacetime. Effectively, this corresponds to $ \epsilon = +1 $ in this work}} {}}\BibitemShut {NoStop}%
\bibitem [{\citenamefont {Flanagan}(2006)}]{flanagan2006fourth}%
  \BibitemOpen
  \bibfield  {author} {\bibinfo {author} {\bibfnamefont {{\'E}.~{\'E}.}\
  \bibnamefont {Flanagan}},\ }\href {https://doi.org/10.1103/PhysRevD.74.023002} {\bibfield  {journal} {\bibinfo
  {journal} {Phys. Rev. D}\ }\textbf {\bibinfo {volume} {74}},\ \bibinfo
  {pages} {023002} (\bibinfo {year} {2006})}\BibitemShut {NoStop}%
\bibitem [{\citenamefont {Oda}(2015)}]{oda2015conformal}%
  \BibitemOpen
  \bibfield  {author} {\bibinfo {author} {\bibfnamefont {I.}~\bibnamefont
  {Oda}},\ }\href@noop {} {\bibfield  {journal} {\bibinfo  {journal} \Eprint{https://arxiv.org/pdf/1505.06760.pdf} {gr-qc/1505.06760}\ }}\BibitemShut {NoStop}%
\bibitem [{\citenamefont {Sbisa}(2014)}]{2014SbisaClassicalquantumghostsEJoP}%
  \BibitemOpen
  \bibfield  {author} {\bibinfo {author} {\bibfnamefont {F.}~\bibnamefont
  {Sbisa}},\ }\href {http://dx.doi.org/10.1088/0143-0807/36/1/015009} {\bibfield  {journal} {\bibinfo  {journal}
  {Eur. J. Phys.}\ }\textbf {\bibinfo {volume} {36}},\ \bibinfo
  {pages} {015009} (\bibinfo {year} {2014})}\BibitemShut {NoStop}%
\bibitem [{\citenamefont
  {Schmidt}(1985{\natexlab{a}})}]{1985SchmidtstaticsphericalsymmetricAN}%
  \BibitemOpen
  \bibfield  {author} {\bibinfo {author} {\bibfnamefont {H.-J.}\ \bibnamefont
  {Schmidt}},\ }\href {https://doi.org/10.1002/asna.2113060206} {\bibfield  {journal} {\bibinfo  {journal}
  {Astron. Nachr.}\ }\textbf {\bibinfo {volume} {306}},\ \bibinfo
  {pages} {67} (\bibinfo {year} {1985}{\natexlab{a}})}\BibitemShut {NoStop}%
\bibitem [{\citenamefont
  {Schmidt}(1985{\natexlab{b}})}]{1985_Schmidt_SolutionsoftheLinearizedBachEinsteinEquationintheStaticSphericallySymmetricCase_AN}%
  \BibitemOpen
  \bibfield  {author} {\bibinfo {author} {\bibfnamefont {H.-J.}\ \bibnamefont
  {Schmidt}},\ }\href {https://doi.org/10.1002/asna.2113060411} {\bibfield  {journal} {\bibinfo  {journal}
  {Astron. Nachr.}\ }\textbf {\bibinfo {volume} {306}},\ \bibinfo
  {pages} {231} (\bibinfo {year} {1985}{\natexlab{b}})}\BibitemShut {NoStop}%
\bibitem [{\citenamefont
  {Teyssandier}(1989)}]{1989_Teyssandier_LinearisedR+R2gravity_anewgaugeandnewsolutions_CaQG}%
  \BibitemOpen
  \bibfield  {author} {\bibinfo {author} {\bibfnamefont {P.}~\bibnamefont
  {Teyssandier}},\ }\href {http://iopscience.iop.org/article/10.1088/0264-9381/6/2/016/meta#artAbst} {\bibfield  {journal} {\bibinfo  {journal}
  {Class. Quant. Grav.}\ }\textbf {\bibinfo {volume} {6}},\ \bibinfo
  {pages} {219} (\bibinfo {year} {1989})}\BibitemShut {NoStop}%
\bibitem [{\citenamefont {Biswas}\ \emph {et~al.}(2013)\citenamefont {Biswas},
  \citenamefont {Koivisto},\ and\ \citenamefont
  {Mazumdar}}]{2013BiswasKoivistoMazumdarNonlocaltheoriesgravityapa}%
  \BibitemOpen
  \bibfield  {author} {\bibinfo {author} {\bibfnamefont {T.}~\bibnamefont
  {Biswas}}, \bibinfo {author} {\bibfnamefont {T.}~\bibnamefont {Koivisto}}, \
  and\ \bibinfo {author} {\bibfnamefont {A.}~\bibnamefont {Mazumdar}},\ \\
  }\href@noop {} {\bibfield  {journal} {\bibinfo  {journal} \Eprint{https://arxiv.org/pdf/1302.0532.pdf} {gr-qc/1302.0532}\ }}\BibitemShut {NoStop}%
\bibitem [{foo()}]{footnoteghost}%
  \BibitemOpen
  \href@noop {} {\bibinfo  {journal} {In
  \citep{2013BiswasKoivistoMazumdarNonlocaltheoriesgravityapa} it is unclear
  which sign conventions have been used and hence it is not apparent if CG or
  MCG is studied. Therefore it is not clear whether the signs in their
  Eq.~$(33)$ are in agreement with our result or not. The difference in the
  sign for the massless and massive part appears nevertheless}\ }\BibitemShut
  {NoStop}%
\bibitem [{\citenamefont {Pais}\ and\ \citenamefont
  {Uhlenbeck}(1950)}]{1950PaisUhlenbeckfieldtheoriesnonPR}%
  \BibitemOpen
\bibfield  {journal} {  }\bibfield  {author} {\bibinfo {author} {\bibfnamefont
  {A.}~\bibnamefont {Pais}}\ and\ \bibinfo {author} {\bibfnamefont
  {G.}~\bibnamefont {Uhlenbeck}},\ }\href {https://doi.org/10.1103/PhysRev.79.145} {\bibfield  {journal}
  {\bibinfo  {journal} {Phys. Rev.}\ }\textbf {\bibinfo {volume} {79}},\
  \bibinfo {pages} {145} (\bibinfo {year} {1950})}\BibitemShut {NoStop}%
\bibitem [{\citenamefont {Mannheim}\ and\ \citenamefont
  {Davidson}(2000)}]{2000MannheimDavidsonFourthordertheoriesaph}%
  \BibitemOpen
  \bibfield  {author} {\bibinfo {author} {\bibfnamefont {P.~D.}\ \bibnamefont
  {Mannheim}}\ and\ \bibinfo {author} {\bibfnamefont {A.}~\bibnamefont
  {Davidson}},\ }\href@noop {} {\bibfield  {journal} {\bibinfo  {journal} \Eprint{https://arxiv.org/pdf/hep-th/0001115.pdf}
  {hep-th/0001115}\ }}\BibitemShut
  {NoStop}%
\bibitem [{\citenamefont {Bender}(2007)}]{2007BenderMakingsensenonRoPiP}%
  \BibitemOpen
  \bibfield  {author} {\bibinfo {author} {\bibfnamefont {C.~M.}\ \bibnamefont
  {Bender}},\ }\href {http://dx.doi.org/10.1088/0034-4885/70/6/R03} {\bibfield  {journal} {\bibinfo  {journal}
  {Rep. Prog. Phys.}\ }\textbf {\bibinfo {volume} {70}},\
  \bibinfo {pages} {947} (\bibinfo {year} {2007})}\BibitemShut {NoStop}%
\bibitem [{\citenamefont {Bender}\ and\ \citenamefont
  {Mannheim}(2008{\natexlab{a}})}]{2008BenderMannheimExactlysolvablePPRD}%
  \BibitemOpen
  \bibfield  {author} {\bibinfo {author} {\bibfnamefont {C.~M.}\ \bibnamefont
  {Bender}}\ and\ \bibinfo {author} {\bibfnamefont {P.~D.}\ \bibnamefont
  {Mannheim}},\ }\href {https://doi.org/10.1103/PhysRevD.78.025022} {\bibfield  {journal} {\bibinfo  {journal}
  {Phys. Rev. D}\ }\textbf {\bibinfo {volume} {78}},\ \bibinfo {pages}
  {025022} (\bibinfo {year} {2008}{\natexlab{a}})}\BibitemShut {NoStop}%
\bibitem [{\citenamefont {Bender}\ and\ \citenamefont
  {Mannheim}(2008{\natexlab{b}})}]{2008BenderMannheimGivingghostJoPAMaT}%
  \BibitemOpen
  \bibfield  {author} {\bibinfo {author} {\bibfnamefont {C.~M.}\ \bibnamefont
  {Bender}}\ and\ \bibinfo {author} {\bibfnamefont {P.~D.}\ \bibnamefont
  {Mannheim}},\ }\href {http://dx.doi.org/10.1088/1751-8113/41/30/304018} {\bibfield  {journal} {\bibinfo  {journal}
  {J. Phys. A}\ }\textbf {\bibinfo
  {volume} {41}},\ \bibinfo {pages} {304018} (\bibinfo {year}
  {2008}{\natexlab{b}})}\BibitemShut {NoStop}%
\bibitem [{\citenamefont {Bender}\ and\ \citenamefont
  {Mannheim}(2008{\natexlab{c}})}]{2008BenderMannheimNoghosttheoremPRL}%
  \BibitemOpen
  \bibfield  {author} {\bibinfo {author} {\bibfnamefont {C.~M.}\ \bibnamefont
  {Bender}}\ and\ \bibinfo {author} {\bibfnamefont {P.~D.}\ \bibnamefont
  {Mannheim}},\ }\href {https://doi.org/10.1103/PhysRevLett.100.110402} {\bibfield  {journal} {\bibinfo  {journal}
  {Phys. Rev. Lett.}\ }\textbf {\bibinfo {volume} {100}},\ \bibinfo
  {pages} {110402} (\bibinfo {year} {2008}{\natexlab{c}})}\BibitemShut
  {NoStop}%
\bibitem [{\citenamefont
  {Mannheim}(2013)}]{2013MannheimPTsymmetryasPTotRSoLAMPaES}%
  \BibitemOpen
  \bibfield  {author} {\bibinfo {author} {\bibfnamefont {P.~D.}\ \bibnamefont
  {Mannheim}},\ }\href {http://doi.org/10.1098/rsta.2012.0060 } {\bibfield  {journal} {\bibinfo  {journal}
  {‎Philos. Trans. Royal Soc. A}\ }\textbf {\bibinfo {volume} {371}},\
  \bibinfo {pages} {20120060} (\bibinfo {year} {2013})}\BibitemShut {NoStop}%
\bibitem [{\citenamefont
  {Mannheim}(2015{\natexlab{a}})}]{2015MannheimAntilinearityRatherthanapa}%
  \BibitemOpen
  \bibfield  {author} {\bibinfo {author} {\bibfnamefont {P.~D.}\ \bibnamefont
  {Mannheim}},\ }\href@noop {} {\bibfield  {journal} {\bibinfo  {journal}
  \href@noop {} \Eprint
  {https://arxiv.org/abs/1512.04915} {hep-th/1512.04915}\ }}\BibitemShut {NoStop}%
\bibitem [{\citenamefont
  {Mannheim}(2016)}]{2016MannheimExtensionCPTtheoremPLB}%
  \BibitemOpen
  \bibfield  {author} {\bibinfo {author} {\bibfnamefont {P.~D.}\ \bibnamefont
  {Mannheim}},\ }\href {https://doi.org/10.1016/j.physletb.2015.12.033} {\bibfield  {journal} {\bibinfo  {journal}
  {Phys. Lett. B}\ }\textbf {\bibinfo {volume} {753}},\ \bibinfo {pages}
  {288} (\bibinfo {year} {2016})}\BibitemShut {NoStop}%
\bibitem [{\citenamefont
  {Stelle}(1977)}]{1977StelleRenormalizationhigherderivativePRD}%
  \BibitemOpen
  \bibfield  {author} {\bibinfo {author} {\bibfnamefont {K.}~\bibnamefont
  {Stelle}},\ }\href {https://doi.org/10.1103/PhysRevD.16.953} {\bibfield  {journal} {\bibinfo  {journal}
  {Phys. Rev. D}\ }\textbf {\bibinfo {volume} {16}},\ \bibinfo {pages}
  {953} (\bibinfo {year} {1977})}\BibitemShut {NoStop}%
\bibitem [{\citenamefont
  {Mannheim}(2015{\natexlab{b}})}]{2015MannheimLivingSupersymmetryConformalapa}%
  \BibitemOpen
  \bibfield  {author} {\bibinfo {author} {\bibfnamefont {P.~D.}\ \bibnamefont
  {Mannheim}},\ }\href {https://doi.org/10.1088/1361-6471/aa888e} {\bibfield  {journal} {\bibinfo  {journal} {J. Phys. G}\ }\textbf {\bibinfo {volume} {44}},\ \bibinfo {pages} {115003} (\bibinfo {year} {2017})}\BibitemShut {NoStop}%
\bibitem [{\citenamefont {Salvio}\ and\ \citenamefont
  {Strumia}(2014)}]{2014_Salvio_Agravity_JoHEP}%
  \BibitemOpen
  \bibfield  {author} {\bibinfo {author} {\bibfnamefont {A.}~\bibnamefont
  {Salvio}}\ and\ \bibinfo {author} {\bibfnamefont {A.}~\bibnamefont
  {Strumia}},\ }\href {https://doi.org/10.1007} {\bibfield  {journal} {\bibinfo  {journal}
  {J. High Energy Phys.}\ }\textbf {\bibinfo {volume} {2014}},\
  \bibinfo {pages} {80} (\bibinfo {year} {2014})}\BibitemShut {NoStop}%
\bibitem [{\citenamefont {Maggiore}(2008)}]{maggiore2008gravitational}%
  \BibitemOpen
  \bibfield  {author} {\bibinfo {author} {\bibfnamefont {M.}~\bibnamefont
  {Maggiore}},\ }\href@noop {} {\emph {\bibinfo {title} {Gravitational waves.
  Vol 1, Theory and Experiments}}}\ (\bibinfo  {publisher} {Oxford university press},\ \bibinfo {year} {2008})\BibitemShut
  {NoStop}%
\bibitem [{\citenamefont {Lazaridis}\ \emph {et~al.}(2009)\citenamefont
  {Lazaridis}, \citenamefont {Wex}, \citenamefont {Jessner}, \citenamefont
  {Kramer}, \citenamefont {Stappers}, \citenamefont {Janssen}, \citenamefont
  {Desvignes}, \citenamefont {Purver}, \citenamefont {Cognard}, \citenamefont
  {Theureau} \emph
  {et~al.}}]{2009LazaridisWexJessnerKramerStappersJanssenDesvignes}%
  \BibitemOpen
  \bibfield  {author} {\bibinfo {author} {\bibfnamefont {K.}~\bibnamefont
  {Lazaridis}}, \bibinfo {author} {\bibfnamefont {N.}~\bibnamefont {Wex}},
  \bibinfo {author} {\bibfnamefont {A.}~\bibnamefont {Jessner}}, \bibinfo
  {author} {\bibfnamefont {M.}~\bibnamefont {Kramer}}, \bibinfo {author}
  {\bibfnamefont {B.}~\bibnamefont {Stappers}}, \bibinfo {author}
  {\bibfnamefont {G.}~\bibnamefont {Janssen}}, \bibinfo {author} {\bibfnamefont
  {G.}~\bibnamefont {Desvignes}}, \bibinfo {author} {\bibfnamefont
  {M.}~\bibnamefont {Purver}}, \bibinfo {author} {\bibfnamefont
  {I.}~\bibnamefont {Cognard}}, \bibinfo {author} {\bibfnamefont
  {G.}~\bibnamefont {Theureau}},  \emph {et~al.},\ }\href {https://doi.org/10.1111/j.1365-2966.2009.15481.x} {\bibfield
  {journal} {\bibinfo  {journal} {Mon. Not. Roy. Astron. Soc.}\ }\textbf {\bibinfo {volume} {400}},\ \bibinfo {pages} {805}
  (\bibinfo {year} {2009})}\BibitemShut {NoStop}%
\bibitem [{\citenamefont {Nicastro}\ \emph {et~al.}(1995)\citenamefont
  {Nicastro}, \citenamefont {Lyne}, \citenamefont {Lorimer}, \citenamefont
  {Harrison}, \citenamefont {Bailes},\ and\ \citenamefont
  {Skidmore}}]{1995NicastroLyneLorimerHarrisonBailesSkidmorePSRJ10125307MNotRAS}%
  \BibitemOpen
  \bibfield  {author} {\bibinfo {author} {\bibfnamefont {L.}~\bibnamefont
  {Nicastro}}, \bibinfo {author} {\bibfnamefont {A.}~\bibnamefont {Lyne}},
  \bibinfo {author} {\bibfnamefont {D.}~\bibnamefont {Lorimer}}, \bibinfo
  {author} {\bibfnamefont {P.}~\bibnamefont {Harrison}}, \bibinfo {author}
  {\bibfnamefont {M.}~\bibnamefont {Bailes}}, \ and\ \bibinfo {author}
  {\bibfnamefont {B.}~\bibnamefont {Skidmore}},\ }\href {https://doi.org/10.1093/mnras/273.1.L68} {\bibfield
  {journal} {\bibinfo  {journal} {Mon. Not. Roy. Astron. Soc.}\ }\textbf {\bibinfo {volume} {273}},\ \bibinfo {pages} {L68}
  (\bibinfo {year} {1995})}\BibitemShut {NoStop}%
\bibitem [{\citenamefont {Callanan}\ \emph {et~al.}(1998)\citenamefont
  {Callanan}, \citenamefont {Garnavich},\ and\ \citenamefont
  {Koester}}]{1998CallananGarnavichKoestermassneutronstarMNotRAS}%
  \BibitemOpen
  \bibfield  {author} {\bibinfo {author} {\bibfnamefont {P.~J.}\ \bibnamefont
  {Callanan}}, \bibinfo {author} {\bibfnamefont {P.~M.}\ \bibnamefont
  {Garnavich}}, \ and\ \bibinfo {author} {\bibfnamefont {D.}~\bibnamefont
  {Koester}},\ }\href {https://doi.org/10.1046/j.1365-8711.1998.01634.x} {\bibfield  {journal} {\bibinfo  {journal}
  {Mon. Not. Roy. Astron. Soc. }\ }\textbf {\bibinfo
  {volume} {298}},\ \bibinfo {pages} {207} (\bibinfo {year}
  {1998})}\BibitemShut {NoStop}%
\bibitem [{\citenamefont
  {Mannheim}(2007)}]{2007MannheimSchwarzschildlimitconformalPRD}%
  \BibitemOpen
  \bibfield  {author} {\bibinfo {author} {\bibfnamefont {P.~D.}\ \bibnamefont
  {Mannheim}},\ }\href {https://doi.org/10.1103/PhysRevD.75.124006} {\bibfield  {journal} {\bibinfo  {journal}
  {Phys. Rev. D}\ }\textbf {\bibinfo {volume} {75}},\ \bibinfo {pages}
  {124006} (\bibinfo {year} {2007})}\BibitemShut {NoStop}%
\bibitem [{\citenamefont {Barabash}\ and\ \citenamefont
  {Shtanov}(1999)}]{1999BarabashShtanovNewtonianlimitconformalPRD}%
  \BibitemOpen
  \bibfield  {author} {\bibinfo {author} {\bibfnamefont {O.}~\bibnamefont
  {Barabash}}\ and\ \bibinfo {author} {\bibfnamefont {Y.~V.}\ \bibnamefont
  {Shtanov}},\ }\href {https://doi.org/10.1103/PhysRevD.75.124006} {\bibfield  {journal} {\bibinfo  {journal}
  {Phys. Rev. D}\ }\textbf {\bibinfo {volume} {60}},\ \bibinfo {pages}
  {064008} (\bibinfo {year} {1999})}\BibitemShut {NoStop}%
\bibitem [{\citenamefont {Barabash}\ and\ \citenamefont
  {Pyatkovska}(2007)}]{2007BarabashPyatkovskaWeakfieldlimitapa}%
  \BibitemOpen
  \bibfield  {author} {\bibinfo {author} {\bibfnamefont {O.}~\bibnamefont
  {Barabash}}\ and\ \bibinfo {author} {\bibfnamefont {H.}~\bibnamefont
  {Pyatkovska}},\ }\href@noop {} {\bibfield  {journal} {\bibinfo  {journal} \Eprint{https://arxiv.org/pdf/0709.1044.pdf}
  {astro-ph/0709.1044}\ }}\BibitemShut
  {NoStop}%
\bibitem [{\citenamefont
  {Faria}(2016)}]{2016FariaQuantummassiveconformalTEPJC}%
  \BibitemOpen
  \bibfield  {author} {\bibinfo {author} {\bibfnamefont {F.}~\bibnamefont
  {Faria}},\ }\href {https://doi.org/10.1140/epjc/s10052-016-4037-5} {\bibfield  {journal} {\bibinfo  {journal} {Eur. Phys. J. C}\ }\textbf {\bibinfo {volume} {76}},\ \bibinfo
  {pages} {1} (\bibinfo {year} {2016})}\BibitemShut {NoStop}%
\bibitem [{\citenamefont {Adelberger}\ \emph {et~al.}(2009)\citenamefont
  {Adelberger}, \citenamefont {Gundlach}, \citenamefont {Heckel}, \citenamefont
  {Hoedl}\ and\ \citenamefont
  {Schlamminger}}]{2009AdelbergerGundlachHeckelHoedlSchlammingerTorsionbalanceexperimentsPiPaNP}%
  \BibitemOpen
  \bibfield  {author} {\bibinfo {author} {\bibfnamefont {E.}~\bibnamefont
  {Adelberger}}, \bibinfo {author} {\bibfnamefont {J.}~\bibnamefont
  {Gundlach}}, \bibinfo {author} {\bibfnamefont {B.}~\bibnamefont {Heckel}},
  \bibinfo {author} {\bibfnamefont {S.}~\bibnamefont {Hoedl}}, \ and\ \bibinfo
  {author} {\bibfnamefont {S.}~\bibnamefont {Schlamminger}},\ }\href {https://doi.org/10.1016/j.ppnp.2008.08.002}
  {\bibfield  {journal} {\bibinfo  {journal} {Prog. Part. Nucl. Phys.}\ }\textbf {\bibinfo {volume} {62}},\ \bibinfo {pages} {102}
  (\bibinfo {year} {2009})}\BibitemShut {NoStop}%
\bibitem [{\citenamefont {Stabile}\ and\ \citenamefont
  {Scelza}(2011)}]{2011StabileScelzaRotationcurvesgalaxiesPRD}%
  \BibitemOpen
  \bibfield  {author} {\bibinfo {author} {\bibfnamefont {A.}~\bibnamefont
  {Stabile}}\ and\ \bibinfo {author} {\bibfnamefont {G.}~\bibnamefont
  {Scelza}},\ }\href {https://doi.org/10.1103/PhysRevD.84.124023} {\bibfield  {journal} {\bibinfo  {journal}
  {Phys. Rev. D}\ }\textbf {\bibinfo {volume} {84}},\ \bibinfo {pages}
  {124023} (\bibinfo {year} {2011})}\BibitemShut {NoStop}%
\bibitem [{\citenamefont
  {Sanders}(1990)}]{1990SandersMassdiscrepanciesgalaxiesTAaAR}%
  \BibitemOpen
  \bibfield  {author} {\bibinfo {author} {\bibfnamefont {R.}~\bibnamefont
  {Sanders}},\ }\href {https://doi.org/10.1007/BF00873540} {\bibfield  {journal} {\bibinfo  {journal} {Astron. Astrophys. Rev.}\ }\textbf {\bibinfo {volume} {2}},\
  \bibinfo {pages} {1} (\bibinfo {year} {1990})}\BibitemShut {NoStop}%
\bibitem [{\citenamefont {Mishra}\ and\ \citenamefont
  {Singh}(2013)}]{2013MishraSinghFourthordergravityPRD}%
  \BibitemOpen
  \bibfield  {author} {\bibinfo {author} {\bibfnamefont {P.}~\bibnamefont
  {Mishra}}\ and\ \bibinfo {author} {\bibfnamefont {T.~P.}\ \bibnamefont
  {Singh}},\ }\href {https://doi.org/10.1103/PhysRevD.88.104036} {\bibfield  {journal} {\bibinfo  {journal}
  {Phys. Rev. D}\ }\textbf {\bibinfo {volume} {88}},\ \bibinfo {pages}
  {104036} (\bibinfo {year} {2013})}\BibitemShut {NoStop}%
\bibitem [{\citenamefont {Stabile}\ and\ \citenamefont
  {Capozziello}(2013)}]{2013StabileCapozzielloGalaxyrotationcurvesPRD}%
  \BibitemOpen
  \bibfield  {author} {\bibinfo {author} {\bibfnamefont {A.}~\bibnamefont
  {Stabile}}\ and\ \bibinfo {author} {\bibfnamefont {S.}~\bibnamefont
  {Capozziello}},\ }\href {https://doi.org/10.1103/PhysRevD.87.064002} {\bibfield  {journal} {\bibinfo  {journal}
  {Phys. Rev. D}\ }\textbf {\bibinfo {volume} {87}},\ \bibinfo {pages}
  {064002} (\bibinfo {year} {2013})}\BibitemShut {NoStop}%
\end{thebibliography}
\end{document}